\documentclass[ALICE,manyauthors]{cernphprep}

\usepackage[comma,square,numbers,sort&compress]{natbib}
\usepackage{hyperref}
\usepackage{relsize}
\usepackage{lineno}
\usepackage{textcomp}
\newcommand{\roots}{\ensuremath{{\sqrt{s}}}~}
\newcommand{\rootsnn}{\ensuremath{{\sqrt{s_{\mathrm{NN}}}}}~}
\newcommand{\pt}{\ensuremath{p_{\rm T}}~}
\newcommand{\pp}{pp~}

\newcommand{\dEdx}{d$E$/d$x$~}
\newcommand{\sDETi}{\ensuremath{n^{\rm DET}_{\sigma_{\rm{i}}}}~}
\newcommand{\sTPCe}{\ensuremath{n^{\rm TPC}_{\sigma_{{\rm{e}}}}}~}
\newcommand{\sTPCpi}{\ensuremath{n^{\rm TPC}_{\sigma_{\pi}}}~}
\newcommand{\sITSe}{\ensuremath{n^{\rm ITS}_{\sigma_{{\rm{e}}}}}~}
\newcommand{\sTOFe}{\ensuremath{n^{\rm TOF}_{\sigma_{{\rm{e}}}}}~}
\newcommand{\mee}{\ensuremath{m_{{\rm ee}}}~}
\newcommand{\ptee}{\ensuremath{p_{\rm T,ee}}~}
\newcommand{\dcaee}{\ensuremath{{\rm{DCA_{\rm ee}}}}~}
\newcommand{\Spm}{{\ensuremath{ S}}~}
\newcommand{\Bpm}{{\ensuremath{ B}}~}
\newcommand{\fgOS}{{\ensuremath{ OS}}~}
\newcommand{\fgSS}{{\ensuremath{ SS}}~}

\newcommand{\GeVovercs}{GeV/\ensuremath{c^{2}}~}
\newcommand{\GeVoverc}{GeV/\ensuremath{c}~}
\newcommand{\ccbar}{\ensuremath{{\rm{c}}{\overline{{\rm{c}}}}}~}
\newcommand{\bbbar}{\ensuremath{{\rm{b}}{\overline{{\rm{b}}}}}~}
\newcommand{\ee}{\ensuremath{{\rm{e}}^{+}{\rm{e}}^{-}}~}
\newcommand{\effrec}{\ensuremath{\epsilon^{\rm ee}_{\rm rec}(m_{{\rm
        ee}},p_{{\rm
        T,ee}})}~}

\begin{document}%

%%%%%%%%%%%%%%%  Title page %%%%%%%%%%%%%%%%%%%%%%%%
\begin{titlepage}
\PHyear{2018}
\PHnumber{102}      % required, will be obtained from PH
\PHdate{9 May}  % required, will be obtained from PH
%

%%% Put your own title + short title here:
\title{Dielectron production in proton--proton collisions at $\sqrt{s}$ = 7\,TeV}
\ShortTitle{Dielectron production in proton--proton collisions at $\sqrt{s}$ = 7\,TeV}   % appears on right page headers

%%% Do not change the next lines
\Collaboration{ALICE Collaboration\thanks{See Appendix~\ref{app:collab} for the list of collaboration members}}
\ShortAuthor{ALICE Collaboration} % appears on left page headers, do not change

\begin{abstract}

The first measurement of e$^+$e$^-$ pair production at mid-rapidity ($|\eta_{{\rm e}}|$ $<$ 0.8) in pp collisions at $\sqrt{s} = 7$ TeV with ALICE at the LHC is presented. The dielectron production is studied as a function of the invariant mass ($m_{\rm ee}$ $<$ 3.3 GeV/$c^{2}$), the pair transverse momentum ($p_{\rm T,ee}$ $<$ 8 GeV/$c$), and the pair transverse impact parameter (DCA$_{{\rm ee}}$), i.e., the average distance of closest approach of the reconstructed electron and positron tracks to the collision vertex, normalised to its resolution. The results are compared with the expectations from a cocktail of known hadronic sources and are well described when PYTHIA is used to generate the heavy-flavour contributions. In the low-mass region (0.14 $<$ $m_{\rm ee}$ $<$ 1.1 GeV/$c^{2}$), prompt and non-prompt e$^+$e$^-$ sources can be separated via the DCA$_{\rm ee}$. In the intermediate-mass region (1.1 $<$ $m_{\rm ee}$ $<$ 2.7 GeV/$c^{2}$), a double-differential fit to the data in $m_{\rm ee}$ and $p_{\rm T,ee}$ and a fit of the DCA$_{\rm ee}$ distribution allow the total ${\rm c\overline c}$ and ${\rm b\overline b}$ cross sections to be extracted. Two different event generators, PYTHIA and POWHEG, can reproduce the shape of the two-dimensional $m_{\rm ee}$ and $p_{\rm T,ee}$ spectra, as well as the shape of the DCA$_{\rm ee}$ distribution, reasonably well. However, differences in the ${\rm c\overline c}$ and ${\rm b\overline b}$ cross sections are observed when using the generators to extrapolate to full phase space. Finally, the ratio of inclusive to decay photons is studied via the measurement of virtual direct photons in the transverse-momentum range 1 $<$ $p_{\rm T}$ $<$ 8 GeV/$c$. This is found to be unity within the statistical and systematic uncertainties and consistent with expectations from next-to-leading order perturbative quantum chromodynamic calculations.

\end{abstract}
\end{titlepage}
\setcounter{page}{2}

\newpage
% !TEX root = alicepreprint_CDS.tex
\section{Introduction}

The main goal of the ALICE~\cite{0954-3899-30-11-001, 0954-3899-32-10-001, 1748-0221-3-08-S08002} Collaboration is to study strongly-interacting matter at the high energy density and temperature reached in ultra-relativistic heavy-ion collisions at the CERN Large Hadron Collider (LHC). In such collisions, the formation of a deconfined state of quarks and gluons, the Quark$-$Gluon Plasma (QGP), is predicted by Quantum ChromoDynamic (QCD) calculations on the lattice~\cite{pQCD1,pQCD2,pQCD3,pQCD4,pQCD5}. Moreover, chiral symmetry is expected to be restored in the QGP phase~\cite{introa,introb}.

Electron-positron pairs are produced at all stages of the collision. Since they are not affected by the strong interaction, they can escape from the dense medium without final-state interaction, and are suited to probe the entire time evolution and dynamics of the system.
At low invariant masses of the dielectron pairs (\mee $<$ 1.1\,GeV/$c^{2}$), \ee spectra are sensitive to the properties of vector mesons $\rho$, $\omega$, and $\phi$ in the medium. The $\rho$ meson has a shorter lifetime ($\approx$ 1.3\,fm/$c$ in its rest frame) than that of the medium \mbox{($\approx$ 10\,fm/$c$~\cite{QGPformationtime})}. Therefore, its spectral function, which can be measured through its dielectron decay channel, is affected by the dense medium and the predicted restoration of chiral symmetry~\cite{chirala,chiralb,chiralberratum,chiralc}.
Thermal radiation emitted by the system, both during the partonic and hadronic phase~\cite{introa,introb}, contributes to the dielectron yield over a broad mass range. In the intermediate-mass region \mbox{(IMR, 1.1 $<$ \mee $<$ 2.7\,GeV/$c^{2}$)}, the measurement of thermal dielectrons from the QGP is very challenging at the LHC due to the dominant contribution of \ee pairs from semileptonic decays of charm and beauty hadrons, correlated through flavour conservation\footnote{These contributions are referred to as charm/beauty or $\rm{c}\overline{\rm{c}}$/$\rm{b}\overline{\rm{b}}$ contributions throughout this paper}.
The continuum produced by these decays is sensitive to the energy loss~\cite{Radiativea,Radiativeb,Colla,Collb,Collc} and the degree of thermalisation of charm and beauty quarks in the medium, as well as the heavy-quark hadronisation mechanism, e.g. recombination of heavy quarks with light quarks from the thermalised medium~\cite{hfflowa,hfflowb,hfflowc}. To single out the interesting signal characteristics of the QGP, it is crucial to understand the dielectron yield from primordial heavy-flavour production. The latter can be studied in proton--proton (pp) collisions, which serve as a reference for the analysis of heavy-ion collisions and provide a test for perturbative QCD (pQCD) calculations and Monte Carlo (MC) event generators. Complementary to single-electron or D meson measurements, the yield of correlated e$^{+}$e$^{-}$ pairs from charm-hadron decays contains information about kinematical correlations between the c and $\overline{{\rm{c}}}$ quarks, i.e.\,the production mechanisms, and is sensitive to soft heavy-flavour production.

At the Relativistic Heavy Ion Collider (RHIC), the PHENIX and STAR Collaborations found that the dielectron production in \pp collisions at \roots $=$ 200\,GeV is well described by a cocktail of expected hadronic sources~\cite{phenixpp,phenixa,starpp}. In addition, PHENIX measured the total \ccbar and \bbbar cross sections in \pp and d-Au collisions at \rootsnn $=$ 200\,GeV by fitting the spectra of \ee pairs from heavy-flavour hadron decays simultaneously in \mee and pair transverse momentum $p_{\rm T,ee}$~\cite{phenixhfee,phenixhfeee}.
At this energy and in the PHENIX acceptance, the yield from correlated pairs from beauty-hadron decays dominates across all mass regions for \ptee $>$ 2.5\,GeV/$c$, whereas the \ccbar contribution is preeminent for \mee $<$ 3\,\GeVovercs and \ptee $<$ 2\,GeV/$c$. The extraction of the heavy-flavour cross sections, in particular the total \ccbar cross section, depends nevertheless on the event generator used to extrapolate the measurements to full phase space.
Finally, at lower masses (\mee $<$ 0.3\,GeV/$c^{2}$) and high \ptee (\ptee $>$ 1\,GeV/$c$), i.e.\,the quasi-real virtual-photon region where the \ptee of the dilepton pair is much larger than its mass ($p_{{\rm T,ee}}^{2}$ $\gg$ $m_{\rm ee}^{{\rm 2}}$), the measured \ee yield was used to study the production of virtual direct photons.
 The corresponding yield of real direct photons in \pp and d-Au collisions is reproduced by next-to-leading order perturbative quantum chromodynamic (NLO pQCD) calculations~\cite{phenixa,phenixdAudirect}. At the LHC, no significant signal of direct photons for $p_{{\rm T}} <$ 16\,GeV/$c$ could be extracted from the inclusive photon measurements in pp collisions at $\sqrt{s} =$ 2.76\,TeV and 8\,TeV by the ALICE Collaboration\,\cite{directphoton2768}. However, the results are consistent with expectations from NLO pQCD calculations, which predict a smaller contribution of direct photons to the inclusive photon spectrum with increasing $\sqrt{s}$.

In heavy-ion collisions, a strong enhancement at low invariant mass of dilepton pairs ($m_{\rm ll} <$ 1\,GeV/$c^{2}$) is observed at different energies, at the Super Proton Synchroton (SPS) by the CERES and NA60 Collaborations~\cite{ceresa,ceresb,ceresc,ceresd,na60a,na60b} and at RHIC energies by the PHENIX and STAR Collaborations~\cite{phenixa,phenixb,stara,starc}. The data can be explained by thermal radiation of the hadronic phase, dominated by the $\rho$ meson, which appears strongly broadened~\cite{explaina,explainb,explainc,explaind,explaine,explainf,explaing,explainh,explaini} with essentially no change of the pole mass. This broadening is consistent with chiral symmetry restoration~\cite{chiralc}. At RHIC, the data show a further excess of the direct-photon yield over the \pp expectation, which is exponential in \pt with an inverse slope $T$ of about 221 MeV~\cite{phenixa}. This excess can be attributed to thermal radiation from the partonic and/or hadronic phase~\cite{explaindd,explainh,explainhh}. At the LHC, a similar enhancement of the direct-photon production, with $T \approx$ 297\,MeV, is observed in central Pb--Pb collisions at $\sqrt{s_{\rm NN}} =$ 2.76\,TeV\,\cite{directphotonPbPb}. In the IMR, a significant excess over the yield from semileptonic decays of heavy-flavour hadrons is found at the SPS~\cite{na60a,na60b,na60c,na60d}, whereas at RHIC the data can be fairly well described by calculations including heavy-flavour contributions estimated in pp collisions and scaled with the number of binary collisions~\cite{phenixa,phenixb,stara,starc}. At the SPS, the NA60 Collaboration showed, by using precise vertex information, that the excess is associated with a prompt source, as opposed to $\mu^{+}\mu^{-}$ pairs from D mesons that decay further away from the interaction point~\cite{na60c}. The analysis of the $p_{{\rm T,\mu\mu}}$-spectra, with the extraction of the slope parameter $T_{{\rm eff}}$ as a function of $m_{{\rm \mu\mu}}$, revealed that the IMR is dominated by an early source of dileptons, i.e.\,partonic radiation, where radial flow is negligible~\cite{na60d}. Models including thermal radiation from the QGP~\cite{explaina,explainc,explaincc,explaind} can reproduce the data in the IMR.

In this paper, the first measurement of the \ee pair production in \pp collisions at $\sqrt{s}$ = 7\,TeV with ALICE at the LHC is presented. The invariant yield is studied within the central barrel acceptance of ALICE ($|\eta_{{\rm{e}}}|$ $<$ 0.8) as a function of $m_{{\rm ee}}$ ($m_{{\rm ee}}$ $<$ 3.3\,GeV/$c^{2}$), $p_{{\rm T,ee}}$ ($p_{{\rm T,ee}}$ $<$ 8\,GeV/$c$), and DCA$_{\rm ee}$ (DCA$_{\rm ee}$ $<$ 10\,$\sigma$), i.e.\,the average distance of closest approach of the reconstructed electron and positron tracks to the collision vertex, normalised to its resolution. The latter allows the prompt and non-prompt dielectron sources to be separated and provides an additional variable to disentangle the contributions from \mbox{\ccbar (with $c\tau \approx$ 150\,\textmu m for D mesons)} and \mbox{\bbbar (with $c\tau \approx$ 470\,\textmu m for B mesons).} The data are compared with a cocktail of expected \ee sources from known hadrons based on measured cross sections. Correlated pairs from heavy-flavour hadron decays are calculated with two different MC event generators, PYTHIA~\cite{PYTHIA6425} and POWHEG~\cite{POWHEGa,POWHEGb,POWHEGboxa,POWHEGboxb}. Finally, the relative contribution of virtual direct photons is shown and compared with NLO pQCD calculations.

This article is organised as follows: the experimental apparatus and data sample used in the analysis are presented in Section~\ref{section2}. The analysis strategy, including the electron identification, the background subtraction, and the efficiency corrections are described in Section~\ref{section3}, together with the associated systematic uncertainties. In Section~\ref{section4}, the procedures used to calculate the expected dielectron cross section from the known hadronic sources are explained.  The results, i.e.\,the invariant mass spectrum, the \ptee and \dcaee distributions, are finally presented and discussed in Section~\ref{results}. In the same section, the charm and beauty total cross sections, as well as the fraction of direct photons to inclusive photons, are extracted from the data.
               %%%%%%%%%%% put the body of the article here
%\newpage

\section{Experimental apparatus and data sample} \label{section2}

The ALICE apparatus and its performance are described in detail in~\cite{0954-3899-30-11-001, 0954-3899-32-10-001, 1748-0221-3-08-S08002,aliceperf}. In the following, only the subsystems relevant for the dielectron analysis are briefly discussed. Electrons\footnote{The term `electron' is used for both electrons and positrons throughout this paper.} are reconstructed and identified at mid-rapidity ($|\eta_{{\rm e}}|$ $<$ 0.8) in the central barrel of ALICE with the Inner Tracking System (ITS), the Time Projection Chamber (TPC), and the Time-Of-Flight system (TOF). These detectors are located inside a large solenoidal magnet that provides a uniform magnetic field of $B=0.5$~T along the beam direction.

The ITS~\cite{1748-0221-5-03-P03003} is the detector closest to the beam axis. It is composed of six cylindrical layers of silicon detectors, with radial distances ranging from 3.9\,cm to 43\,cm. The two innermost layers are equipped with Silicon Pixel Detectors (SPD). The two intermediate layers consist of Silicon Drift Detectors (SDD), and the two outermost layers are made of Silicon Strip Detectors (SSD). The high spatial resolution of the silicon sensors allows the Distance-of-Closest-Approach (DCA) of the track to the reconstructed collision vertex (primary vertex) to be measured. The DCA resolution in the plane transverse to the beam direction is better than 75\,\textmu m for charged particles with transverse momenta \pt $>$ 1\,GeV/$c$. Moreover, the four SDD and SSD layers provide charged-particle identification via the measurement of their specific energy loss d$E$/d$x$.

At larger radii (85 $<$ $r$ $<$ 247\,cm), a 500\,cm long cylindrical TPC~\cite{Alme:2010ke} provides identification of charged particles and reconstruction of their trajectories. Up to 159 three-dimensional space points per track, which corresponds to the number of pad rows in one TPC sector out of 18 in azimuth, are recorded and used to estimate the \dEdx of charged particles in the gas. The \dEdx resolution in \pp collisions is about 5.2\% for minimum-ionising particles passing through the full detector~\cite{aliceperf}.

The charged-particle identification capability of the TPC and ITS is supplemented by the TOF~\cite{tofperf}, which is located at a radial distance of 3.7\,m from the beam axis. It provides a measurement of the time of flight for particles from the interaction point up to the detector itself. The event collision time is either measured with the T0 detector, which consists of two arrays of Cherenkov counters located at $z = +375$\,cm and $z = -72.7$\,cm from the nominal interaction point, or estimated using the particle arrival times at the TOF for events with sufficiently large multiplicity~\cite{aliceperf}. Due to their curved paths in the magnetic field of the solenoidal magnet, charged particles need a minimum \pt of about 300\,MeV/$c$ to reach the TOF detector. Since the TOF matching efficiency is of the order of 30\% at a \pt of 500\,MeV/$c$, the TOF information is used in this analysis only if the particle has an associated hit in the TOF detector, otherwise the particle is identified with the ITS and TPC only.

The V0 detector~\cite{vZero}, used for triggering, consists of two arrays of 32 scintillators each, placed around the beam vacuum tube on either side of the interaction region at $z = -90$\,cm and $z = +340$\,cm. The two arrays cover the pseudorapidity ranges $2.8 < \eta < 5.1$ (V0A) and $-3.7 < \eta < -1.7$ (V0C), respectively.

The data used in this paper were recorded with ALICE at the LHC during the \pp run at \roots = 7\,TeV in 2010. Minimum bias (MB) collisions were triggered by requiring at least one hit in the SPD or in one of the two forward scintillator systems V0A and V0C. In addition, the timing information from the V0 and the correlation between the number of hits and track segments in the SPD detector were used offline to remove background from beam-gas interactions. The primary vertex is reconstructed by extrapolating charged-particle tracks in the TPC and ITS to the beam line. It is required to be within $\pm10$\,cm of the nominal interaction point along the beam direction in order to provide a uniform pseudo-rapidity acceptance of the detectors. A total of 370 million \pp events at  \roots = 7\,TeV pass the offline event selection criteria, corresponding to an integrated luminosity of $L_{\rm int}$ = (6.0$\pm$0.2)\,nb$^{-1}$~\cite{lumia}.
               %%%%%%%%%%% put the body of the article here
%\newpage
% !TEX root = alicepreprint_CDS.tex
%\chapter*{Introduction}

%\markboth{\MakeUppercase{Introduction}}{\MakeUppercase{Introduction}}
\section{Data analysis} \label{section3}

\subsection{Electron identification}

The strategy to identify electrons relies on a combination of tracking and particle identification
(PID) information from different detectors in the central barrel.
Reconstructed charged-particle tracks in the ITS and TPC are selected in $|\eta|$ $<$ 0.8 and \pt$>$ 0.2\,GeV/$c$. For the \dcaee analysis, the tracks must have a \pt $>$ 0.4\,GeV/$c$ to assure a sufficient separation between prompt and non-prompt e$^{+}$e$^{-}$ sources.
The DCA resolution worsens at low \pt and is larger than 150\,\textmu m for tracks reconstructed in the ITS and TPC with \pt $<$ 0.35\,GeV/$c$~\cite{aliceperf}, which is of the same order of magnitude as the decay length of the D$^{0}$ and D$^{\pm}_{\rm s}$ mesons ($c\tau$ $\approx$\,122.9 and 149.9\,\textmu m, respectively). The tracks are required to have at least 100 out of a maximum of 159 reconstructed space points in the TPC with at least 100 crossed pad-rows, while the ITS track segments must have a hit in at least 5 of the 6 detector layers.
The maximum $\chi^2$ per space point in the TPC (ITS) from the track fit must be less than 4 (4.5). Only tracks with a DCA to the primary vertex smaller than 1\,cm in the $xy$-plane and 3\,cm along the $z$-axis are accepted. To suppress electron tracks from photon conversions in the detector material at large radii, a hit in the first layer of the SPD is required. This rejects about 63\% of the conversion electron tracks, keeping 83\% of the signal electrons from light and heavy-flavour hadron decays. A small fraction of electrons from photon conversions in the second ITS layer may still have a hit in the first layer associated wrongly to their reconstructed track. Such cases are further removed from the sample by requiring that the reconstructed track does not share any ITS cluster with other tracks (see below). This requirement also reduces the amount of conversion electrons from the first ITS layer, rejecting 38\% of the remaining conversion contamination after the requirement of a hit in the first SPD layer and keeping 97\% of the signal electrons. For e$^{+}$e$^{-}$ pairs from photon conversions, where both electrons pass the default track selection, the rejection factor is even higher, about 92\%.

 The PID information is based on the measurement of the specific ionisation energy loss
 (d$E$/d$x$) in the TPC and ITS, and the time-of-flight information from TOF.
 The cut values for electron selection and hadron rejection  are expressed in
 terms of the deviation of the respective PID signal from its expectation value
 for a particle species $i$.  The PID variables \sDETi  are normalised to units of standard deviations of the respective detector (DET) resolution.

\begin{figure}[ht!]
\begin{minipage}{18pc}
\begin{center}
\includegraphics[scale=0.4]{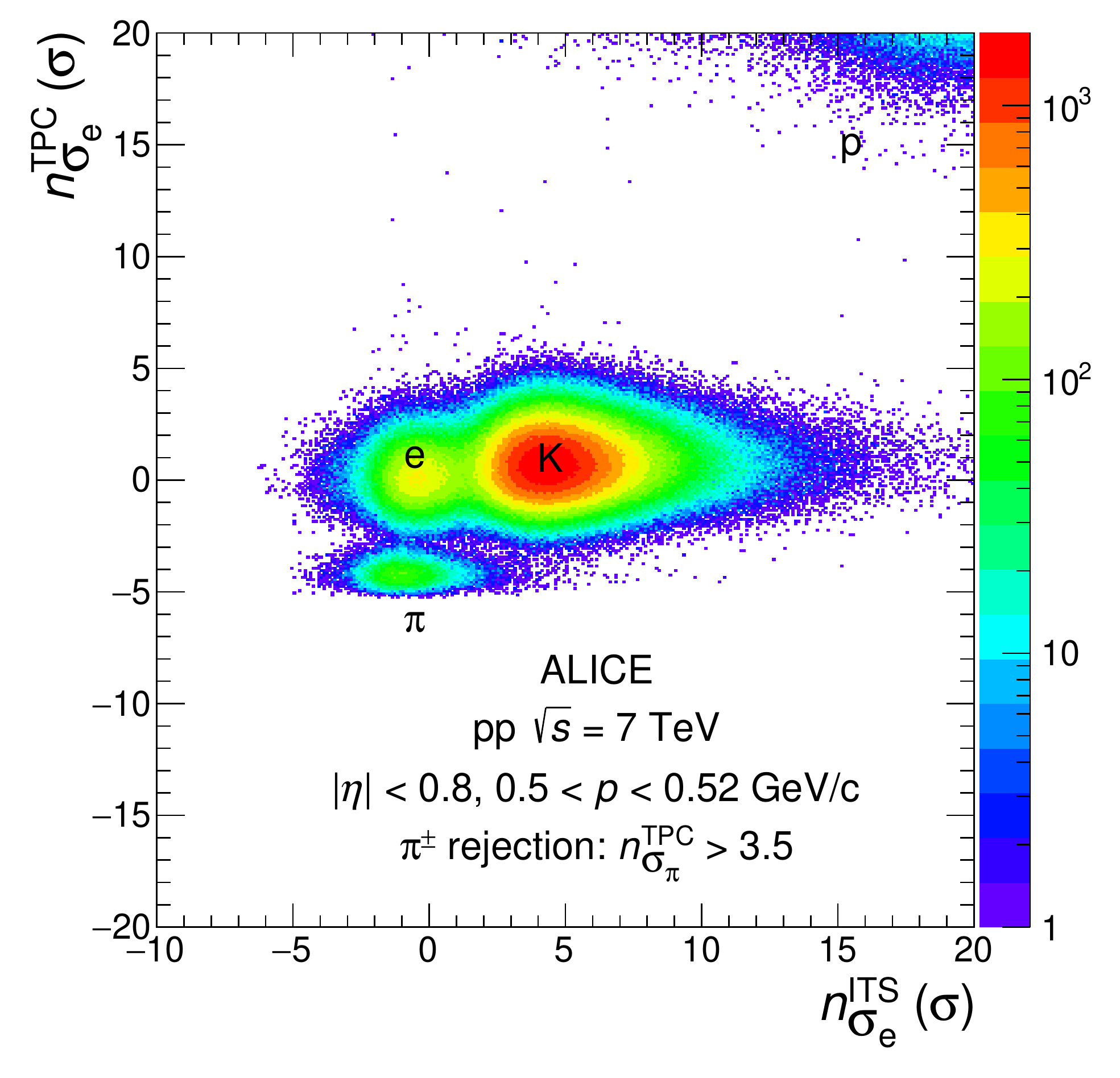}
\end{center}
\end{minipage}\hspace{1.5pc}%
\begin{minipage}{18pc}
\begin{center}
\includegraphics[scale=0.4]{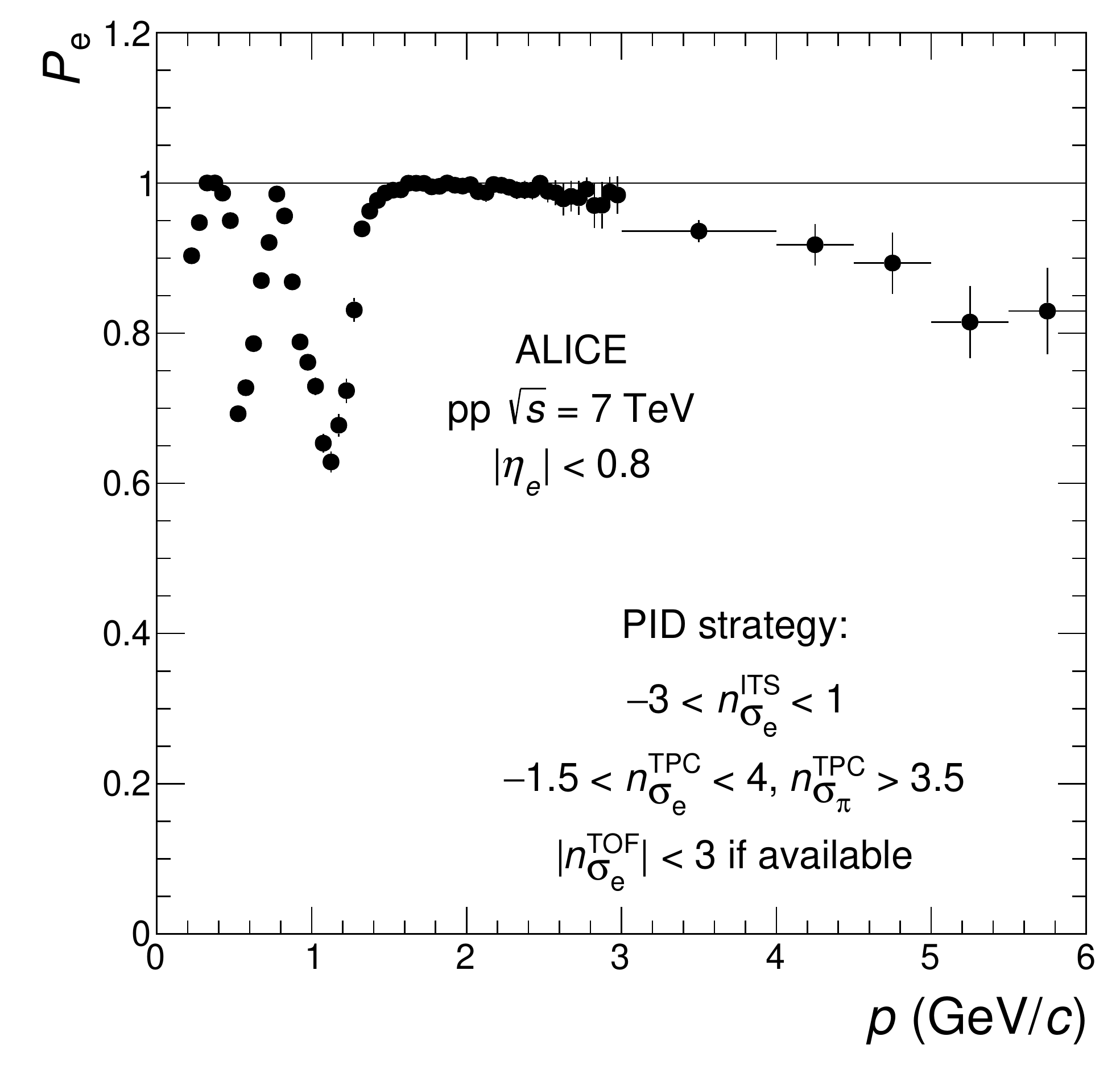}
\end{center}
\end{minipage}
\caption{(Colour online) TPC PID signal expressed as \sTPCe (see text) as a function of the ITS PID signal ($n^{\rm ITS}_{\sigma_{{\rm{e}}}}$) for selected tracks with 0.5 $<$ $p$ $<$ 0.52~\GeVoverc after applying the pion rejection in the TPC (left). Electron purity $P_{\rm e}$ as a function of momentum (right). Only statistical uncertainties are shown.}
\label{fig:pid}
\end{figure}

 In the TPC, electrons are selected in the interval  {\mbox{$-1.5<\sTPCe<4.0$}}. Additionally, pions are rejected by requiring that the measured TPC \dEdx of the track is far from the expectation value for pions: {\mbox{$\sTPCpi > 3.5$}}. Since electrons have a larger energy loss in the TPC than pions for momenta above 0.25\,GeV/$c$, the $\sTPCpi$ requirement is asymmetric. The remaining contamination by kaons and protons occurs mainly in the crossing regions of the expected \dEdx values in the TPC for these particle species and for electrons, i.e.\,around a momentum of 0.5 and 1.1 GeV/$c$, respectively. This contamination can be reduced by using the ITS information, where the crossings occur at higher momenta around 0.7 and 1.5\,GeV/$c$ for kaons and protons, respectively.
 In the left panel of Fig.~\ref{fig:pid}, the variable \sTPCe is shown as a function of \sITSe for selected tracks with $0.5 < p < 0.52$\, \GeVoverc after the pion rejection in the TPC was applied. In this momentum interval, kaons and electrons have a very similar energy loss in the TPC, whereas they are still separated in the energy loss measurements of the ITS. The electron selection criterion in the ITS is {\mbox{$-3<\sITSe<1$}}. Further reduction of the hadron contamination can be achieved using the TOF information with $|\sTOFe|<3$ (in case the selected tracks have an associated hit in the TOF detector). The electron purity $P_{\rm e}$ is estimated at low momenta ($p$ $<$ 3\,GeV/$c$) by fitting the \sTPCe distribution in momentum slices after the ITS and TOF selection, as well as the \sTPCpi rejection, following a procedure explained in~\cite{Abelev:2012xe}.
 At higher momenta, the $\sTPCpi$ distribution is fitted after the requirements on $n^{\rm ITS}_{\sigma_{\rm e}}$,  $n^{\rm TOF}_{\sigma_{\rm e}}$, and $n^{\rm TPC}_{\sigma_{\rm e}}$, are fulfilled. The result is shown in the right panel of Fig.~\ref{fig:pid}. The purity ranges from 60\% to 99\%, depending on the particle momentum $p$. The lowest purity is observed where  kaons ($p$ $\approx$ 0.5~GeV/$c$) or protons  ($p$ $\approx$ 1.2~GeV/$c$) have similar \dEdx as electrons in the TPC.

\subsection{Dielectron spectrum}

 All electron candidates from the same event are combined into pairs, characterised by their $m_{\rm ee}$, $p_{\rm T,ee}$, and DCA$_{\rm ee}$. The latter is calculated from the single-electron DCAs as:
\begin{equation}
\dcaee = \sqrt{\frac{({\rm DCA}_{xy,1}/\sigma_{xy,1})^{2}+({\rm DCA}_{xy,2}/\sigma_{xy,2})^{2}}{2}},
\end{equation}
where DCA$_{xy,i}$ is the DCA of the electron $i$ in the transverse plane and $\sigma_{xy,i}$ is its resolution estimated from the covariance matrix of the track reconstruction parameters obtained with the Kalman filter technique~\cite{kalman,aliceperf}. The absolute DCA resolution worsens at low \pt due to multiple scattering in the detector material. Therefore, the analysis is performed using the DCA normalised to its resolution, which decreases the sensitivity to the particle momentum.

The distribution of same-event pairs of opposite sign ($OS$) is composed of true signal pairs ($S$) as well as background pairs ($B$).
The  background pairs are mainly combinatorial but contain also residual correlations such as from  jets and from conversions of correlated decay photons originating from the same mother particle. Typical values of $S$/$B$ range between $O(1)$ and $O(10^{-1})$, depending on \mee~and $p_{\rm{T,ee}}$ (see below). Therefore, the minimisation of $B$ and a careful subtraction of the remaining background are key aspects of this analysis.

\begin{figure}[ht!]
\begin{minipage}{18pc}
\begin{center}
\includegraphics[scale=0.4]{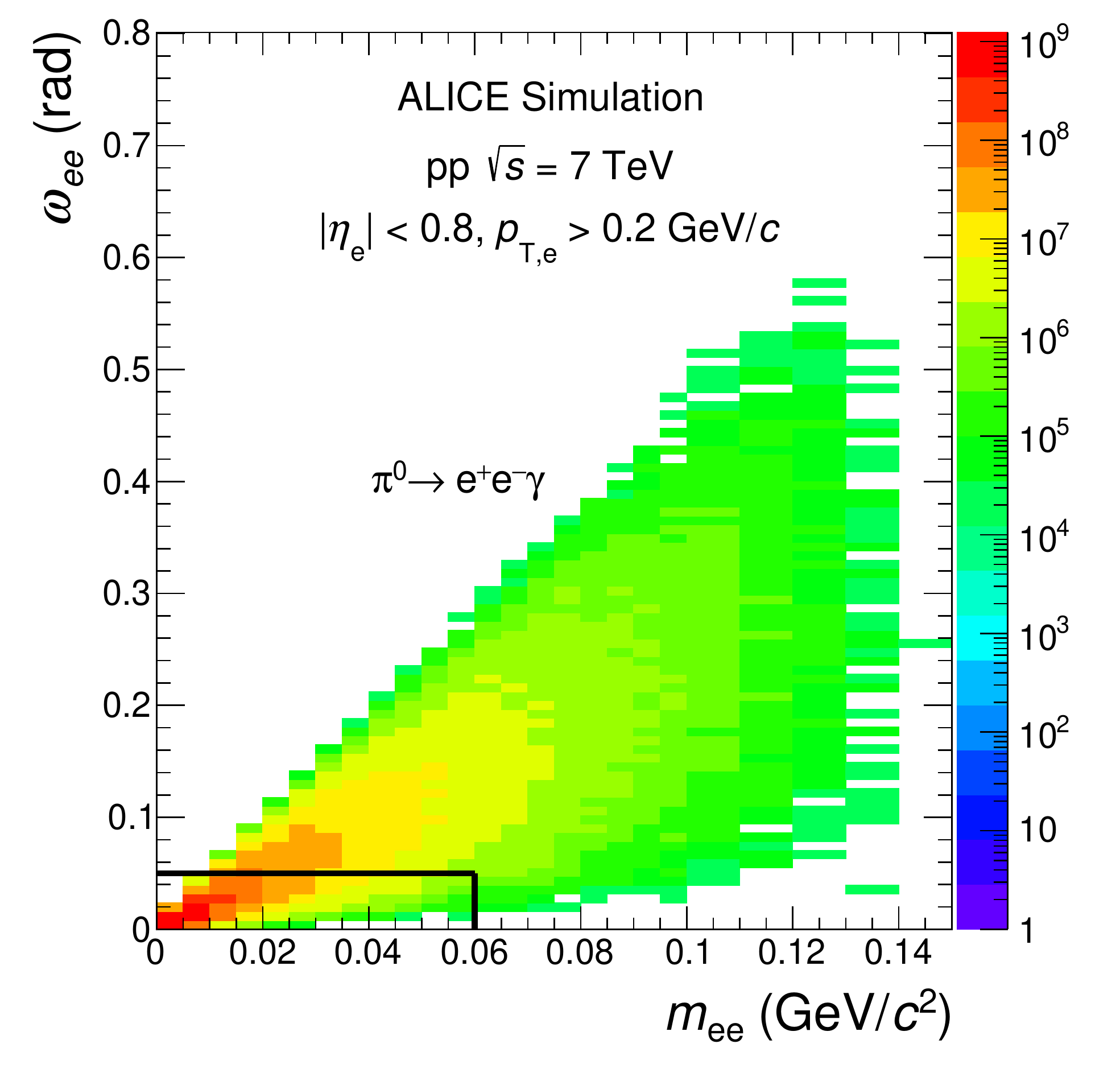}
\end{center}
\end{minipage}\hspace{1.5pc}%
\begin{minipage}{18pc}
\begin{center}
\includegraphics[scale=0.4]{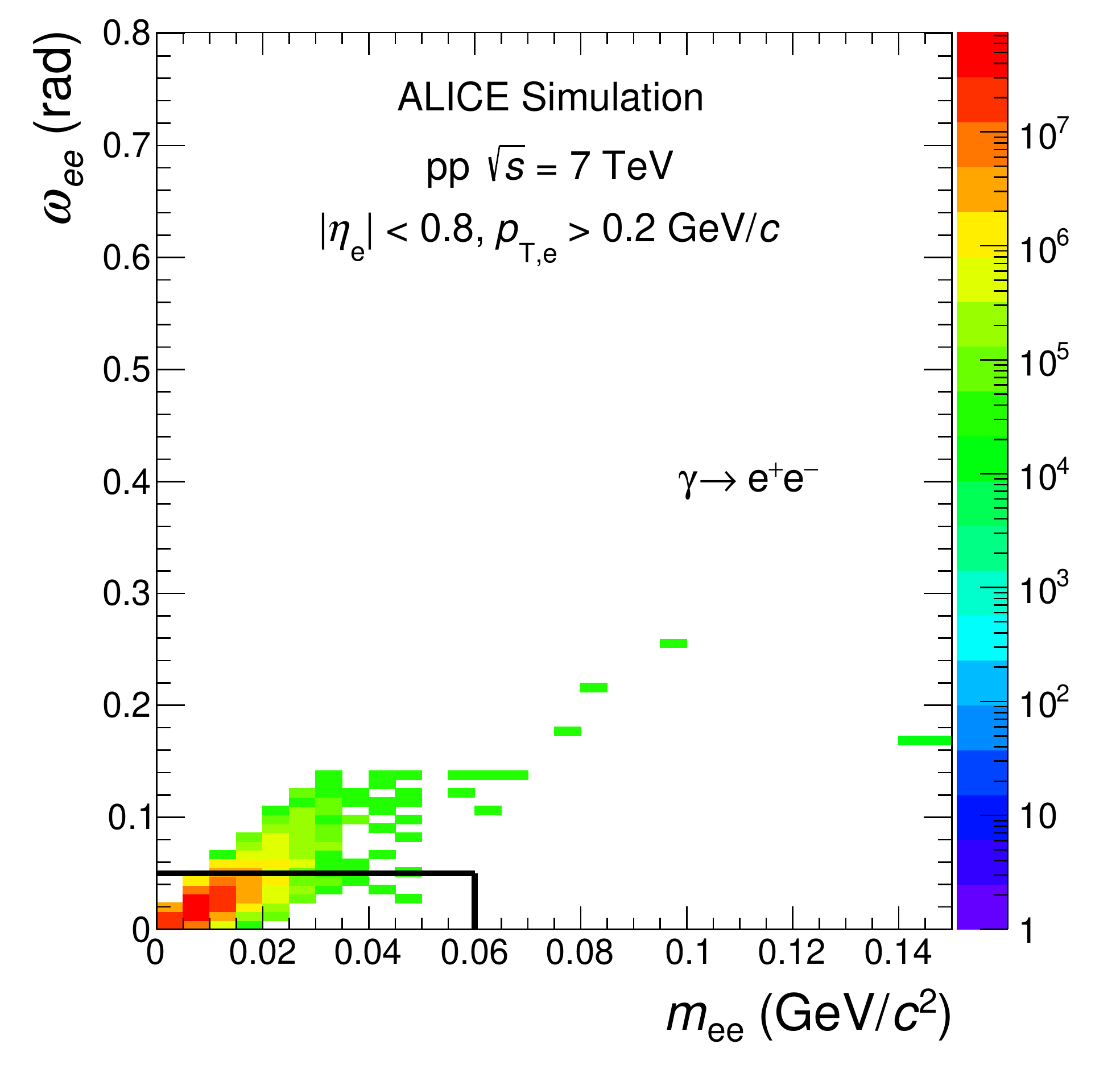}
\end{center}
\end{minipage}
\caption{(Colour online) Opening angle vs invariant mass of e$^{+}$e$^{-}$ pairs from $\pi^{0}$ Dalitz decays (left) and from photon conversions (right) in MC simulations after the single track selection criteria.  The lines indicate the prefilter requirement.}
\label{fig:openingangle}
\end{figure}

\begin{figure}[ht!]
\begin{minipage}{18pc}
\begin{center}
\includegraphics[scale=0.4]{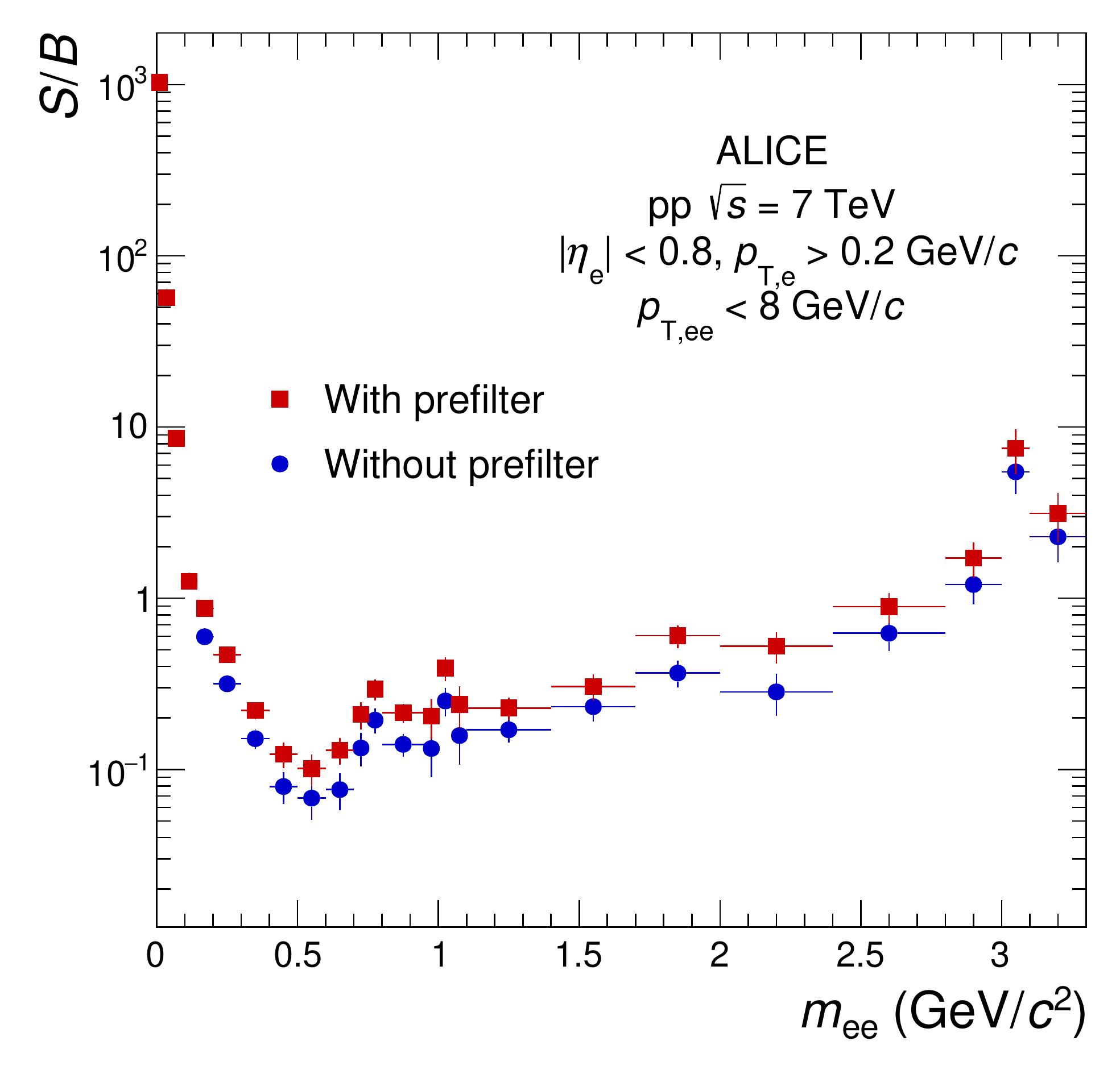}
\end{center}
\end{minipage}\hspace{1.5pc}%
\begin{minipage}{18pc}
\begin{center}
\includegraphics[scale=0.4]{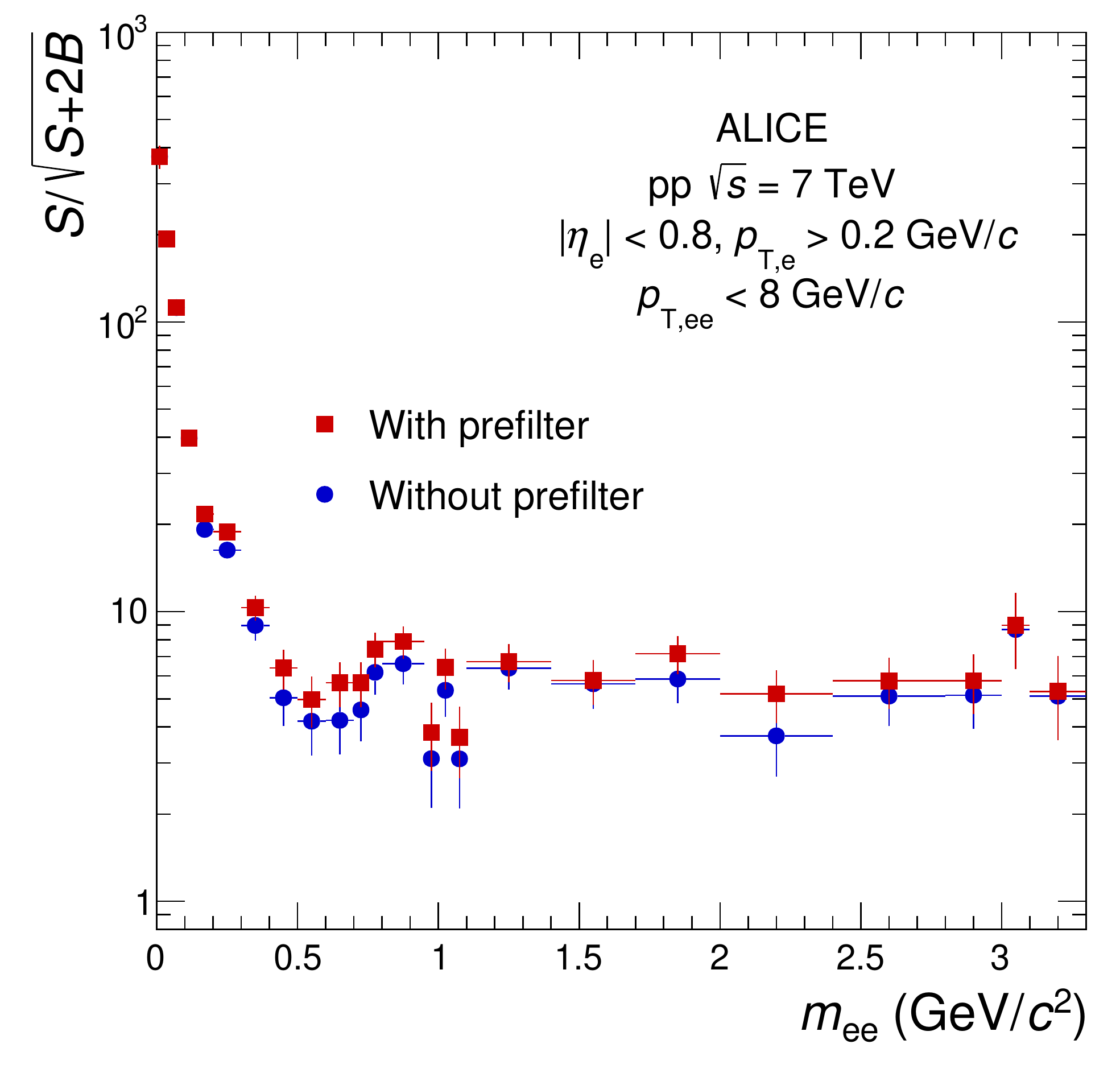}
\end{center}
\end{minipage}
\caption{Signal to background ratio (left) and significance (right) obtained with and without applying the prefilter. Only statistical uncertainties are shown. The background is estimated as explained in the text later.}
\label{fig:prefilter}
\end{figure}

The main sources of electrons contributing to $B$ are $\pi^0$ Dalitz decays and photon conversions. To reject these most efficiently, a prefilter algorithm is applied where tracks
from the selected electron candidate sample are combined with charged-particle tracks from a sample
with relaxed tracking selection criteria and no PID. Dielectron pairs originating from $\pi^0$ Dalitz decays and photon
conversions have small invariant masses and opening angles ($\omega_{{\rm ee}}$), as shown in Fig.~\ref{fig:openingangle}. Therefore, if an opposite-sign pair with small invariant mass and opening angle is formed with a track $h$ of the sample with relaxed selection criteria, the electron candidate is rejected and not used for further pairing. The cut values applied in the prefilter algorithm are $m_{{\rm eh}}$ $<$ 0.06 GeV/$c^2$ and $\omega_{{\rm eh}}$ $<50$~mrad.
These selection criteria lead to an improvement of the $S$/$B$ by a factor of about 1.5 and an increase of the significance $S$$/$$\sqrt{S+{\rm{2}}B}$ by a factor  of about 1.2 for $m_{\rm ee}$ $<$ 1\,GeV/$c^2$, as can be seen in the left and right panels of Fig.~\ref{fig:prefilter}, respectively. For $m_{\rm ee}$ $<$ 0.06\,GeV/$c^2$, the prefilter algorithm cuts systematically into the signal acceptance. Since the $S$/$B$ is large in the low-mass region, no prefilter is applied for \mee$<0.14~$GeV/$c^2$. The random rejection probability, caused by accidental combinations of electron candidates with an uncorrelated track, is small (about 3\%) and taken into acount in the efficiency corrections.

\begin{figure}[ht!]
\begin{minipage}{18pc}
\begin{center}
\includegraphics[scale=0.4]{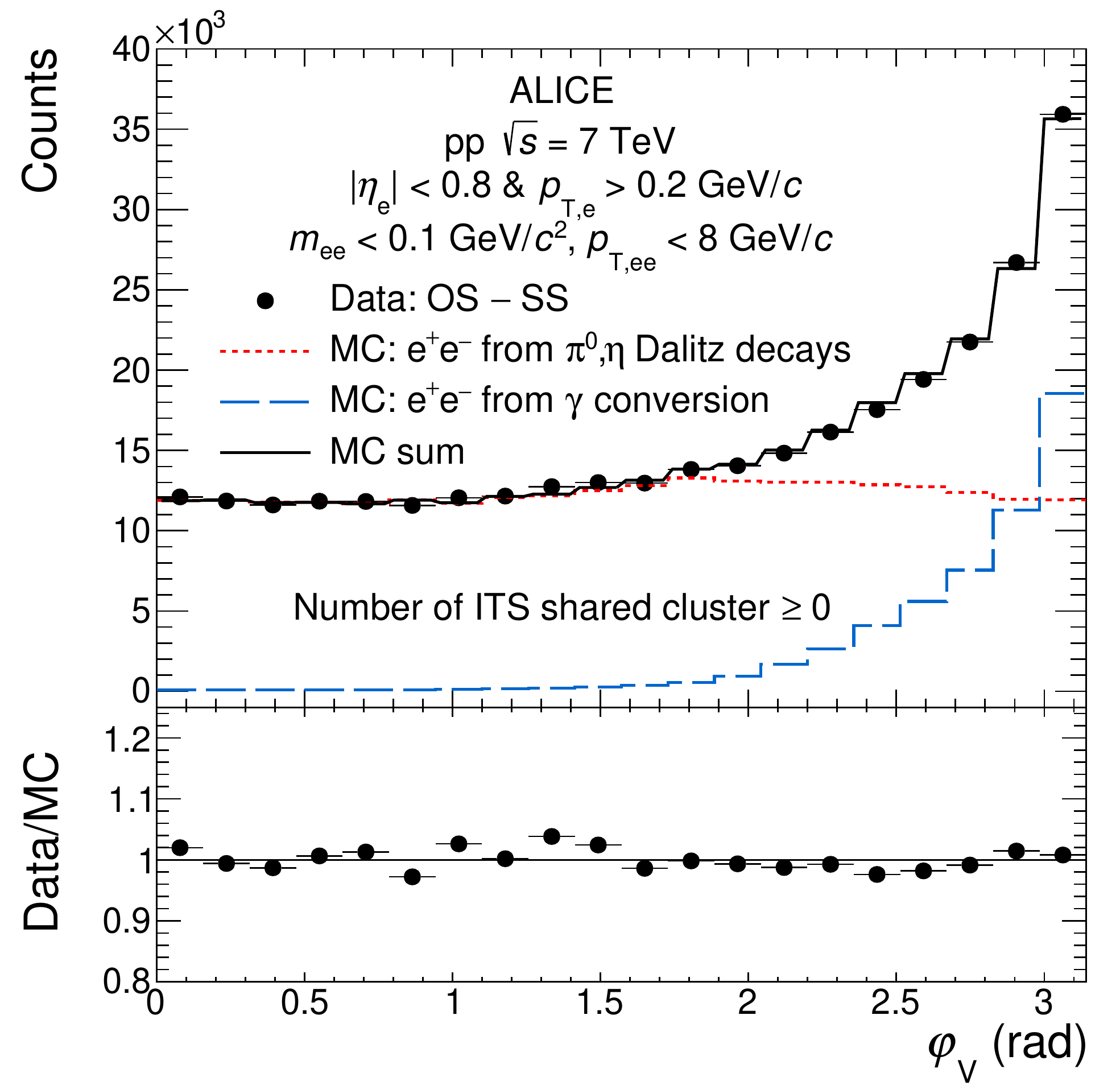}
\end{center}
\end{minipage}\hspace{1.5pc}%
\begin{minipage}{18pc}
\begin{center}
\includegraphics[scale=0.4]{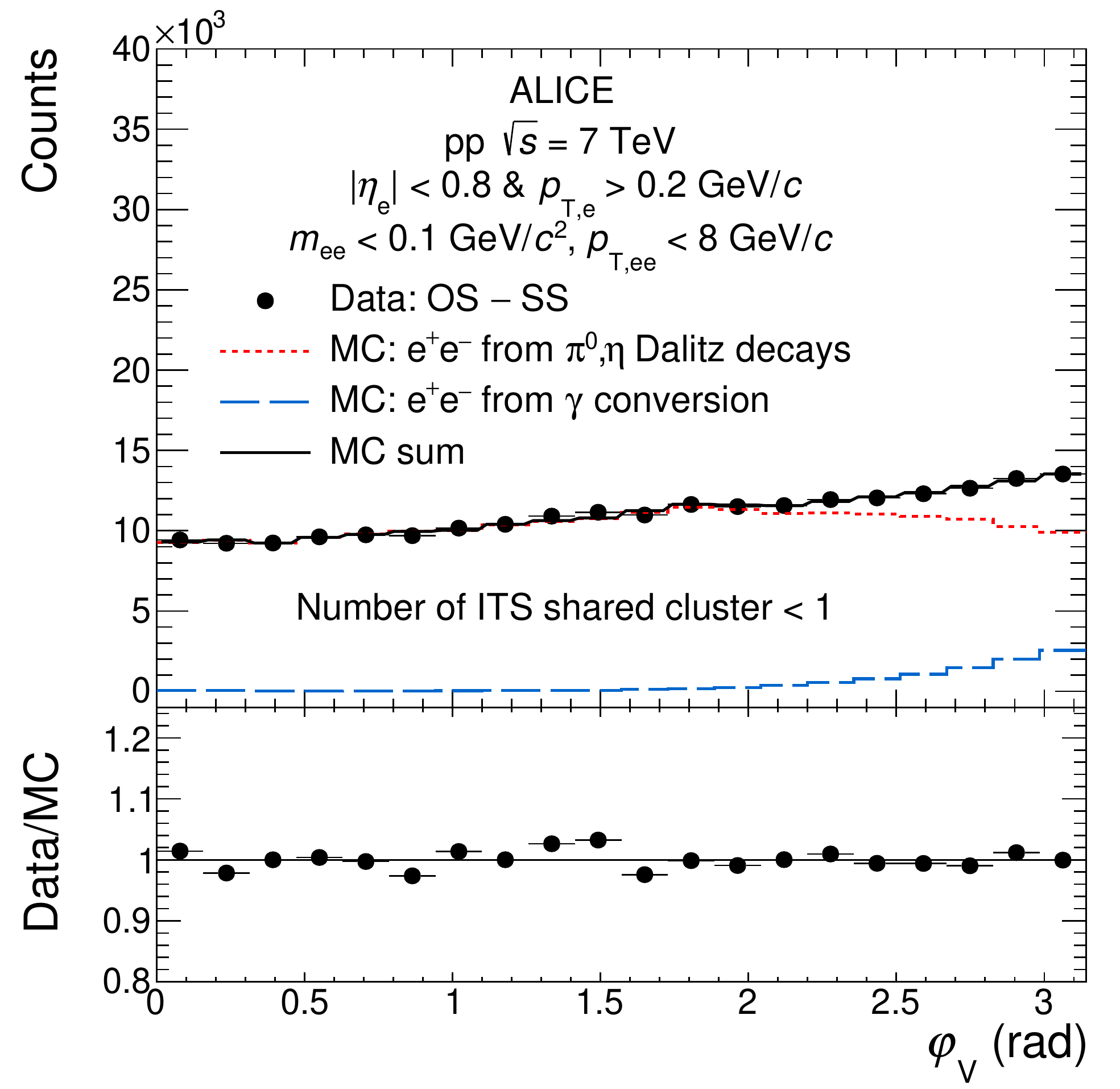}
\end{center}
\end{minipage}
\caption{Measured $\varphi_{\rm{v}}$ distribution of correlated e$^{+}$e$^{-}$ pairs with \mee $<$ 0.1~GeV/$c^2$ and \ptee $<$ 8~GeV/$c$ compared with a sum of MC templates for different dielectron sources. The distributions are shown for all tracks including those that share some ITS clusters with other tracks (left) and with such tracks removed (right), as in the analysis. Only statistical uncertainties are shown for the data points.}
\label{fig:phiv}
\end{figure}

To further suppress the contamination by dielectron pairs from photon conversions in the dielectron yield, two additional selection criteria are applied.
Conversions occur in the beam pipe or in the detector material, mainly of the ITS, and are characterised by a common secondary vertex of the dielectron pair. Any electron candidate found to form such a secondary vertex with another track is rejected from the analysis. In addition, dielectron pairs from photon conversions are characterised by a finite apparent invariant mass. The extrapolation of displaced conversion electron tracks to the collision point results in a non-vanishing artificial opening angle that is caused by the deflection of the tracks in the magnetic field. The opening angle is preferentially in the  plane perpendicular to the magnetic field direction, which can be used to further reject such conversion dielectron pairs~\cite{phenixa}. To this end, the angle $\varphi_{\rm v}$  which measures the orientation of the opening angle relative to the magnetic field is calculated according to:
\begin{equation}
\cos(\varphi_{\rm v}) = \frac{\bf{w} \cdot \bf{u_{a}}}{|{\bf{w}}| |{\bf{u_{a}}}|}.
\label{phiv}
\end{equation}
The two vectors $\bf{w}$ and $\bf{u_{a}}$ are given by:
\begin{equation}
\bf{w} = {\bf{u}}\times{\bf{v}},
\label{phiva}
\end{equation}
\begin{equation}
\bf{u_{a}} = \frac{{\bf{u}}\times{\bf{z}}}{|{\bf{u}}\times{\bf{z}}|},
\label{phivb}
\end{equation}
\begin{equation}
\bf{u} = \frac{{\bf{p_{\rm e^{+}}}}+{\bf{p_{\rm e^{-}}}}}{|{\bf{p_{\rm e^{+}}}}+{\bf{p_{\rm e^{-}}}}|},
\label{phivc}
\end{equation}
\begin{equation}
\bf{v} = \frac{{\bf{p_{\rm e^{+}}}}\times{\bf{p_{\rm e^{-}}}}}{|{\bf{p_{\rm e^{+}}}}\times{\bf{p_{\rm e^{-}}}}|},
\label{phivd}
\end{equation}
where $\bf{p_{\rm e^{+}}}$, $\bf{p_{\rm e^{-}}}$, and $\bf{z}$ are the 3-momentum vectors of the positron, electron, and the orientation of the magnetic field parallel to the beam axis, respectively.
In the left panel of Fig.~\ref{fig:phiv}, the measured $\varphi_{\rm{v}}$ distribution without ITS shared-cluster cut for \mee $<$ 0.1\,GeV/$c^2$ and \ptee $<$ 8\,GeV/$c$ is compared with the sum of two MC templates, one for pairs from $\pi^{0}$ and $\eta$ Dalitz decays and one for pairs from photon conversions, fitted to the data.
Prompt pairs with finite invariant mass have an almost uniform $\varphi_{\rm{v}}$ distribution in this kinematic domain, while conversion pairs show a peak around {\mbox{$\varphi_{\rm{v}}$ $=$ $\pi$}}. To reject these conversions, reconstructed electron tracks that share at least one ITS cluster with another track are not used in the analysis. The measured $\varphi_{\rm {v}}$ distribution after this requirement is shown in the right panel of Fig.~\ref{fig:phiv}.
The conversion peak around {\mbox{$\varphi_{\rm{v}}$ $=$ $\pi$}} is clearly suppressed. The MC simulations describe the data very well. Moreover, dielectron pairs with \mbox{$\varphi_{\rm {v}}$ $>$ 2~rad} and \mbox{$\mee$ $<$ 0.1 GeV/$c^{2}$} are removed from the selected pairs to further reduce the amount of conversion electrons. From MC studies, their final contribution is expected to be below 1\% down to \mee $=$ 0.

\begin{figure}[h]
\centering
\includegraphics[scale=.8]{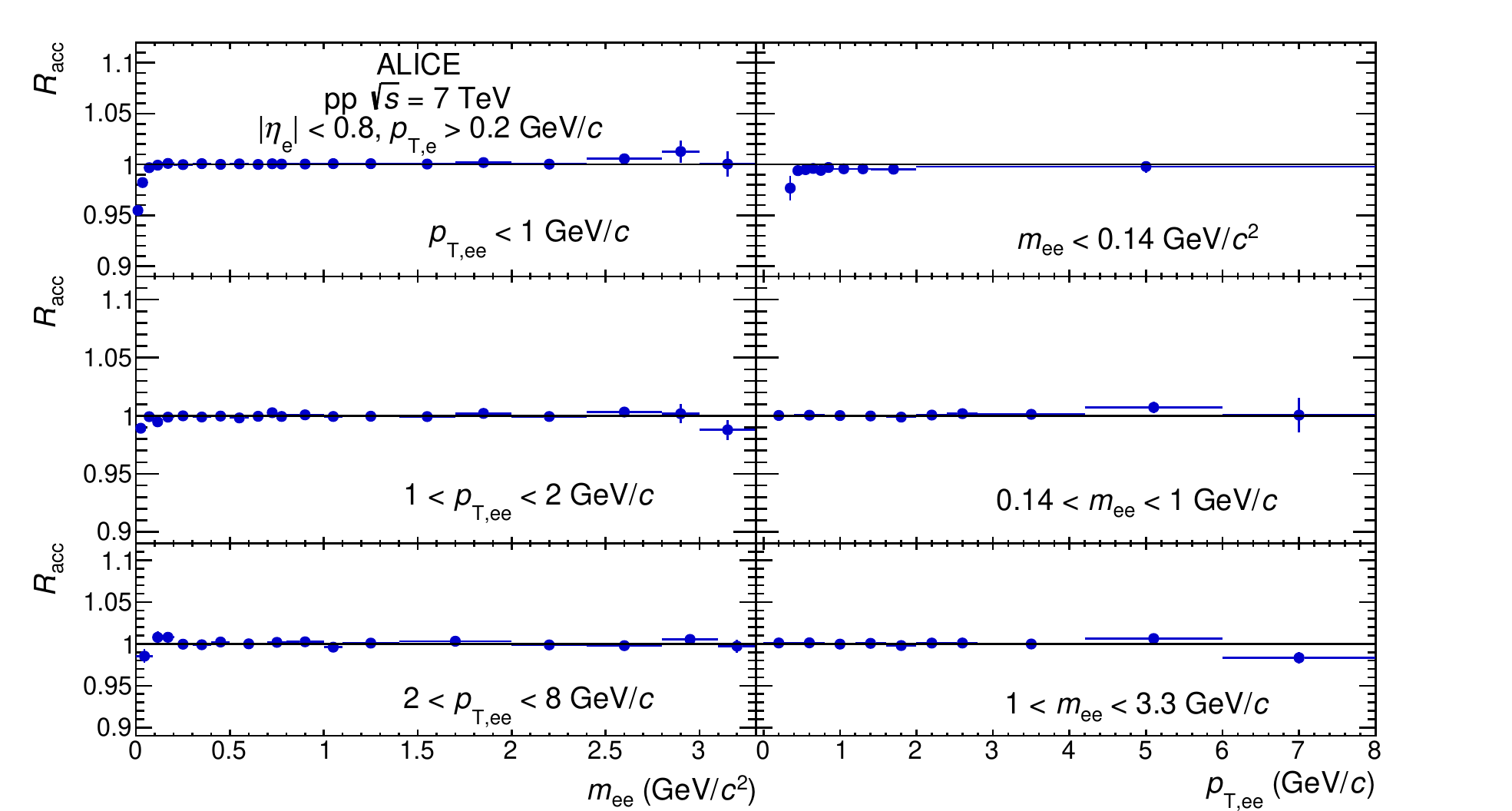}
\caption{Relative acceptance correction factor $R_{\rm acc}$ as a function of $m_{\rm ee}$ (left) and \ptee (right). Statistical uncertainties are represented by vertical bars.}
\label{fig:Rfactor}
\end{figure}

The remaining background \Bpm is estimated from the distribution of same-sign pairs, $SS$,  from the same event. In comparison to a mixed-event approach, the same-sign approximation of the combinatorial background has the advantage of containing all residual correlations arising from charge-symmetric processes such as jet fragmentation or conversions of decay photons from the same mother particle that are present in $B$, but the disadvantage of suffering from the limited statistics available in the analysed data sample. The same-sign distribution \fgSS is computed in the same bins of $m_{\rm ee}$, $p_{\rm T,ee}$, and \dcaee as the \fgOS distributions by forming in each bin the geometric mean \fgSS $=$ 2$\cdot \sqrt{N_{++}N_{--}} $  of the number of $(++)$ and $(--)$ pairs,  $N_{++}$ and $N_{--}$, respectively. The geometric mean is robust against charge asymmetries in the electron sample, as they may arise from acceptance differences of positive and negative tracks, and from charge asymmetries of the hadronic background. In the present data set, charge asymmetries of up to 5\% are observed, depending on $p_{\rm T}$. MC simulations confirm that such asymmetries do not lead to a bias in the estimate of $\Bpm$ if the geometric mean is used for the same-sign background calculation. In a few bins with low pair statistics, however, $N_{++}$ or $N_{--}$ is zero. In such bins, the arithmetic sum \mbox{\fgSS $=$ $N_{++}$$+$$N_{--}$} is used instead, to avoid underestimation of the combinatorial background.

\begin{figure}[h]
\centering
\includegraphics[scale=0.4]{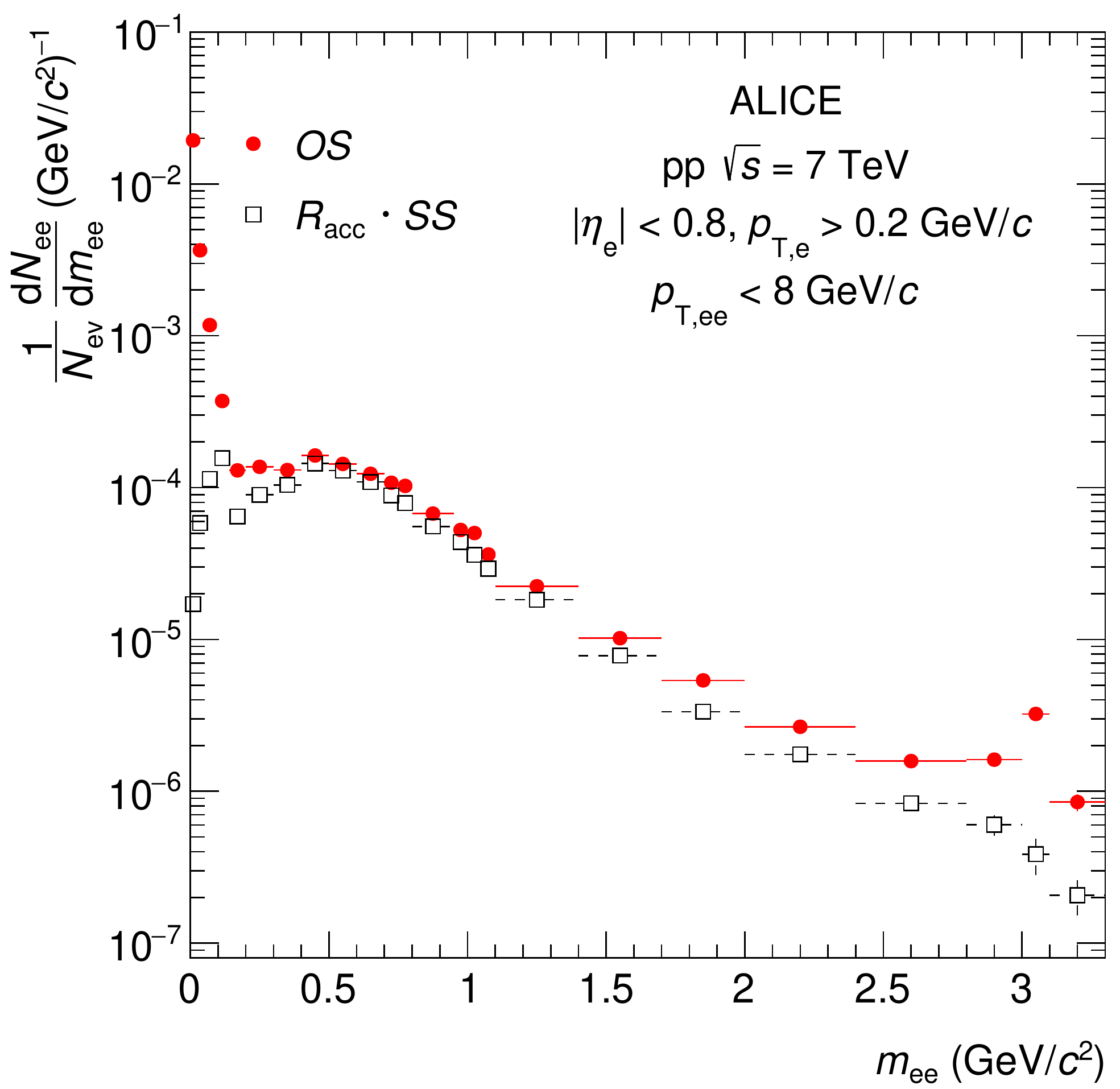}
\caption{Opposite-sign $m_{{\rm ee}}$-differential yield integrated in \ptee and \dcaee overlaid with the same-sign spectrum corrected by the acceptance correction factor $R_{\rm acc}$. Statistical uncertainties are represented by vertical bars.}
\label{fig:rawspectra}
\end{figure}

A bias in the estimate of $\Bpm$ using the same-sign technique can occur as a consequence of differences of the detector acceptance for same-sign and opposite-sign pairs. Due to the full  coverage of the ALICE central barrel in azimuth, i.e.\,in the bending plane of the spectrometer, such acceptance differences are small. Residual effects arise due to malfunctioning detector segments and can be estimated by event mixing. The relative acceptance correction factor  $R_{\rm acc}=M_{+-}$/(2$\cdot \sqrt{M_{++}M_{--}}$) is calculated, where $M_{+-}$ and $M_{\pm\pm}$ are the mixed-event opposite-sign and same-sign pair distributions. The relative acceptance correction factor $R_{\rm acc}$ as a function of \mee and \ptee is shown in Fig.~\ref{fig:Rfactor}. \mbox{For $\sqrt{(m_{{\rm{ee}}}c)^{2}+p_{{\rm T,ee}}^{2}}$ $>$ 1~GeV/$c$},  $R_{\rm acc}$ is consistent with unity and no correction is applied, while at smaller \mee and \ptee deviations of up to 5\% are observed. The relative acceptance correction factor is applied differentially in $m_{\rm ee}$, $p_{\rm T,ee}$, and DCA$_{\rm ee}$.

In Fig.~\ref{fig:rawspectra},  the opposite-sign and relative-acceptance corrected same-sign \mee spectra, i.e.\,$OS$ and $R_{\rm acc} \cdot SS$, are shown integrated over \ptee and DCA$_{\rm ee}$. The raw pair signal $\Spm$ is obtained with the formula:
\begin{equation}
\Spm = \fgOS - R_{\rm acc} \cdot SS.
\end{equation}

\subsection{Efficiency corrections}

\begin{figure}[h]
\centering
\includegraphics[scale=0.4]{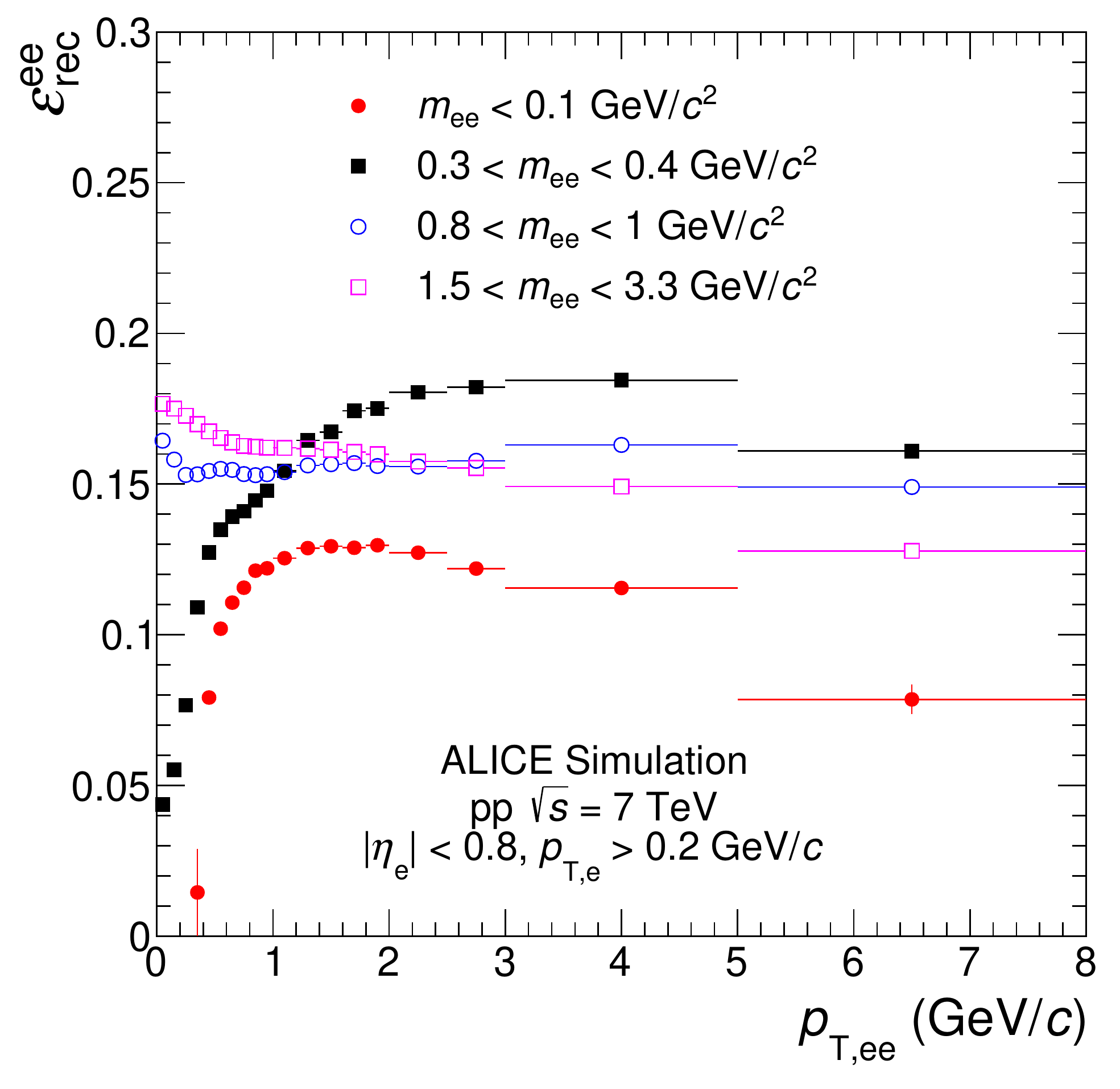}
\caption{Pair efficiency as a function of \ptee for different \mee intervals for the default electron selection criteria.}
\label{fig:pairefficiency}
\end{figure}

 The single-electron and pair efficiencies, including all tracking and PID selection criteria, are calculated using a detailed MC simulation. The event generator PYTHIA 6.4.25~\cite{PYTHIA6425} with the Perugia 2011 tune~\cite{PYTHIAtune} is used to generate \pp events. A realistic detector response is modelled using GEANT3~\cite{Brun:118715}, with the same detector configurations as in data. The reconstruction efficiency ($\epsilon_{\rm e}$) for single-electron tracks does not show any dependence on the electron DCA, for which loose selection criteria were applied (DCA$_{{\rm xy}}$ $<$ 1\,cm and DCA$_{{\rm z}}$ $<$ 3\,cm). Also no strong $\eta$ dependence of $\epsilon_{\rm e}$ is observed within {\mbox{$|\eta| <$ 0.8}} as well, whereas the dead zones of the first ITS layer can be seen in the $\varphi$ distribution of the electron candidates due to the requirement of a hit in the first pixel. The random rejection probability of the prefilter algorithm is estimated by embedding test particles in real data events. It is found to be about 3\% independent of $p_{{\rm T}}$. The resulting pair efficiency $\epsilon^{\rm ee}_{\rm rec}(m_{{\rm ee}},p_{{\rm T,ee}})$, shown in a few selected intervals of $m_{\rm ee}$ in Fig.~\ref{fig:pairefficiency}, is calculated and applied to the data as a function of \mee and $p_{\rm T,ee}$. The efficiency of the $\varphi_{\rm v}$ requirement for dielectron pairs with $m_{{\rm ee}}$ $<$ 0.1 GeV/$c^2$ is estimated assuming that the $\varphi_{\rm v}$ distribution of the signal dielectron pairs is flat (see Fig.~\ref{fig:phiv}). For \mee $>$ 0.8~GeV/$c^{\rm 2}$, \effrec reaches about 15\%. At lower $m_{{\rm ee}}$, the pair efficiency drops at low $p_{\rm T,ee}$.

Electrons suffer from Bremsstrahlung in the detector material, for which no correction is applied during the tracking procedure. This results in a smaller reconstructed momentum and distorts the shape of the \mee distributions, which develop tails towards lower $m_{{\rm ee}}$. Moreover, the reconstructed momentum of the electrons is also affected by the finite detector resolution. Such effects, i.e.\,$p_{{\rm T}}$, $\theta$, and $\varphi$ single-track resolution and Bremsstrahlung, are not accounted for by the efficiency corrections, which do not contain any unfolding procedure. However, the detector reponses are folded into the particle spectrum generated by the hadronic cocktail, as explained in detail in~\cite{publicanalysisnote}.

The corrected number of dielectron pairs is expressed as:
\begin{equation}
\frac{{\rm {d}}^{3}N_{{\rm e}^{+}{\rm e}^{-}}}{{\rm{d}}\mee {\rm{d}}\ptee {\rm{d}}\dcaee} = \frac{1}{\Delta \ptee}\frac{1}{\Delta \mee}\frac{1}{\Delta \dcaee}\frac{S(m_{{\rm ee}},p_{{\rm T,ee}},\rm{DCA}_{{\rm ee}})}{\effrec},
\end{equation}
where $\Delta p_{\rm T,ee}$, $\Delta m_{\rm ee}$, and $\Delta \dcaee$ are the width of the $p_{{\rm T,ee}}$, $m_{{\rm ee}}$, and DCA$_{{\rm ee}}$ intervals, respectively. The spectra are finally normalised by the number of minimum bias \pp collisions corrected for the primary vertex reconstruction efficiency, which is about 88\%. The invariant dielectron cross section is obtained by multiplying the yield by the minimum bias \pp cross section at \roots = 7\,TeV, of $\sigma_{\rm MB}=62.4 \pm 2.2$\,mb, which is estimated from the cross section $\sigma_{\rm V0AND}$ of the coincidence V0AND between signals in the two VZERO detectors, measured in a van der Meer scan~\cite{lumia}. The relative factor $\sigma_{\rm
  V0AND}$/$\sigma_{\rm MB}$ is given by the fraction of MB events where the L0 trigger input corresponding to the V0AND condition has fired. Its value is 0.87, and is stable within 0.5\% over the analyzed data sample. The corresponding normalisation uncertainty is $\pm 3.5\%$.

\subsection{Systematic uncertainties}

\begin{table}[ht!]
\begin{center}
\begin{tabular}{ll}
  \hline \hline
  Requirements &Variations \\
  \hline
  % \cline{2-3} & AliAODTrack::kTrkGlobalNoDCA& Bit 4\\
  Hits required in the SPD & \textbf{in the first layer}, in both layers\\
  Minimum number of ITS clusters & 4, \textbf{5}, 6 \\
  Maximum $\chi^{2}$ per ITS cluster & \textbf{4.5}, 3.5, 2.5\\
  Maximum number of  ITS shared clusters & \textbf{0}, 1, 2, 3, 4, 6\\
  Minimum number of TPC clusters & 80, \textbf{100}, 120\\
  Minimum number of crossed rows in TPC & 80, \textbf{100}, 130\\
  Minimum ratio of crossed pad-rows to findable TPC clusters & 0.5, \textbf{0.7}, 0.9\\
  Maximum $\chi^{2}$ per TPC cluster & \textbf{4}, 3, 2.5\\
  \hline \\[-1.0em]
  TOF electron identification & $|$\sTOFe$|$ $<$ 2, \textbf{3}, 4\\
  TPC electron identification & \textbf{$-$1.5}, $-$1, $-$0.5 $<$ \sTPCe $<$ 2, 3, \textbf{4}\\
  TPC pion rejection & \sTPCpi $<$ 3, \textbf{3.5}, 4\\
  ITS electron identification & $-$4, $-$3.5, \textbf{$-$3} $<$ \sITSe $<$ 0, 0.5, \textbf{1}\\
  \hline \hline
\end{tabular}
\centering
\caption{Summary of the single-track selection criterion variations to determine the systematic uncertainties. The default values are shown in bold.}
\label{table:sys}
\end{center}
\end{table}

The systematic uncertainties arise from limitations in the determination of the background, the relative acceptance correction factor $R_{\rm acc}$,  the electron selection efficiency, the prefilter efficiency, and the pair-cut efficiency. These uncertainties are evaluated by varying all the electron and pair selection criteria simultaneously and by comparing the results with and without prefilter. Table~\ref{table:sys} summarises the single-track selection criteria variations. The signal is extracted and corrected for 22 random combinations of selection criteria, which probe different but still reasonable single-electron efficiencies and $S/B$ ratios, ranging from 0.22 to 0.42 at $p_{\rm T}$ $=$ 1\,GeV/$c$ and from about 0.05 to 0.15 at \mee $=$ 0.5~GeV/$c^2$, respectively. More than one selection criteria are varied at the same time to take into account possible correlations between them. The final systematic uncertainty is calculated as the root mean square of the variation of the final data points. These extracted systematic uncertainties contain not only systematic effects from the signal efficiency, but also from the  background estimation. The maximum $\varphi_{\rm v}$ requirement for pairs with \mee $<$ 0.1\,GeV/$c^2$ is also varied, around the default value of 2\,rad, from 1.57 to 2.5\,rad. Deviations from a flat $\varphi_{\rm v}$ distribution for the signal dielectron pairs are estimated with a MC simulation and found to lead to a systematic uncertainty below 1\% for the default $\varphi_{\rm v}$ requirement. The resulting systematic uncertainties from the selection criterion variation is listed in Table~\ref{table:sysdca} in the case of the DCA analysis with $p_{{\rm T,e}}$ $>$ 0.4\,GeV/$c$.

An additional source of systematic uncertainty is considered for the DCA$_{{\rm ee}}$-differential dielectron cross section. The electron efficiency is found to be independent on the single-track DCA within the range under study by checking the fraction of reconstructed electrons as a function of the distance of their production vertex to the reconstructed primary vertex in MC. However, some correlations still remain between \ptee and DCA$_{{\rm ee}}$. In the $\pi^{0}$ mass region (0.08 $<$ \mee $<$ 0.14\,GeV/$c^2$), the mean \ptee is approximately constant as a function of DCA$_{{\rm ee}}$, which is expected since the electron tracks should always point to the primary vertex for a prompt source, and the finite DCA$_{{\rm ee}}$ values are only due to the detector resolution.
This is not the case for the $J/\psi$ region (2.7 $<$ \mee $<$ 3.3\,GeV/$c^2$), where the mean \ptee exhibits an increase as a function of DCA$_{{\rm ee}}$. The reasons are twofold: first, non-prompt $J/\psi$ from feed-down from B-mesons have a harder \pt spectrum than prompt $J/\psi$, and second, high-\pt non-prompt $J/\psi$ decay farther away than low-\pt non-prompt $J/\psi$ so that the decay electrons have larger DCAs and larger \pt compared to electrons from low-\pt non-prompt $J/\psi$.
The possible remaining uncertainty from this correlation is estimated by half the difference of the pair efficiency at the maximum and minimum mean-$p_{\rm T,ee}$, seen as a function of \dcaee in a given mass region. This systematic uncertainty is found to be less than 5\%, increasing from low to high $m_{{\rm ee}}$. Table~\ref{table:sysdca} summarises the systematic uncertainties arising from the DCA analysis.

\begin{table}[ht!]
\begin{center}
\centering
\begin{tabular}{lll}
  \hline \hline
  Mass region & Uncertainty from & Uncertainty from \\
              & DCA$_{{\rm ee}}$-\ptee correlation & selection criterion variation \\
  \hline
  \mee $<$ 0.14\,\GeVovercs & $-$ & 8.7\% \\
  0.14 $<$ \mee $<$ 1.1\,\GeVovercs & 1.5\% & 11\% \\
  1.1 $<$ \mee $<$ 2.7\,\GeVovercs & 3.0\% & 17\% \\
  2.7 $<$ \mee $<$ 3.3\,\GeVovercs & 4.9\% & 17\% \\
\hline \hline
\end{tabular}
\caption{Summary of the systematic uncertainties for the DCA$_{\rm ee}$ analysis ($p_{\rm T,e} > 0.4 {\rm\,GeV}$/$c$ and $|\eta_{\rm e}| < 0.8$).}
\label{table:sysdca}
\end{center}
\end{table}
               %%%%%%%%%%% put the body of the article here
%\newpage
%\input{NHFelectron.tex}               %%%%%%%%%%% put the body of the article here
%\newpage
% !TEX root = alicepreprint_CDS.tex
%\chapter*{Introduction}

%\markboth{\MakeUppercase{Introduction}}{\MakeUppercase{Introduction}}
\section{Cocktail of hadronic sources} \label{section4}
To allow for a detailed interpretation of the data, the contribution from all known hadronic sources must be estimated.
%The expected dielectron cross section of the known hadronic sources is calculated in order to interpret the data.
The so-called hadronic cocktail contains contributions from pseudoscalar and vector-meson decays as well as from semileptonic decays of heavy-flavour hadrons.

\subsection{e$^{+}$e$^{-}$ pairs from light-flavour hadrons and $J/\psi$ mesons}

The Dalitz decays of light neutral mesons, $\pi^{0} \to\ee\gamma$,
$\eta\to\ee\gamma$, $\eta^{\prime} \to\ee\gamma$, $\eta^{\prime}
\to\ee\omega$, $\omega \to\ee\pi^{0}$, $\phi \to\ee\eta$, and $\phi \to\ee\pi^{0}$, and the dielectron decays of the vector mesons,
$\rho$, $\omega$, $\phi$, and $J/\psi$ are simulated with the
phenomenological event generator EXODUS~\cite{phenixpp}. The radiative decay of
$J/\psi$ ($J/\psi \to \ee \gamma$) is also included. The pair mass distribution
from the Dalitz decays follows the Kroll-Wada expression~\cite{KrollWada:formula} multiplied by
the electromagnetic form factors measured by the Lepton-G
Collaboration~\cite{formfactora,formfactorb} and more recently by the
NA60 Collaboration~\cite{na60eta,na60omega}. The $\rho$ line shape has
been studied in detail in p$-$A collisions at 400\,GeV by the NA60
Collaboration~\cite{na60eta}, who confirmed the need for a Boltzmann
term beyond the standard description~\cite{rhoshape} and
provided a precise measurement of the effective temperature parameter. For the decay of the other vector mesons, which are assumed to be unpolarised, the Gounaris-Sakurai
expression~\cite{goumarissakurai} is used.

The rapidity distribution of the mesons is assumed to be flat at
mid-rapidity. The momentum distributions of $\pi^{0}$, $\eta$, $\phi$, and $J/\psi$ are obtained by fitting the spectra
measured by the ALICE Collaboration~\cite{chargedpi,pi0eta,phi,jpsi} with a modified
Hagedorn function~\cite{Hagedorn}. The measured
$\pi^{\pm}$ and $\pi^{0}$ spectra agree within their systematic
uncertainties. Since the $\pi^{\pm}$ measurement extends to lower
$p_{{\rm T}}$, and exhibits smaller uncertainties than the $\pi^{0}$,
charged pions are used to approximate neutral pions. For the other
mesons, $\rho$, $\omega$, and $\eta^{\prime}$, the shape of their \pt spectra is
derived from the $\pi^{\pm}$ spectrum. The $\eta^{\prime}$ mesons are
generated assuming $m_{\rm T}$ scaling~\cite{mtscalinga,mtscalingb,mtscalingconstantin}, implying that the
spectra of all light mesons as a function of $m_{\rm
  T} = \sqrt{m^{2}+p_{\rm T}^{2}}$ are the same and only differ by a
normalisation factor. The normalisation factors are based on the ratio
of the $p_{{\rm T}}$ spectra of the given hadron to the $p_{{\rm T}}$ spectrum of the
$\pi^{\pm}$ at high $p_{{\rm T}}$: $0.4 \pm 0.08$ for $\eta^{\prime}$
from PYTHIA 6 calculations of pp collisions at $\sqrt{s} =$ 7\,TeV,
$0.85 \pm 0.17$ for $\omega$ obtained from
measurements in pp collisions at $\sqrt{s} =$ 7\,TeV~\cite{ALICE-PUBLIC-2018-004}, and $1.0 \pm 0.2$ for
$\rho$ obtained from measurements in pp collisions at $\sqrt{s} =$ 2.76\,TeV~\cite{ALICECollaboration:2316135}. The momentum distributions of $\omega$ and
$\rho$ are obtained from the $\omega / \pi^{\pm}$ and $\rho / \pi^{\pm}$
ratios in simulated pp collisions at $\sqrt{s} =$ 7\,TeV with the
Monash 2013 tune of PYTHIA 8~\cite{pythia8intro,pythia8monash13}. This tune describes the measured $\omega / \pi^{0}$ and $\rho / \pi^{0}$
ratios in pp collisions at $\sqrt{s} =$ 7\,TeV and 2.76\,TeV,
respectively. Since the $\omega$ measurement does not extend to low $p_{{\rm
  T}}$ (the $\omega$ meson is measured for $p_{{\rm T}} > 2$\,GeV/$c$), fits of the data are used only to estimate the systematic uncertainties.

The expected dielectron yield as a function of \mee and \ptee is
computed in a fast simulation by filtering the generated hadronic cocktail through the ALICE
acceptance, while applying the detector responses including the
momentum and opening angle resolutions, and the Bremsstrahlung
effect~\cite{publicanalysisnote}, since no unfolding procedure is
applied to the data. The momentum transformation matrices are determined with
full GEANT3~\cite{Brun:118715} simulations of the interactions of the
primary electrons produced in pp collisions with the material of
the ALICE apparatus.

The main systematic uncertainties on the hadronic cocktail arise from
the uncertainties of the measured $\pi^{\pm}$, $\eta$,
$\omega$, $\phi$, and $J/\psi$ \pt spectra and those of the $m_{{\rm
  T}}$ normalisation factors. The first is evaluated by
parameterising the data along the upper and lower ends of their
statistical and systematic uncertainties added in quadrature. The complete cocktail of \ee pairs is then generated again based on
these new parametrisations. For the $\rho$ mesons, $m_{\rm T}$ scaling is used to
estimate the systematic uncertainties originating from the $\rho$ \pt spectrum. The uncertainties from the different
decay branching ratios~\cite{branchingratios} are also taken into account.

\subsection{Open-charm and open-beauty contributions to the dielectron
  yield}

Electron pairs that originate from the semileptonic decays of \ccbar
and \bbbar are simulated with two different generators, the leading-order (LO) event generator PYTHIA 6.4.25~\cite{PYTHIA6425}, and the
next-to-leading order (NLO) event generator POWHEG~\cite{POWHEGa,POWHEGb}. The \ccbar and
\bbbar pairs are produced at leading order through pair creation,
predominantly $gg\to Q\overline{Q}$ with a small contribution of
$q\overline{q}\to Q\overline{Q}$, where
$g$, $q$, and $Q$ are gluons, up or down quarks, and charm or beauty quarks, respectively. At higher order, flavour excitations and
gluon splitting give rise to further contributions.

 PYTHIA simulations utilise LO-pQCD matrix elements for $2 \to 2$ processes together with a
leading-logarithmic $p_{{\rm T}}$-ordered parton shower, and an underlying-event
simulation including multiple parton interactions. The fragmentation
and hadronisation of the charm and beauty quarks are based on the Lund
string model. In this paper, the Perugia-2011 tune~\cite{PYTHIAtune}
is used, for which the first LHC data, mainly from multiplicity and underlying-event related
measurements, have been considered. In this tune, the parton distribution
functions are parametrised with the CTEQ5L~\cite{CTEQ5L} functions.

POWHEG is a NLO-pQCD generator that can be interfaced to a parton shower
MC (e.g. from PYTHIA or HERWIG~\cite{herwig}) to provide final-state particles. The
calculations presented in this paper (POWHEG) are obtained with the POWHEG BOX
framework~\cite{POWHEGboxa,POWHEGboxb} and the tune Perugia-2011 of PYTHIA
6.4.25. The CTEQ6.6~\cite{CTEQ66} functions are used for the input
parton distribution functions. To be consistent with the PYTHIA
simulations, the mass of the charm and beauty quarks are set to
1.5\,\GeVovercs and 4.75\,GeV/$c^{2}$, respectively.

The simulations are normalised to the measured total charm and beauty
cross section, i.e.\,$\sigma_{\ccbar}^{{\rm REF}} = 7.44 \pm 0.14 {\rm (stat.)} \pm 0.58 {\rm (syst.)}$\,mb~\cite{totalccbar}
and $\sigma_{\bbbar}^{{\rm REF}} = 288 \pm 4 {\rm (stat.)} \pm 48 {\rm
  (syst.)}$\,\textmu b~\cite{totalbbbar} and passed through the ALICE acceptance
after applying the detector responses including the
momentum and opening angle resolutions, and the Bremsstrahlung effect~\cite{publicanalysisnote}. The
systematic uncertainties of the  $\sigma_{\ccbar}^{{\rm REF}}$ and $\sigma_{\bbbar}^{{\rm REF}}$
measurements are propagated to the final hadronic cocktail. Whereas the effective beauty-to-electron branching ratio is taken from PYTHIA
(${\rm BR}_{\rm b (\to c) \to e} = {\rm BR}_{\rm b \to c \to e} +{\rm BR}_{\rm  b\to
  e} = 11.7  + 10.2 = 21.9\%$ consistent with~\cite{branchingratios}), the one for charm-to-electron
(BR$_{{\rm {c\to e}}}$) is assumed to be $(9.6 \pm 0.4)$\% as reported
in~\cite{branchingratios}, which is slightly smaller than what has been estimated with PYTHIA (BR$_{{\rm c \to e}} = 10.6$\%). An
additional uncertainty of 9.3\% is added in quadrature for the BR$_{{\rm c
    \to e}}$ to take into account differences in the $\Lambda_{\rm
  c}^{+}$/D$^{0}$ ratio measured by the ALICE Collaboration, $0.543 \pm 0.061 {\rm
  (stat.)} \pm 0.160 {\rm (syst.)}$ (for $p_{\rm T} >$ 1 GeV/$c$) in pp collisions at
$\sqrt{s} =$ 7\,TeV ~\cite{lambdac}, and the LEP average of $0.113 \pm
0.013 {\rm (stat.)} \pm 0.006 {\rm (syst.)}$~\cite{LEPaverage}. This translates into a 22\%
uncertainty at the pair level. The uncertainties of the effective beauty- and charm-to-electron branching ratios are propagated to the final hadronic cocktail. For both generators, the $p_{\rm{T}}$-differential cross-section of single electrons from charm-
and beauty-hadron decays at mid-rapidity is found to be consistent with FONLL
calculations~\cite{FONLL} and to reproduce the
measurements reasonably well within the theoretical and experimental
uncertainties~\cite{charmbeautyhfepaper}. To
obtain the dielectron yield of correlated \ee pairs from heavy-flavour hadron
decays, the distribution of same-sign pairs is subtracted from the \ee
spectrum, as in data.

\subsection{\dcaee template distributions}\label{dcatemplates}

\begin{figure}[h]
\centering
\includegraphics[scale=0.4]{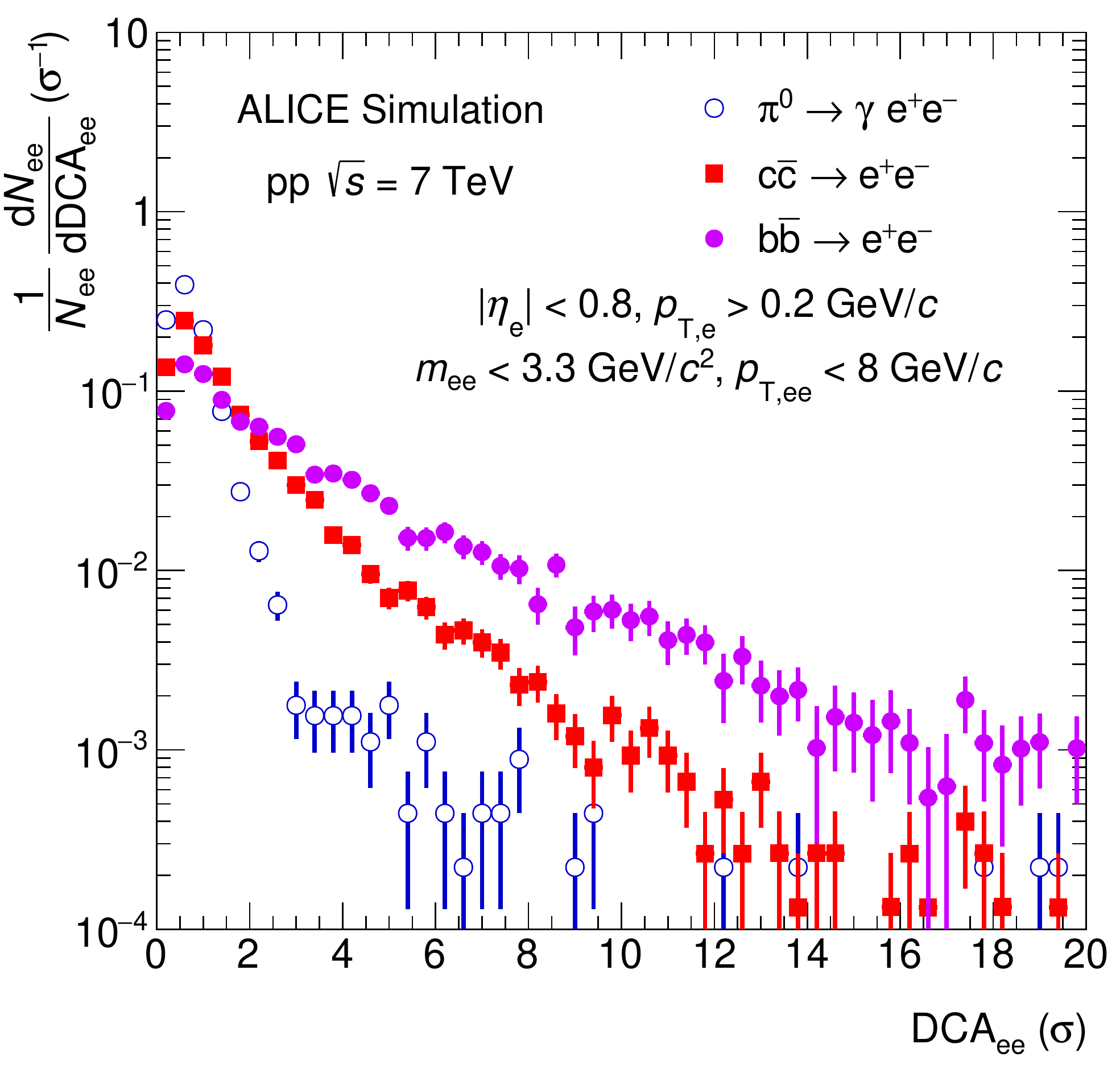}
\caption{\dcaee distributions of \ee pairs from $\pi^{0}$ Dalitz
  decays, from semileptonic decays of charm and beauty hadrons in MC
  simulations (see text), integrated over
  \mee and $p_{{\rm T,ee}}$.}
\label{fig:dcaeetemplatesintegrated}
\end{figure}

 Whereas the differential \ptee and \mee distributions of the hadronic
 cocktail are estimated from a fast simulation,
the \dcaee distributions are determined with a full GEANT3~\cite{Brun:118715}
simulation of the ALICE detector. For this purpose, PYTHIA 6.4.25 events are passed through
the full detector simulation tuned to describe the performance of each
detector subsystem. In particular, all relevant charactereristics of
the SPD, such as a map of dead channels, are included in the
simulation. The same analysis selection criteria
as in data are applied. Since the various charm hadrons have quite
different decay lengths, ranging from about 59.9\,\textmu m for $\Lambda_{\rm
  c}^{\pm}$ to 311.8\,\textmu m for D$^{\pm}$
mesons~\cite{branchingratios}, their relative yields are relevant to build the \dcaee template of
correlated \ee pairs from charm-hadron decays. The measured production
ratios of charm hadrons~\cite{Dmeson7tev,lambdac} and their semileptonic decay branching ratios~\cite{branchingratios}
are used to obtain the \ccbar \dcaee distribution. Finally, \dcaee templates
are extracted for \ee pairs from $\pi^{0}$ Dalitz
decays, from charm- and beauty-hadron decays, and from the decays
of prompt and non-prompt $J/\psi$. In
Fig.~\ref{fig:dcaeetemplatesintegrated}, the $\pi^{0}$, $\rm{c}\overline{\rm{c}}$, and
\bbbar templates are shown integrated over \ptee and $m_{{\rm ee}}$. Whereas the
distributions for prompt sources, like $\pi^{0}$, directly reflect the detector DCA resolution,
those of non-prompt sources, like heavy-flavour hadrons,
are characterised by the convolution of the DCA resolution and the
decay length of the mother particle ($c\tau_{{\rm D}}$ $\approx$
150\,\textmu m and $c\tau_{{\rm B}}$ $\approx$
470\,\textmu m~\cite{branchingratios}). The \dcaee spectrum of \ee pairs from $\pi^{0}$
Dalitz decays is taken as an approximation for all prompt
light-flavour decays into dielectrons.

Each contribution is normalised to its expected yield from the
hadronic cocktail in the same \mee and \ptee range and after the same
fiducial selection criteria ($|\eta_{{\rm e}}|$ $<$ 0.8 and $p_{{\rm T,e}} > 0.4$\,GeV/$c$). Since \ee pairs from prompt and non-prompt
$J/\psi$ have different \dcaee distributions, the measured fraction of
non-prompt $J/\psi$ in pp collisions at 7\,TeV~\cite{totalbbbara}, $f_{\rm B}$, is used to
scale the templates accordingly. To evaluate the uncertainty originating
from $f_{\rm B}$, the \dcaee distributions are shifted to the upper and
lower bounds of the statistical and systematic uncertainties added in quadrature.

Additional sources of systematic uncertainties are considered. First, the
resolution of the single-track DCA is found to be better in MC
as compared to data by about 15\%. This affects the \dcaee distributions,
in particular those of \ee pairs from prompt
sources like $\pi^{0}$ and prompt $J/\psi$, at around 1-3\,$\sigma$. Second, the uncertainties on
the charm-hadron production ratios~\cite{Dmeson7tev,lambdac} and on their semileptonic decay
branching ratios~\cite{branchingratios} are propagated to the final \dcaee
distribution of correlated \ee pairs from charm-hadron decays. Third, the PYTHIA simulations used to create the \dcaee
templates for the heavy-flavour contributions have been performed with the PYTHIA
Perugia-0 tune~\cite{PYTHIAtune}, which does not reproduce well the
measured $p_{{\rm T}}$ distribution of electrons
from charm-hadron decays at high \pt \mbox{(\pt $>$ 3~GeV/$c$)}~\cite{charmbeautyhfepaper}. This was found to have a negligible
effect on the final \dcaee template, by varying the maximum \pt requirement on the single
electron track \mbox{(\pt$<$ 3~GeV/$c$)}. Moreover, the charm and beauty \dcaee templates do not exhibit any
strong dependence on \ptee and $m_{{\rm ee}}$, as well as on the minimum electron \pt
requirement. The latter is varied from 0.4~\GeVoverc to
0.7~GeV/$c$. Therefore, the shape of the heavy-flavour MC
templates is assumed to be model independent, whereas their absolute
normalisation, i.e.\,the dielectron yields from charm- and beauty-hadron
decays in the given \ptee and
\mee range, is not. The same \dcaee
distributions are used for the two event generators, PYTHIA and
POWHEG, and normalised to the dielectron yield predicted in each mass
interval by the corresponding event generator.
               %%%%%%%%%%%
%\newpage
% !TEX root = alicepreprint_CDS.tex
\section{Results and discussion} \label{results}

\subsection{Comparison of the data to the cocktail}

\begin{figure}[h]
\centering
\includegraphics[scale=0.4]{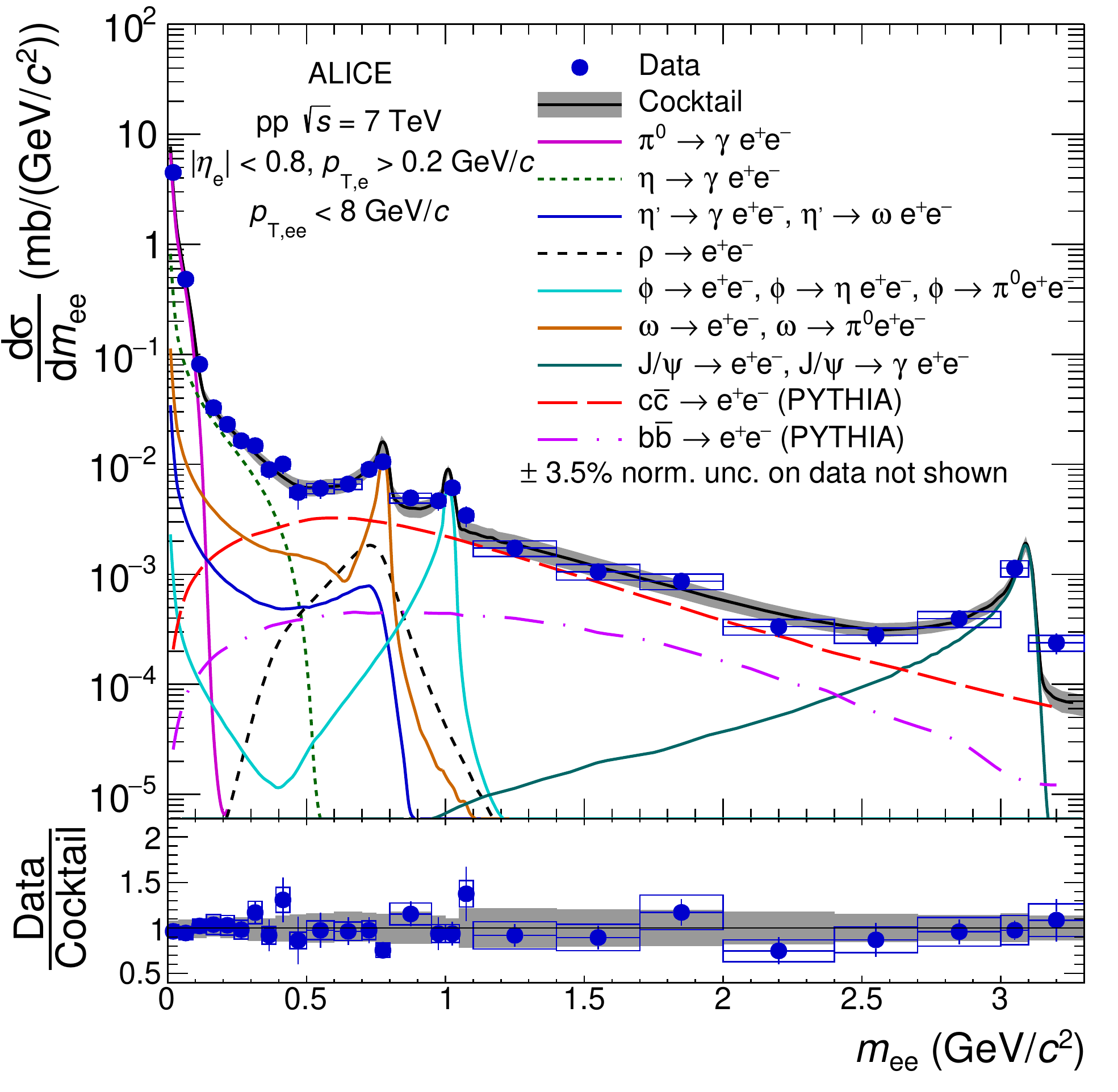}
\caption{(Colour online) Inclusive \ee cross section in \pp collisions at \roots = 7\,TeV in the ALICE acceptance as a function of mass. The data are compared with a cocktail of expected
  sources. In the lower panel, the ratio of data to cocktail is
  shown. Statistical and systematic uncertainties of the data are
  plotted as vertical bars and boxes, respectively. The total uncertainty of the
  cocktail is represented as a grey band.}
\label{fig:invmassintegrated}
\end{figure}

The differential \ee cross section d$\sigma/{\rm{d}}m_{{\rm ee}}$ in minimum bias \pp collisions at \roots =
7\,TeV is presented in Fig.~\ref{fig:invmassintegrated} in the ALICE acceptance ($|\eta_{{\rm e}}|$ $<$ 0.8
and $p_{{\rm T,e}}$ $>$ 0.2\,GeV/$c$) as a function of $m_{{\rm ee}}$.
Statistical and systematic uncertainties of the data are indicated by vertical
bars and boxes, respectively. The measured spectrum is compared with the
cocktail of expected \ee sources, where PYTHIA is used to
calculate the correlated pairs from heavy-flavour decays. The total
systematic uncertainty of the cocktail is shown by the grey band. The
bottom panel shows the ratio of data to cocktail. Good
agreement is observed over the full mass range (\mee $<$ 3.3\,GeV/$c^{2}$).

For a more detailed discussion, the results are presented differentially below
in \ptee and \dcaee in four different mass regions, i.e.\,the
$\pi^{0}$ region (\mee $<$ 0.14\,GeV/$c^{2}$), the low-mass region
(0.14 $<$ \mee $<$ 1.1\,GeV/$c^{2}$), the IMR (1.1 $<$ \mee $<$
2.7\,GeV/$c^{2}$), and the $J/\psi$ region (2.7 $<$ \mee $<$ 3.3\,GeV/$c^{2}$).

\subsubsection{$\pi^{0}$ mass region}

\begin{figure}[ht!]
\begin{minipage}{18pc}
\begin{center}
\includegraphics[scale=0.4]{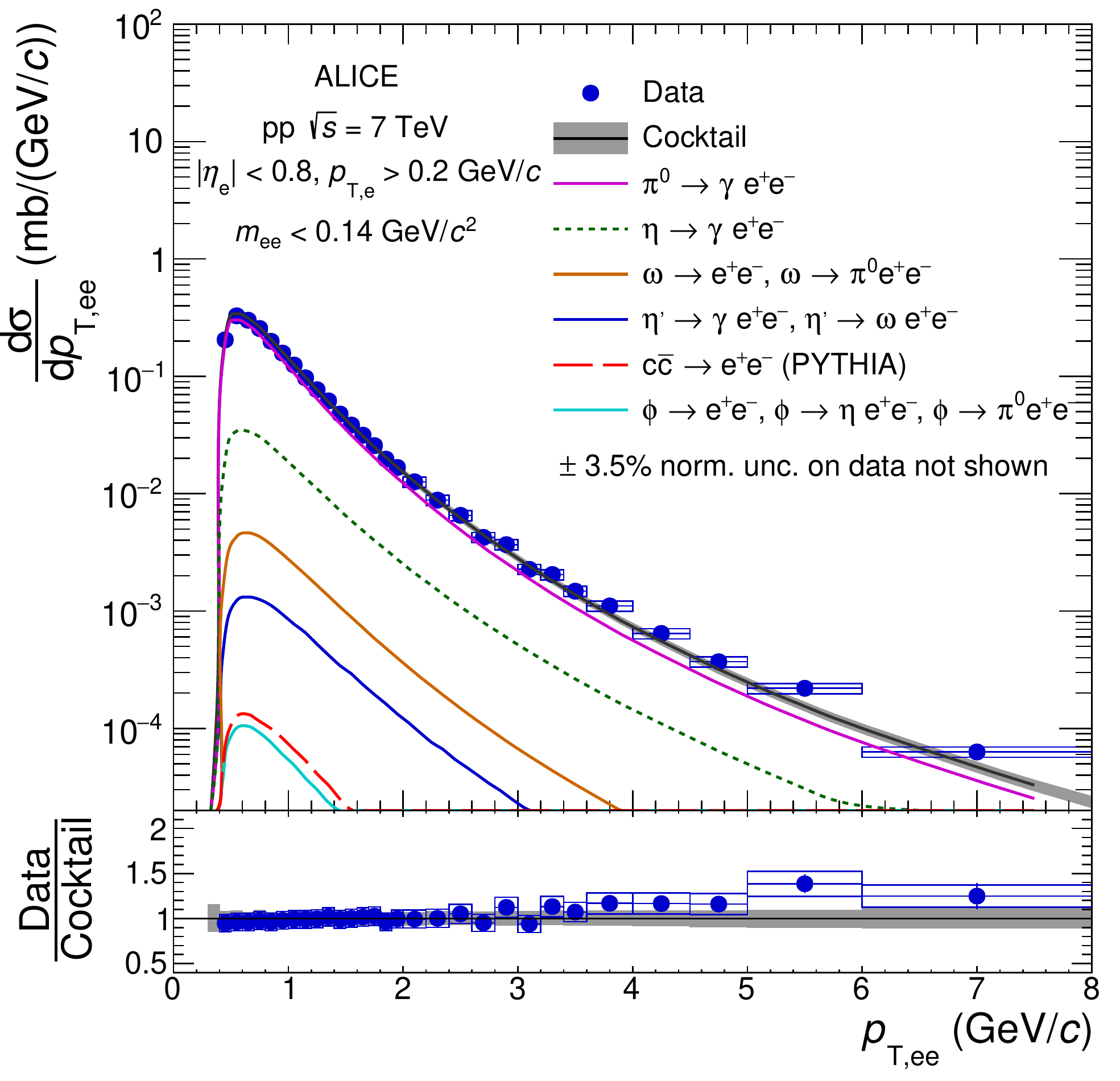}
\end{center}
\end{minipage}\hspace{1.5pc}%
\begin{minipage}{18pc}
\begin{center}
\includegraphics[scale=0.4]{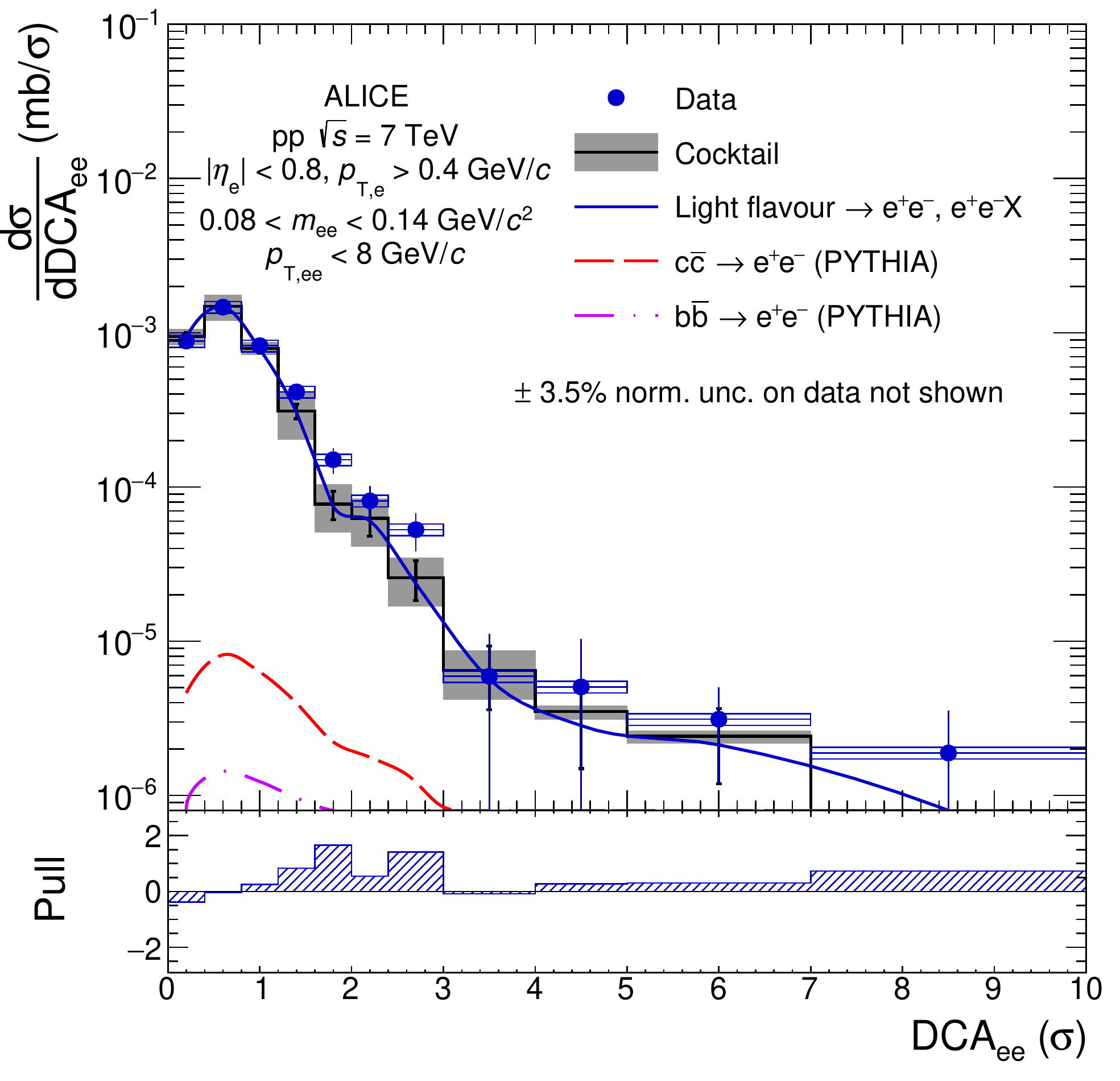}
\end{center}
\end{minipage}
\caption{(Colour online) Inclusive \ee cross section in \pp collisions
  at \roots = 7\,TeV in the ALICE acceptance as a function of \ptee (left) and \dcaee (right) for \mee $<$ 0.14~\GeVovercs and
  0.08 $<$ \mee $<$ 0.14~GeV/$c^{2}$, respectively. The data are compared
  with a cocktail of expected sources. In the bottom panels, the ratio
  of data to cocktail as a function of \ptee (left) and the pull distribution as a function of \dcaee (right) are shown. Statistical and
  systematic uncertainties on the data are plotted as vertical bars and boxes, respectively. The total uncertainty of the
  cocktail is represented as a grey band.}
\label{fig:pi0massregion}
\end{figure}

The mass region \mee $<$ 0.14 \GeVovercs is dominated by $\pi^{0}$
Dalitz decays ($\pi^{0} \to \rm{e}^{+}\rm{e}^{-} \gamma$), with a small contribution
from $\eta$ Dalitz decays ($\eta \to \rm{e}^{+}\rm{e}^{-} \gamma$) of
about 10\%. In the left panel of Fig.~\ref{fig:pi0massregion}, the
measured $p_{{\rm T,ee}}$-differential cross section of
\ee pairs is shown in comparison with the hadronic
cocktail. Good agreement between data and cocktail is observed within the systematic
uncertainties. This confirms that the dielectron analysis is consistent with the
previous light-meson measurements~\cite{chargedpi,pi0eta,phi,ALICE-PUBLIC-2018-004} taken as input for the calculations of
the expected \ee cross section.

In the right panel of Fig.~\ref{fig:pi0massregion}, the measured
\ee cross section is shown as a function of \dcaee in the mass range 0.08
$<$ \mee $<$ 0.14 \GeVovercs for \ptee $<$ 8\,GeV/$c$. The results are
compared with the expectations from MC. The low-mass cut-off at 0.08~GeV/$c^{2}$ is chosen such that residual
contaminations of \ee pairs from photon conversions with
large \dcaee values are suppressed. The blue line
represents the expected cross section of all prompt light-flavour sources, for which the
$\pi^{0}$ \dcaee template is used as an approximation. Some small
statistical fluctuations can be seen in the tail of the distribution,
which would require a very large amount of fully simulated pp
collisions to be removed. In this mass
range, the contributions from non-prompt sources (\ccbar and ${\rm{b}}{\overline{{\rm{b}}}}$) are
negligible, which allows the \dcaee resolution in
data and in MC to be directly compared. To quantify the agreement between data and the
expected \dcaee distribution from MC, the pull distribution is shown
in the bottom right panel of Fig.~\ref{fig:pi0massregion}.
It is defined as the difference between data and MC normalised by the quadratic sum of their statistical and systematic uncertainties.
The \dcaee distribution obtained from
the full simulations of the ALICE detector describes the data well. A
slight excess of the data is observed in 1 $<$ \dcaee $<$ 3
$\sigma$ which is the range mostly affected by discrepancies
in the DCA resolution between data and MC.

\subsubsection{Low-mass region}

\begin{figure}[ht!]
\begin{minipage}{18pc}
\begin{center}
\includegraphics[scale=0.4]{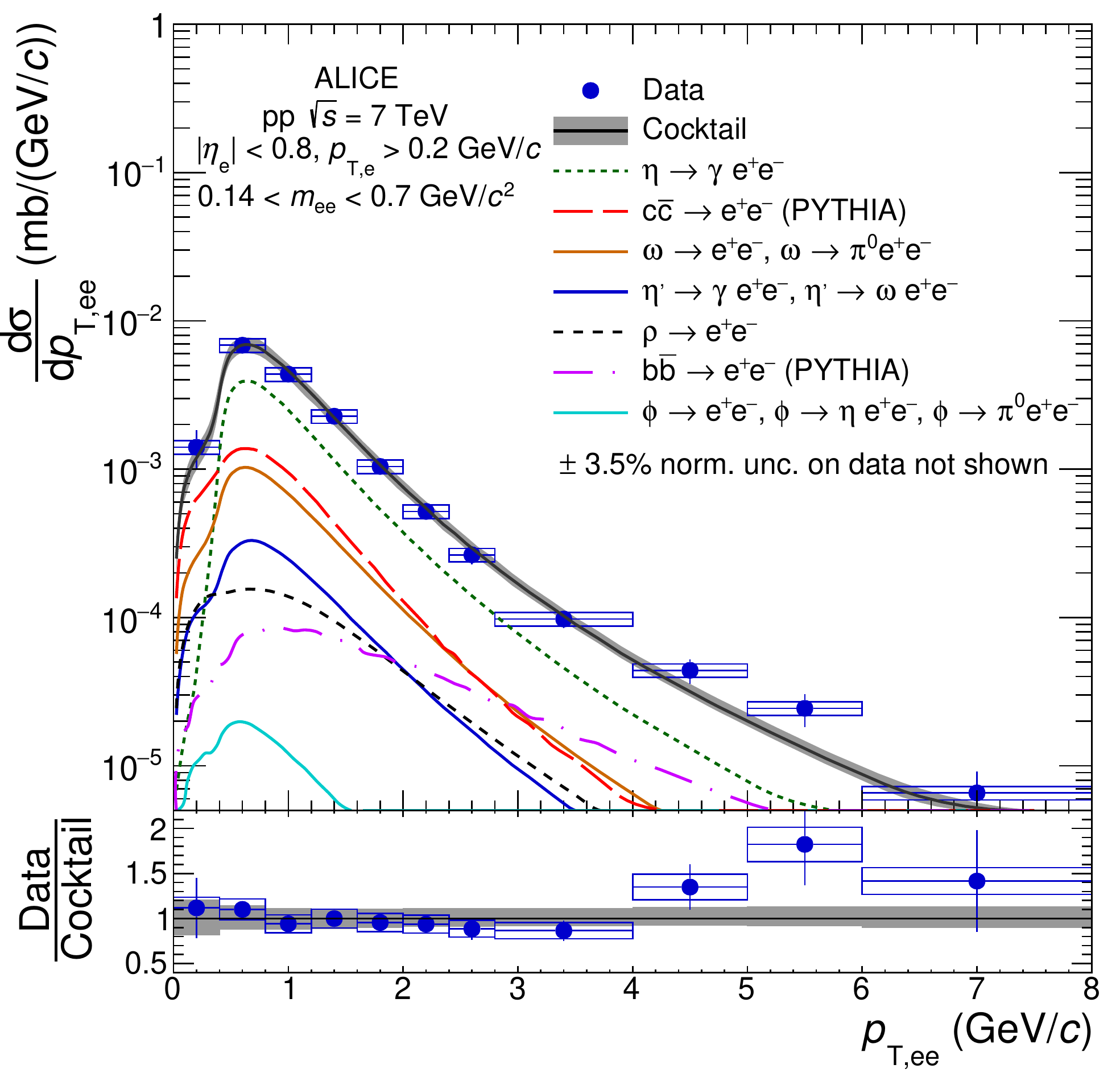}
\end{center}
\end{minipage}\hspace{1.5pc}%
\begin{minipage}{18pc}
\begin{center}
\includegraphics[scale=0.4]{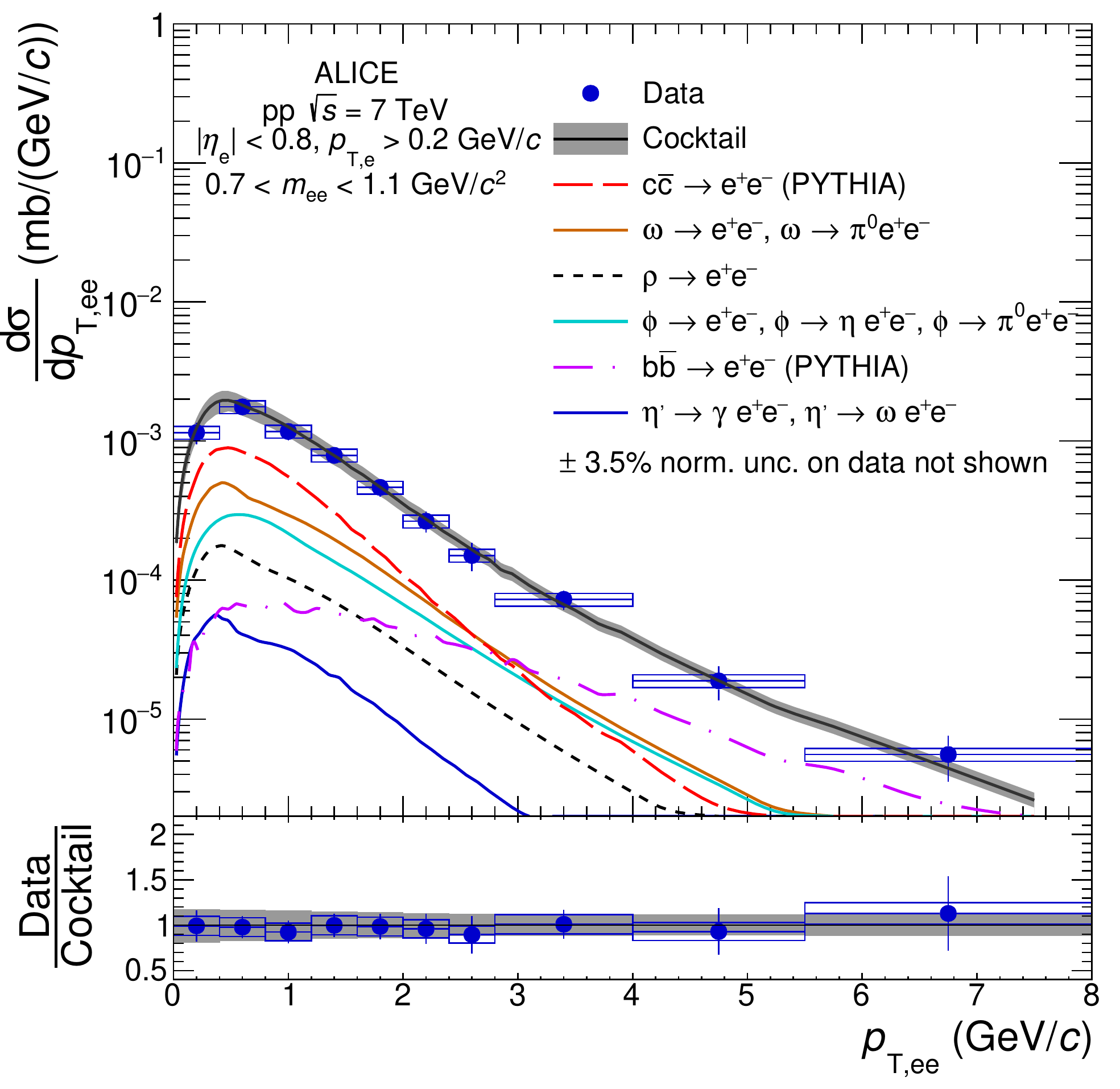}
\end{center}
\end{minipage}
\caption{(Colour online) Inclusive \ee cross section in \pp collisions at \roots = 7\,TeV in the ALICE acceptance as a function of \ptee in the mass range
  0.14 $<$ \mee $<$ 0.7~\GeVovercs (left) and
  0.7 $<$ \mee $<$ 1.1~\GeVovercs (right). The data are compared with
  the hadronic cocktail. In the bottom panels, the ratios
  of data to cocktail as a function of \ptee are shown. Statistical and
  systematic uncertainties on the data are plotted as vertical bars and boxes, respectively. The total uncertainty of the
  cocktail is represented as a grey band.}
\label{resonanceregion}
\end{figure}

The low-mass region, $0.14 < m_{\rm ee} < 1.1$ GeV/$c^{2}$, is expected to be dominated by the light mesons $\eta$, $\eta^{'}$,
$\rho$, $\omega$, $\phi$, and the contribution of
correlated \ee pairs from semileptonic decays of charm hadrons. A very
small contribution of virtual direct photons is also expected (3--5\%
of the total measured yield in 0.14 $<$ \mee $<$ 0.7\,GeV/$c^{2}$ and
4 $<$ \ptee $<$ 8\,GeV/$c$). The
latter is not included in the hadronic cocktail and will be discussed
in section~\ref{sectionvirtualphoton}. At low
\mee (\mee $\le$ $m_{\eta}$), the $\eta$ Dalitz decay is the
main source of \ee pairs for all $p_{{\rm T,ee}}$,
as shown in the left panel of Fig.~\ref{resonanceregion},
whereas at larger $m_{\rm ee}$, the heavy-flavour contributions start to dominate, followed by the $\omega$, $\rho$, and $\phi$ contributions (see right panel of Fig.~\ref{resonanceregion}). The requirement on the single
electron track of \pt~$>$~0.2\,\GeVoverc produces an acceptance hole at low \mee and
$p_{\rm{T,ee}}$, which can be seen in the data in the \mee interval \mbox{0.14~$<$~\mee$<$~0.7}\,GeV/$c^{2}$ for the \ptee range \mbox{0~$<$~\ptee$<$~0.4}\,GeV/$c$ (see left panel of Fig.~\ref{resonanceregion}). Due to their characteristic \mee and \ptee
distributions, the \ee pairs of the various
expected sources are differently affected. The hadronic cocktail is
well in agreement with the data within the statistical and systematic uncertainties.

\begin{figure}[h]
\centering
\includegraphics[scale=0.4]{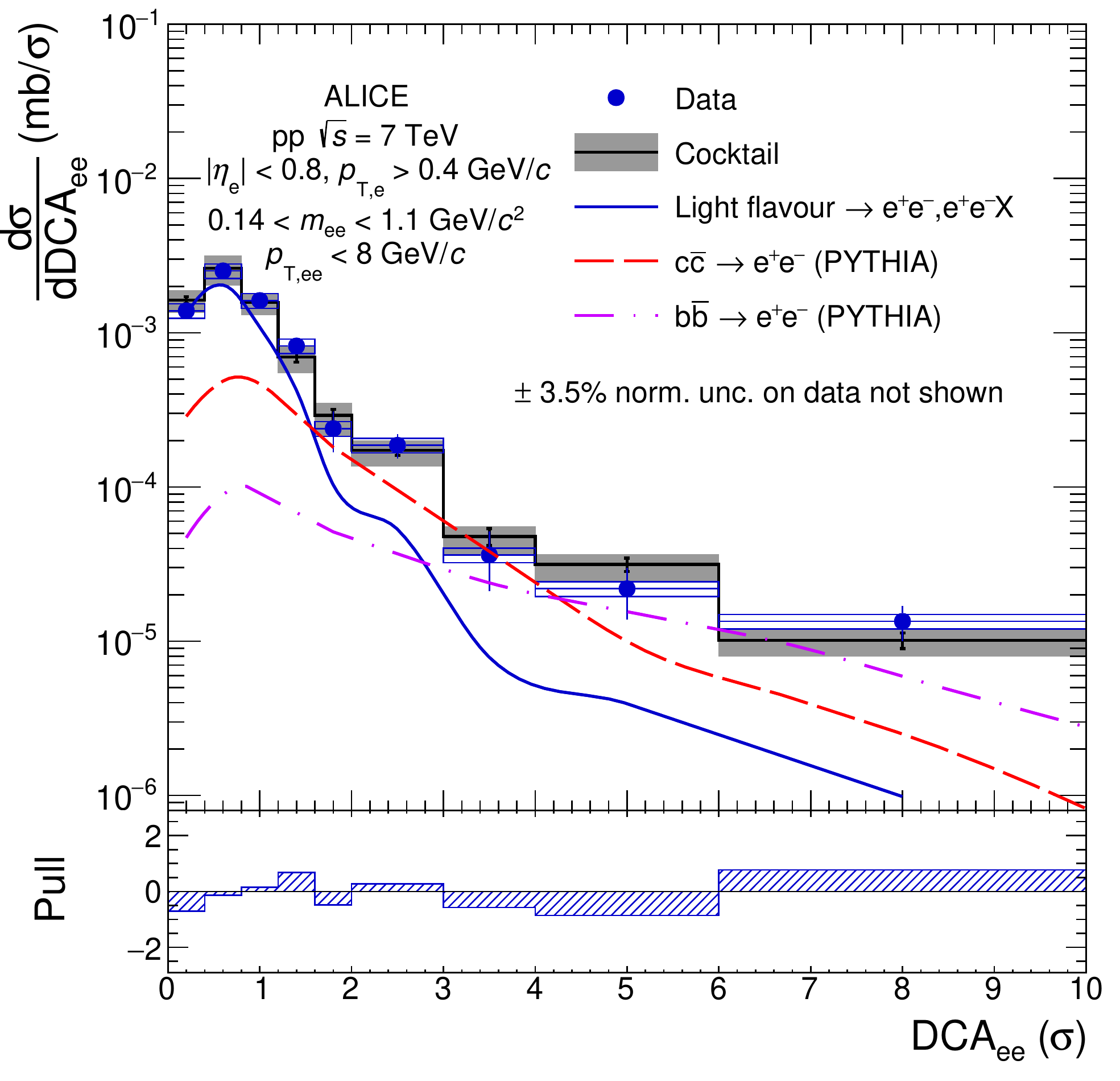}
\caption{(Colour online) Inclusive \ee cross section in \pp collisions at \roots = 7\,TeV in the ALICE acceptance as a function of \dcaee in the mass range 0.14 $<$ \mee $<$
  1.1~GeV/$c^{2}$. The data are compared with a cocktail of expected sources. In the bottom panel, the pull distribution is shown. Statistical and
  systematic uncertainties on the data are plotted as vertical bars and boxes, respectively. The total uncertainty of the
  cocktail is represented as a grey band.}
\label{fig:resonanceregiondca}
\end{figure}

The mixture of prompt and non-prompt sources in the low-mass region
makes it well suited to test the feasibility to separate prompt and
non-prompt contributions via the \dcaee
variable. In Fig.~\ref{fig:resonanceregiondca}, the
DCA$_{{\rm{ee}}}$-differential cross section
of \ee pairs is shown integrated over \ptee in the mass range \mbox{0.14~$<$~\mee$<$~1.1}\,GeV/$c^{2}$. The template for \ee pairs from
prompt light-flavour hadron decays cannot describe the tail of the
\dcaee distribution. The latter is well reproduced by additional contributions from
correlated pairs of heavy-flavour hadron decays. The good agreement
between data and MC validates the possibility to
separate prompt from non-prompt dielectron sources via this observable.

\subsubsection{Intermediate-mass region}

\begin{figure}[ht!]
\begin{minipage}{18pc}
\begin{center}
\includegraphics[scale=0.4]{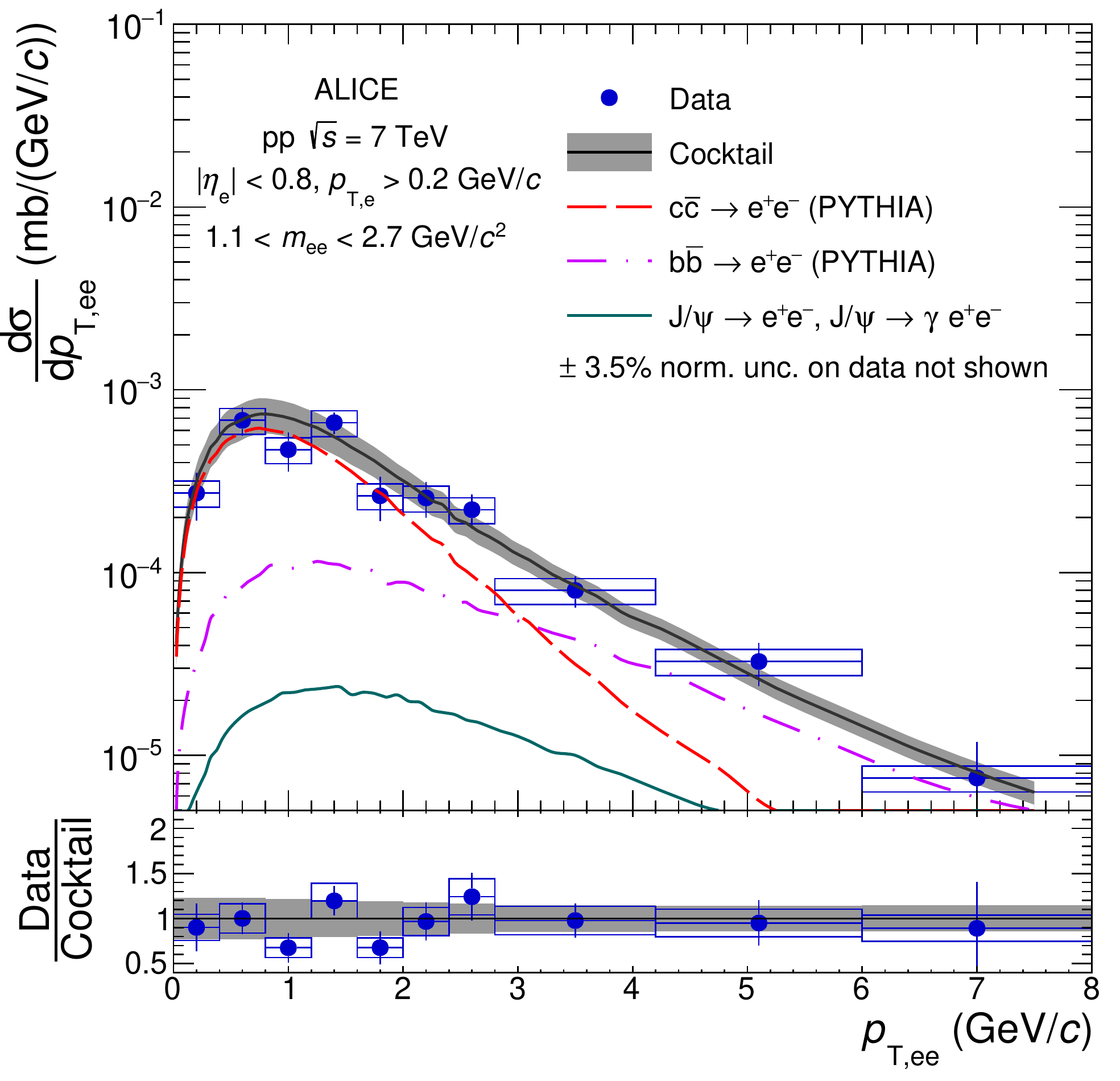}
\end{center}
\end{minipage}\hspace{1.5pc}%
\begin{minipage}{18pc}
\begin{center}
\includegraphics[scale=0.4]{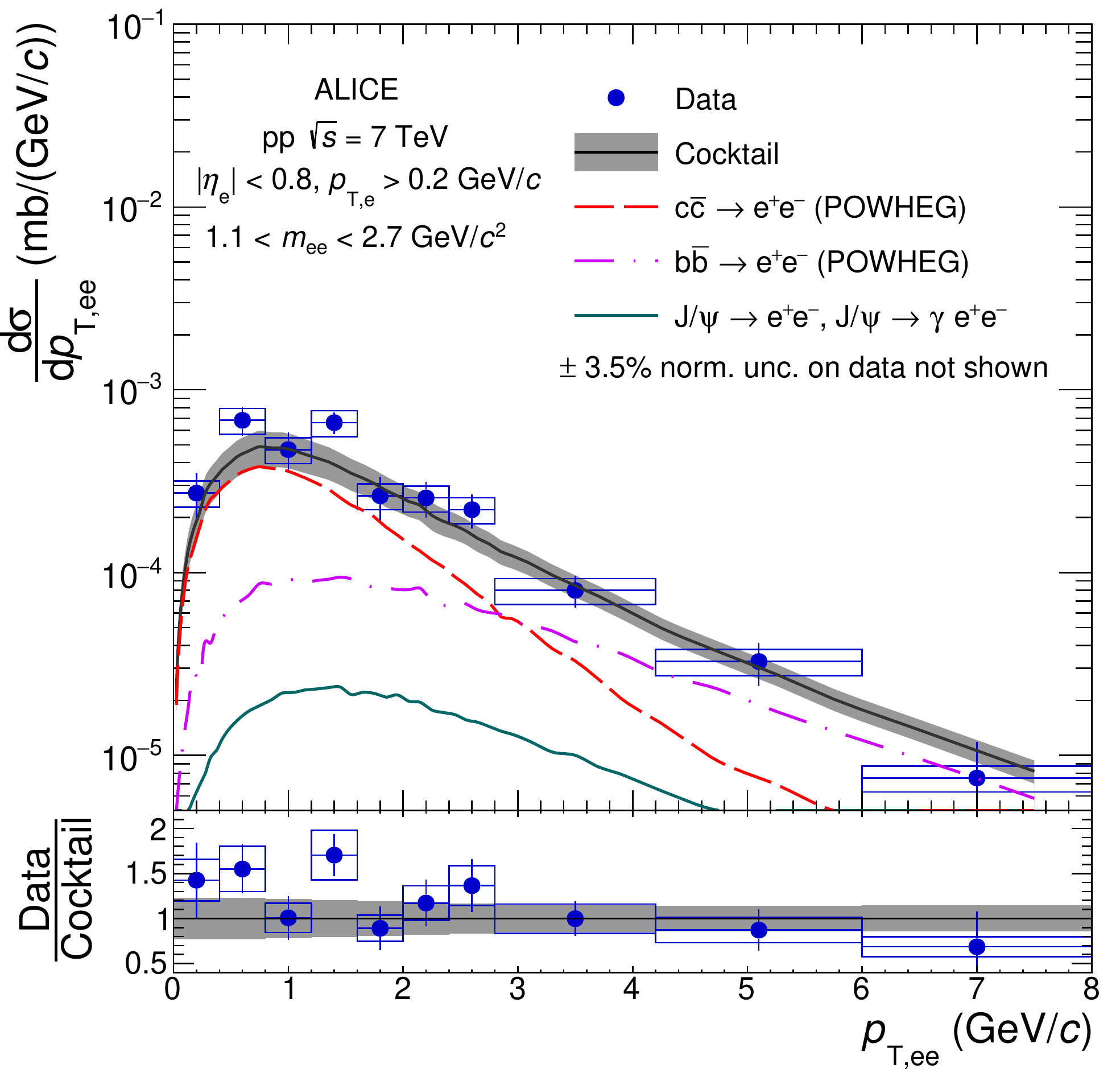}
\end{center}
\end{minipage}
\caption{(Colour online) Inclusive \ee cross section in \pp collisions at \roots =
  7\,TeV in the ALICE acceptance as a function of \ptee in the mass range 1.1 $<$ \mee $<$
  2.7~GeV/$c^{2}$. The data are compared with the hadronic cocktail, where PYTHIA~\cite{PYTHIA6425}  (left) and
  POWHEG~\cite{POWHEGa,POWHEGb,POWHEGboxa,POWHEGboxb} (right) are used to calculate the \ccbar and
  \bbbar contributions. In the bottom panels, the corresponding ratios
  of data to cocktail as a function of \ptee are shown. Statistical and
  systematic uncertainties on the data are plotted as vertical bars and boxes, respectively. The total uncertainty of the
  cocktails is represented as a grey band.}
\label{fig:heavyflavourregion}
\end{figure}

The IMR, 1.1 $<$ \mee $<$ 2.7 GeV/$c^{2}$, is dominated by correlated \ee pairs from semileptonic decays of charm and  beauty
hadrons. The $p_{{\rm{T,ee}}}$-differential cross section of \ee pairs
measured in this \mee region is shown in comparison with the hadronic
cocktail in the left panel of Fig.~\ref{fig:heavyflavourregion}. The cross section of \ee pairs from \ccbar is the dominant dielectron source for {\mbox{\ptee $<$ 3\,GeV/$c$}}, whereas
most of the \ee pairs originate from \bbbar for \ptee $>$ 4\,GeV/$c$. This
allows the correlated pairs from semileptonic decays of
charm and  beauty hadrons to be separated. Reasonable
agreement between data and cocktail is
seen in the bottom left
panel of Fig.~\ref{fig:heavyflavourregion}. The data are compared
with a hadronic cocktail in the
right panel of Fig.~\ref{fig:heavyflavourregion} where POWHEG is used to calculate the
\ccbar and \bbbar contributions. The NLO event generator predicts
harder \ptee spectra for the \ccbar and \bbbar contributions. For the same
global normalisation to $\sigma_{\ccbar}^{{\rm REF}}$ and
$\sigma_{\bbbar}^{{\rm REF}}$ as for the PYTHIA cocktail, the POWHEG calculations
tend to underestimate the data, in particular in
the region where the \ccbar contribution dominates.
This indicates a sensitivity of the present data to the underlying
heavy-quark production mechanism implemented in the two models. The
latter can result in different kinematic correlations of the $Q\overline{Q}$ pair and therefore different probabilities for the \ee pair to enter the detector acceptance.

\begin{table}[ht!]
\begin{center}
\centering
\begin{tabular}{llll}
  \hline \hline
  \ccbar & PYTHIA  & POWHEG & PYTHIA/POWHEG \\
\hline
4$\pi$ & 1. & 1. &\\\hline
$|\eta_{\rm e}|$ $<$ 0.8  & 0.0754 & 0.0518 & 1.46 \\
 (uncorrelated e$^{+}$e$^{-}$) & (0.0428) & (0.0299) & (1.43) \\\hline

$|\eta_{\rm e}|$ $<$ 0.8  & 0.0148 & 0.0100 & 1.48\\
1.1 $<$ \mee $<$ 2.7\,\GeVovercs &  & & \\\hline

$|\eta_{\rm e}|$ $<$ 0.8 $\&$ $p_{{\rm T,e}} >$ 0.2\,\GeVoverc & 0.0146 & 0.0098 & 1.49 \\
1.1 $<$ \mee $<$ 2.7\,\GeVovercs  &  & & \\\hline

$|\eta_{\rm e}|$ $<$ 0.8 $\&$ $p_{{\rm T,e}} >$ 0.4\,\GeVoverc & 0.0115 & 0.0077 & 1.49 \\
1.1 $<$ \mee $<$ 2.7\,\GeVovercs  &  & & \\
\hline \hline \\[-1.0em]
\bbbar & PYTHIA & POWHEG & PYTHIA/POWHEG\\
\hline
4$\pi$ & 1 & 1  &\\\hline
$|\eta_{\rm e}|$ $<$ 0.8 & 0.1250 & 0.1167 & 1.07 \\
(uncorrelated e$^{+}$e$^{-}$) & (0.0581) & (0.0506) & (1.15) \\\hline

$|\eta_{\rm e}|$ $<$ 0.8 & 0.0495 & 0.0472 & 1.05\\
1.1 $<$ \mee $<$ 2.7\,\GeVovercs &  & & \\\hline

$|\eta_{\rm e}|$ $<$ 0.8 $\&$ $p_{{\rm T,e}} >$ 0.2\,\GeVoverc & 0.0484 & 0.0460 & 1.05 \\
1.1 $<$ \mee $<$ 2.7\,\GeVovercs  &  & & \\\hline

$|\eta_{\rm e}|$ $<$ 0.8 $\&$ $p_{{\rm T,e}} >$ 0.4\,\GeVoverc & 0.0413 & 0.0390 & 1.06 \\
1.1 $<$ \mee $<$ 2.7\,\GeVovercs  &  & & \\

\hline \hline
\end{tabular}
\caption{Fraction of correlated \ee pairs in 4$\pi$ and after consecutive
  acceptance selection criteria (left column) for two different event generators (PYTHIA/POWHEG) and their relative difference (right column).}
\label{table:acceptancepp}
\end{center}
\end{table}

Table~\ref{table:acceptancepp} summarises the fraction of correlated \ee
pairs in the full phase space (4$\pi$) and after consecutive acceptance selection criteria for the \ccbar and \bbbar
contributions. The fraction of dielectron pairs from charm-hadron
decays  where both electrons are found at mid-rapidity
($|\eta_{{\rm e^{\pm}}}|$ $<$ 0.8) is about 5.2\% and 7.5\% for the POWHEG and PYTHIA
simulation, respectively. Since the hadronisation of the charm and  beauty quarks, as well
as the decay kinematics of the heavy-flavour hadrons, are the same in
both calculations, this difference results from different treatments of the various
production processes of the \ccbar pair by the two event
generators. First, the rapidity distribution of charm quarks predicted by
POWHEG is slighly broader than the one from PYTHIA, leading to a
smaller probability for single electrons to fall into the acceptance
at mid-rapidity in POWHEG (17.3\%) as compared to PYTHIA (20.7\%). Second, the pseudorapidity
correlation between the electron and positron from charm-hadron decays
gives rise to a larger acceptance for \ee pairs at
mid-rapidity than from a purely random
correlation. The pseudorapidity correlation is model-dependent which increases the difference in
acceptance between the two generators to about 46\%. For electrons from
 beauty-hadron decays, the model dependences are smaller, on the order of 7\%. The rapidity distributions of  beauty quarks predicted by
POWHEG and PYTHIA are quite similar. Moreover, about 50\% of the correlated \ee pairs from  beauty
hadron decays originate from a single B hadron (${\rm B}\to\overline{\rm
    D}{\rm e}^{+} {\rm X}\to{\rm e}^{+}{\rm e}^{-}{\rm X}$) and are insensitive to the
correlations between the B and $\overline{\rm{B}}$ hadrons. Due to the large mass of the B
hadrons, the correlation between the decay electron and the parent
meson is diluted and the pseudorapidity correlation of the
\ee pairs originating from different B hadrons is less related to the
correlation between the b and $\overline{\rm b}$ but more driven by
decay kinematics.

\begin{figure}[h]
\centering
\includegraphics[scale=0.4]{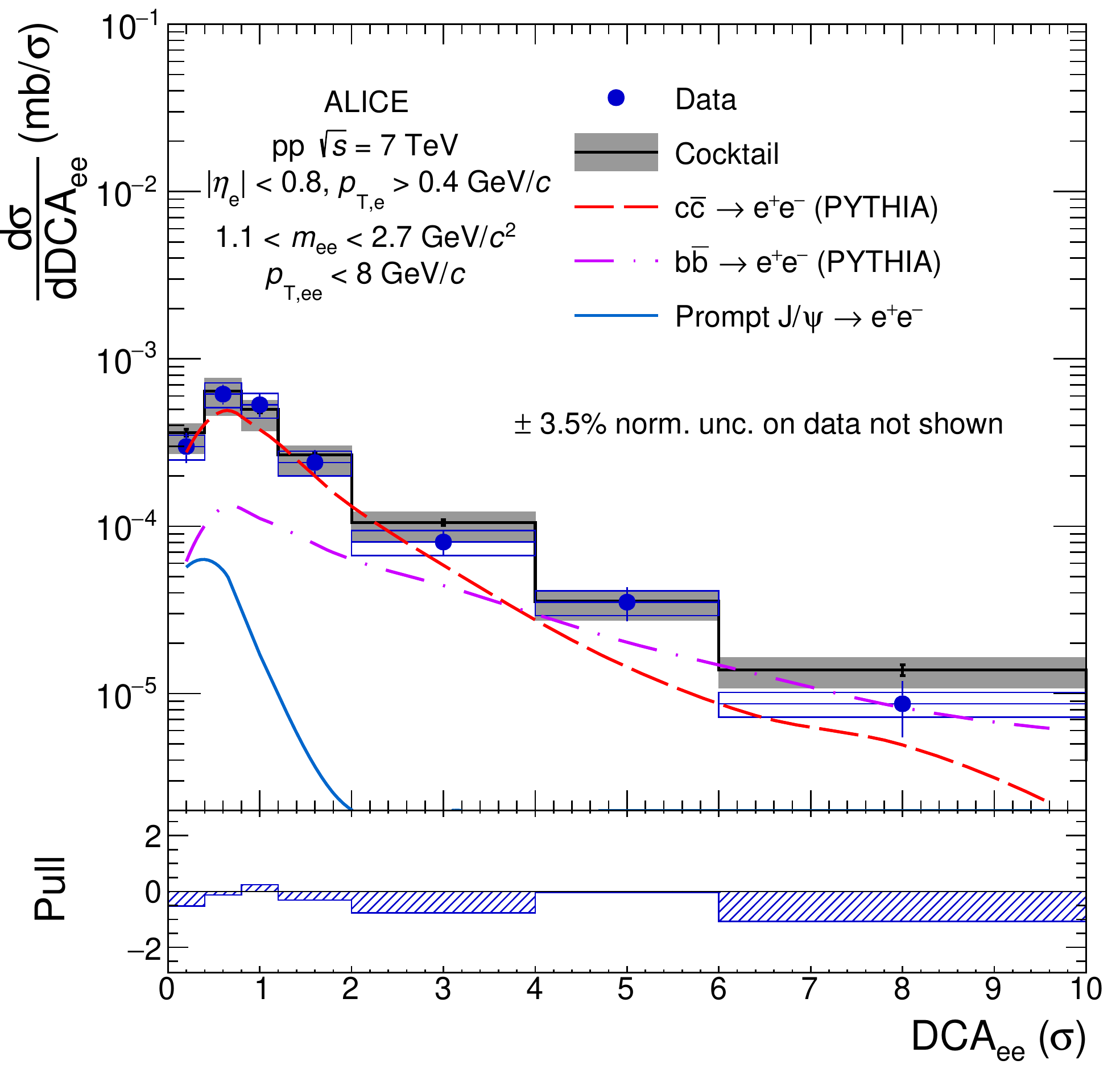}
\caption{(Colour online) Inclusive \ee cross section in \pp collisions at \roots =
  7\,TeV in the ALICE acceptance as a function of \dcaee in the mass range 1.1 $<$ \mee $<$ 2.7~GeV/$c^{2}$. The data are compared
  with a cocktail of expected sources. In the bottom panel, the pull
  distribution is shown. Statistical and
  systematic uncertainties on the data are plotted as vertical bars and boxes, respectively. The total uncertainty of the
  cocktail is represented as a grey band.}
\label{fig:heavyflavourregiondca}
\end{figure}

The measured \dcaee distribution of \ee pairs is shown in Fig.~\ref{fig:heavyflavourregiondca} integrated over \ptee in the mass range
1.1 $<$ \mee $<$ 2.7\,GeV/$c^{2}$. The results are compared with the MC
templates normalised to the PYTHIA cocktail. The shape of the MC \dcaee distribution
for correlated \ee pairs from charm-hadron decays deviates
from the data at large DCA$_{{\rm{ee}}}$. The additional contribution from
\ee pairs from  beauty-hadron decays allows the data to be well described, as can be seen with the pull distribution presented in the bottom panel
of Fig.~\ref{fig:heavyflavourregiondca}. The \dcaee variable gives
additional constraints to separate \ee pairs from charm- and  beauty-hadron decays. No indication for a prompt
source is observed in the IMR.

The total \ccbar and \bbbar cross sections, $\sigma_{\ccbar}$ and $\sigma_{\bbbar}$, can be extracted from the
data by fitting the measured \ee cross section of heavy-flavour hadron decays in the IMR with the sum of two contributions:
\begin{equation}
f^{{\rm{GEN}}}= S_{\ccbar} f^{{\rm{GEN}}}_{\ccbar} +  S_{\bbbar} f^{{\rm{GEN}}}_{\bbbar},
\label{fitfunctionhv}
\end{equation}
where $f^{{\rm{GEN}}}_{\ccbar}$ and $f^{{\rm{GEN}}}_{\bbbar}$ are the cross sections for dielectron pairs from charm and  beauty-hadron decays, calculated with
the event generator GEN and normalised to  $\sigma_{\ccbar}^{{\rm REF}}$ and $\sigma_{\bbbar}^{{\rm REF}}$~\cite{totalccbar,totalbbbar}. The two fit parameters are
the scaling factors $S_{\ccbar}$ and $S_{\bbbar}$, defined also as:
\begin{equation}
\sigma_{\ccbar} = S_{\ccbar} \cdot \sigma_{\ccbar}^{{\rm  REF}},
\end{equation}
\begin{equation}
\sigma_{\bbbar} = S_{\bbbar} \cdot \sigma_{\bbbar}^{{\rm REF}},
\end{equation}
 The \ee spectra from heavy-flavour hadron decays are obtained by
 subtracting the expected cross section of \ee pairs originating from
 vector meson and $J/\psi$ decays, from the measured \ee
 distributions. In the mass range 1.1--2.7~GeV/$c^{2}$ these contributions
are small, of the order of 4\%. The fit is performed separately in \dcaee and in ($m_{{\rm
    ee}}$, $p_{{\rm T,ee}}$). For each combination of scaling factors, $S_{\ccbar}$ and $S_{\bbbar}$, the
$\chi^{2}$ value is calculated:
\begin{equation}
\chi^{2}=\mathlarger{\sum_{i=1}^{n}}\left(\frac{x_{{\rm i}}-\mu_{\rm i}}{\sqrt{(\sigma_{x_{\rm i}}^{{\rm{stat}}})^2+(\sigma_{\mu_{\rm i}}^{{\rm{stat}}})^2}}\right)^{2}.
\label{chi2}
\end{equation}
The values of the data points and MC calculations in bin $i$
are given by $x_{\rm i}$ and $\mu_{\rm i}$, respectively, while
$\sigma_{x_{\rm i}}^{{\rm{stat}}}$ and $\sigma_{\mu_{\rm i}}^{{\rm{stat}}}$ represent their
statistical uncertainties. The result of the fit is determined by the
minimum of $\chi^{2}$.
\begin{figure}[ht!]
\begin{minipage}{18pc}
\begin{center}
\includegraphics[scale=0.4]{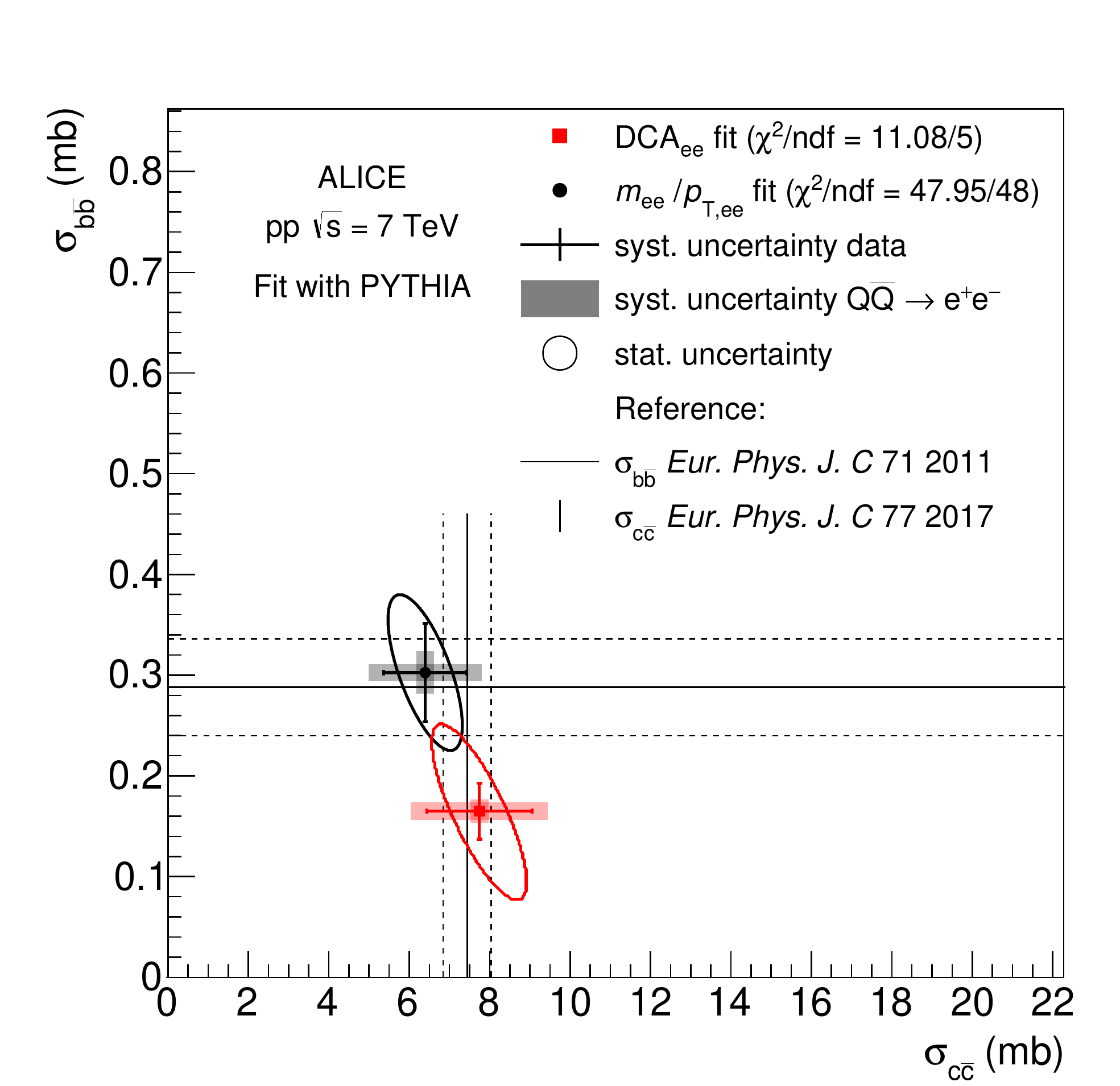}
\end{center}
\end{minipage}\hspace{1.5pc}%
\begin{minipage}{18pc}
\begin{center}
\includegraphics[scale=0.4]{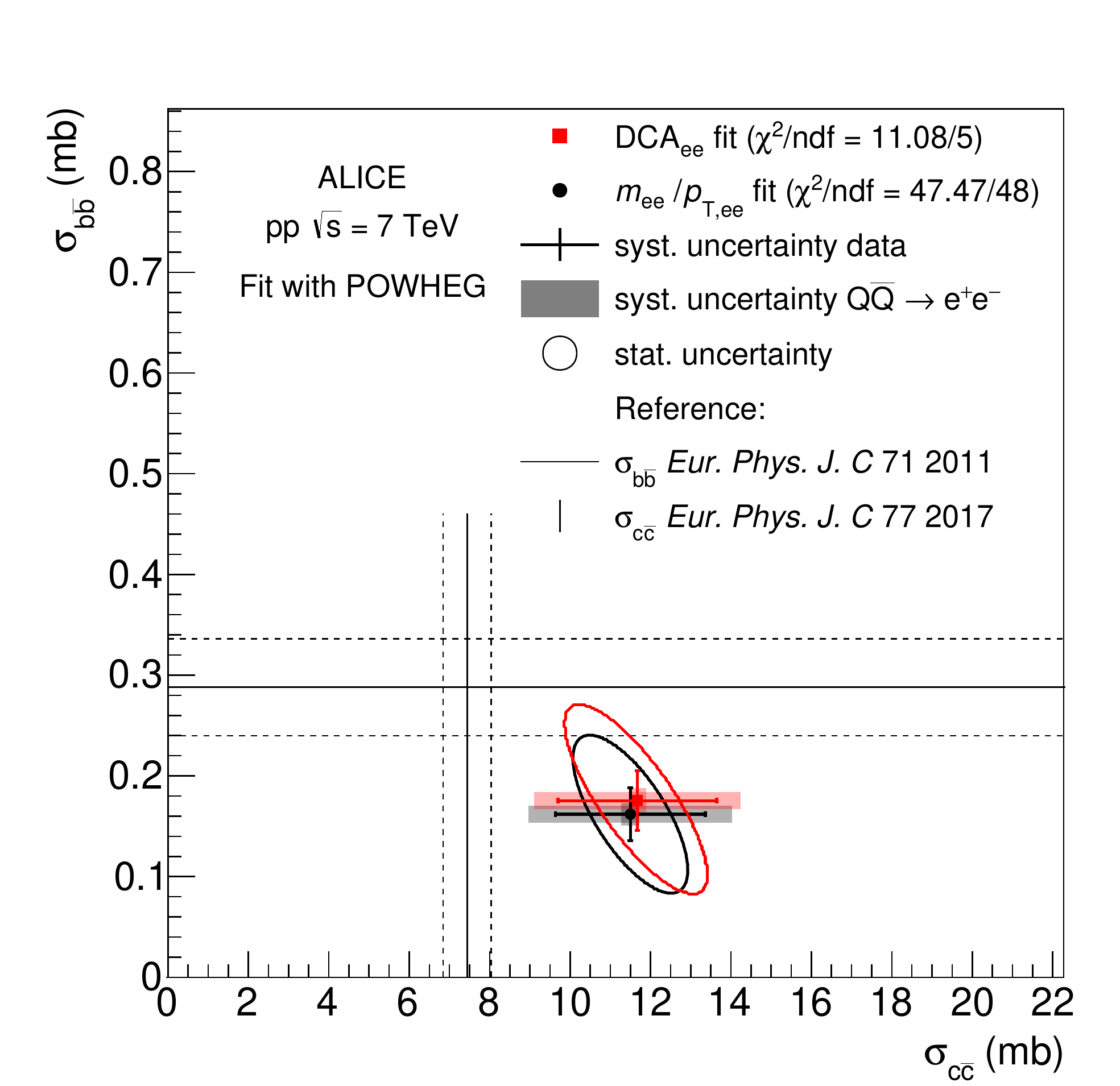}
\end{center}
\end{minipage}
\caption{(Colour online) Total \ccbar and \bbbar cross sections extracted from a fit
  of the measured dielectron yield from heavy-flavour hadron decays
  in \mbox{( $m_{{\rm ee}}$, $p_{{\rm T,ee}}$)} and in \dcaee with
  PYTHIA (left) and POWHEG (right). The results and their uncertainties (see text) are compared to
  published cross sections, for which the total uncertainty is
  represented by dashed lines.}
\label{fig:chi2fit}
\end{figure}
The extracted $\sigma_{\ccbar}$ and $\sigma_{\bbbar}$ cross sections are shown in Fig.~\ref{fig:chi2fit} for the  ($m_{{\rm ee}}$, $p_{{\rm T,ee}}$) and the \dcaee
analysis when PYTHIA (left) or POWHEG (right) are used to calculate
$f^{{\rm{GEN}}}_{\ccbar}$  and $f^{{\rm{GEN}}}_{\bbbar}$. The statistical uncertainties are plotted as ellipses and
correspond to a confidence level of 68.3\% (1$\sigma$) for each
parameter (at $\chi^{2} = \chi^{2}_{{\rm{min}}} + 1$~\cite{chi2error}). The error bars
represent the systematic uncertainties determined by the fit result after moving the data
points coherently up- and downward by their systematic uncertainties. The
uncertainties of the effective  beauty- and charm-to-electron branching
ratios are shown as coloured bands. Finally the full and dashed lines show $\sigma_{\ccbar}^{{\rm REF}}$~\cite{totalccbar} and $\sigma_{\bbbar}^{{\rm REF}}$~\cite{totalbbbar} with their total
uncertainties. The statistical and systematic uncertainties are fully
correlated between the PYTHIA- and POWHEG-based results, whereas they are
partially correlated between the ($m_{{\rm ee}}$, $p_{{\rm T,ee}}$) and \dcaee fits. Both calculations, PYTHIA and POWHEG, are able to
reproduce the ($m_{{\rm ee}}$, $p_{{\rm T,ee}}$) and \dcaee spectra reasonably well and give similar
minimum $\chi^{2}$ per number of degree of freedom (0.999 for POWHEG
and 0.989 for PYTHIA for the  \mbox{($m_{{\rm ee}}$, $p_{{\rm T,ee}}$)} fit). The fit results of the  ($m_{{\rm ee}}$, $p_{{\rm T,ee}}$) and \dcaee spectra are in agreement
within the statistical and systematic uncertainties. The \dcaee
distribution is slightly less sensitive to the total \bbbar cross section. The \ee pairs
from  beauty-hadron decays only dominate the last \dcaee bin
(see Fig.~\ref{fig:heavyflavourregiondca}). The shapes of the MC \dcaee
templates are driven by the decay kinematics and assumed to be model
independent. Therefore the extracted $\sigma_{\ccbar}$ and
$\sigma_{\bbbar}$ directly reflect the different probabilities for the
${\rm e}^+{\rm e}^-$ pair to enter the detector acceptance calculated with PYTHIA or POWHEG (see
Table~\ref{table:acceptancepp}). The ($m_{{\rm ee}}$, $p_{{\rm T,ee}}$) fit depends in
addition on the  $p_{\rm{T,ee}}$ distributions of correlated \ee pairs from charm and  beauty-hadron decays, which are harder in POWHEG
compared to PYTHIA. The total \ccbar and \bbbar cross sections show model
dependences of about a factor of two. The fitted cross sections are
summarised in Tables~\ref{table:crosssectioncc} and \ref{table:crosssectionbb}. For comparison, the total \ccbar
cross section obtained by extrapolating in rapidity the D$^{0}$
\pt spectrum measured by the ALICE Collaboration~\cite{totalccbar} is
of the order of 8.6\,mb with POWHEG, and 7\,mb with PYTHIA. The dielectron measurements can give further
constraints on the MC event generators aiming to reproduce the
heavy-flavour production mechanisms, once the uncertainties, which
are fully correlated between the PYTHIA- and POWHEG-based results, are reduced.

\begin{table}[ht!]
\begin{center}
  \centering
  \begin{tabular}{lll}
    \hline \hline
\ccbar & (\mee, $p_{{\rm T,ee}}$) fit  & \dcaee fit
     \\\hline
POWHEG & 11.6 $\pm$1.4 (stat.) $\pm$1.9 (syst.)\,mb & 11.7
                                                                $\pm$1.8
                                                                (stat.)
                                                                $\pm$2.0 (syst.)\,mb
     \\\hline
PYTHIA & 6.4 $\pm$0.9 (stat.) $\pm$1.1 (syst.)\,mb  &  7.7
                                                                $\pm$1.2
                                                                (stat.)
                                                                $\pm$1.3
                                                                (syst.)\,mb
     \\
\hline \hline
\end{tabular}
\caption{Summary of the total \ccbar cross sections extracted from a fit
    of the measured dielectron spectra from heavy-flavour hadron decays
    in \mbox{($m_{{\rm ee}}$, $p_{{\rm T,ee}}$)} and in \dcaee with PYTHIA and
    POWHEG. The uncertainty of 22\% on the branching fractions and
    fragmentation functions (BR$_{{\rm c \to e}}$) is not listed.}
\label{table:crosssectioncc}
\end{center}
\end{table}

\begin{table}[ht!]
\begin{center}
\centering
\begin{tabular}{lll}
  \hline \hline  \\[-1.0em]
  \bbbar & (\mee, $p_{{\rm T,ee}}$) fit  & \dcaee fit \\\hline
  POWHEG &  0.162 $\pm$0.078 (stat.) $\pm$0.026 (syst.)\,mb &  0.175 $\pm$0.092 (stat.) $\pm$0.030 (syst.)\,mb  \\\hline
  PYTHIA & 0.303 $\pm$0.077 (stat.) $\pm$0.050 (syst.)\,mb & 0.165 $\pm$0.086 (stat.) $\pm$0.028 (syst.)\,mb \\
  \hline  \hline
\end{tabular}
\caption{Summary of the total \bbbar cross sections extracted from a fit
  of the measured dielectron spectra from heavy-flavour hadron decays
  in \mbox{($m_{{\rm ee}}$, $p_{{\rm T,ee}}$)} and in \dcaee with
  PYTHIA and POWHEG. The uncertainty of 6\%~\cite{branchingratios} on
  the branching fractions and fragmentation functions (BR$_{{\rm b (\to c) \to e}}$) is not listed.}
\label{table:crosssectionbb}
\end{center}
\end{table}

\subsubsection{$J/\psi$ mass region}

\begin{figure}[ht!]
\begin{minipage}{18pc}
\begin{center}
\includegraphics[scale=0.4]{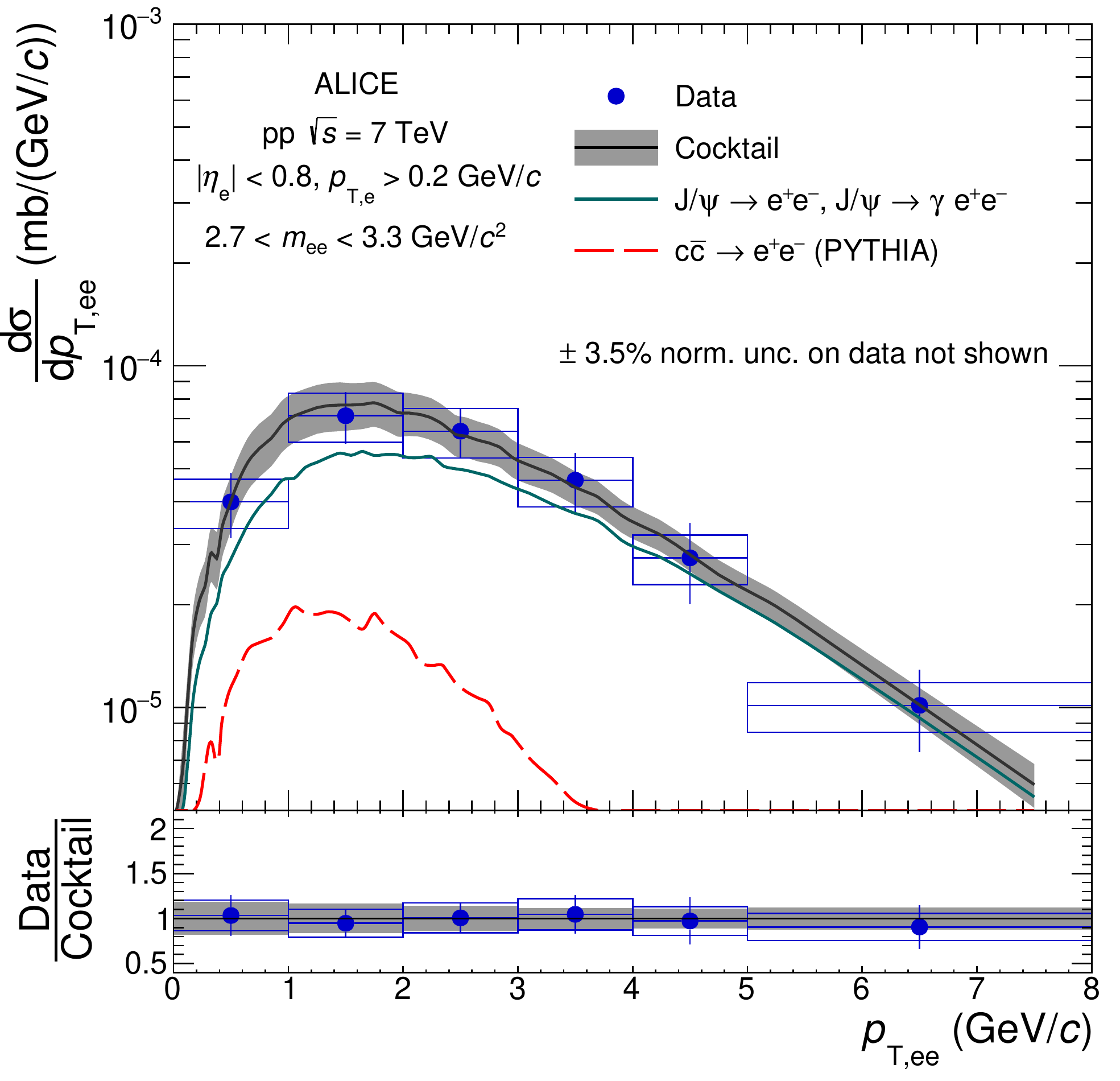}
\end{center}
\end{minipage}\hspace{1.5pc}%
\begin{minipage}{18pc}
\begin{center}
\includegraphics[scale=0.4]{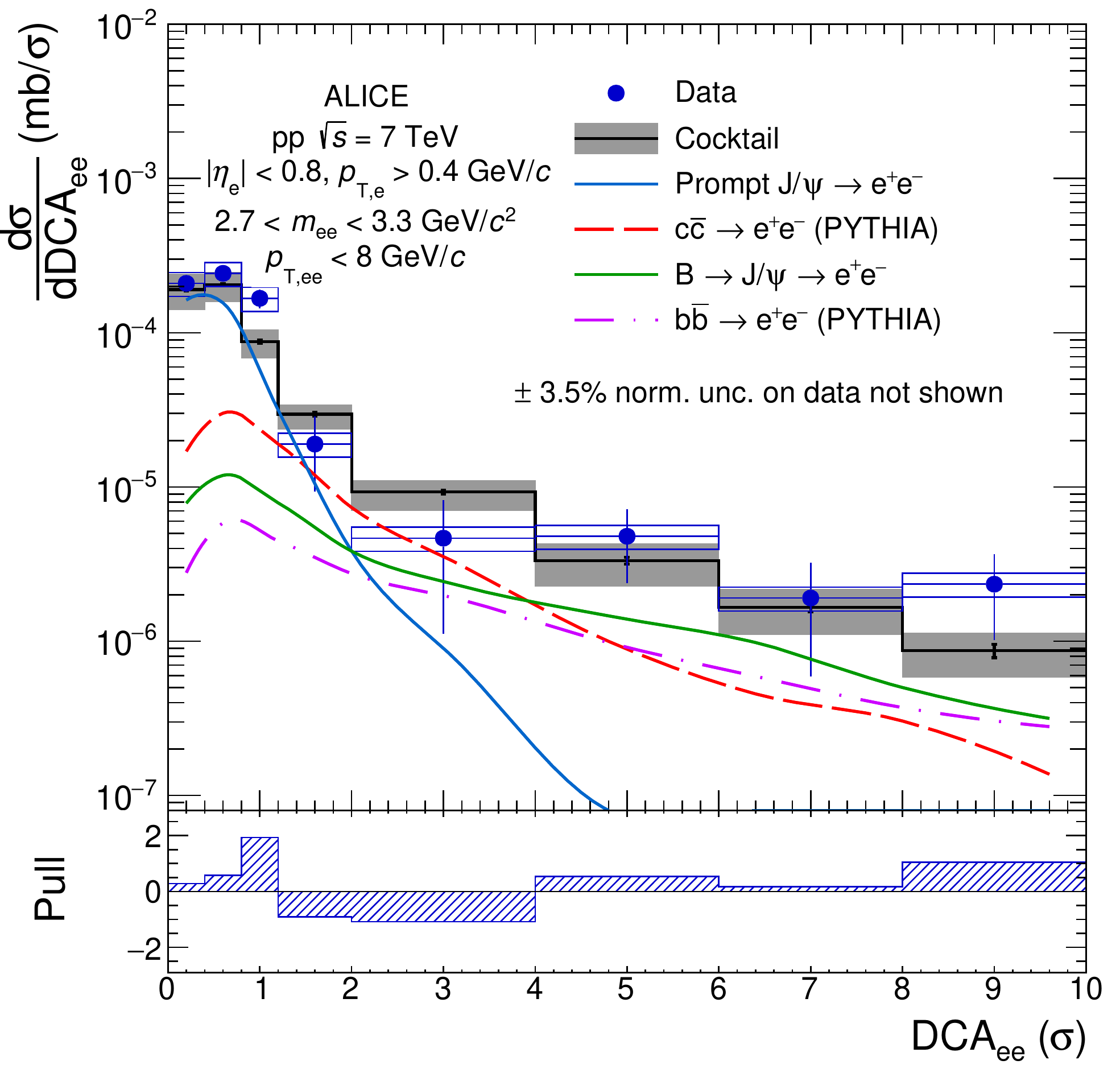}
\end{center}
\end{minipage}
\caption{(Colour online) Inclusive \ee cross section in \pp collisions at \roots = 7\,TeV in the ALICE acceptance as a function of \ptee (left panel) and \dcaee (right panel) in the
  mass range 2.7 $<$ \mee $<$ 3.3~GeV/$c^{2}$. The data are compared
  with a cocktail of expected sources. In the bottom panels, the ratio
  of data to cocktail as a function of \ptee (left) and the pull distribution as a
  function of \dcaee (right) are shown. Statistical and
  systematic uncertainties on the data are plotted as vertical bars and boxes, respectively. The total uncertainty of the
  cocktail is represented as a grey band.}
\label{fig:jpsiregion}
\end{figure}

The mass region 2.7 $<$ \mee $<$ 3.3 \GeVovercs is dominated by
$J/\psi$ decays with a small contribution from charm-hadron decays. In
the left panel of Fig.~\ref{fig:jpsiregion}, the corresponding measured
\ee cross section as a function of \ptee is shown in comparison with the hadronic cocktail. Good agreement between
data and cocktail is observed, as can be seen on the bottom left panel of
Fig.~\ref{fig:jpsiregion} in the ratio of data to cocktail. The \dcaee distribution of \ee pairs is sensitive to the large decay
length of B mesons ($c\tau_{{\rm B}}$ $\approx$ 470\,\textmu m) and
the contribution from $J/\psi$ originating from their decays. The measured
\dcaee spectrum shown in the right panel of Fig.~\ref{fig:jpsiregion}
cannot be reproduced with the MC template of the prompt $J/\psi$ alone. The contribution from non-prompt $J/\psi$, together with those from correlated \ee pairs from heavy-flavour hadron decays, leads
to a reasonable description of the data by the MC calculations. The
data are consistent with the fraction $f_{\rm B}$ of non-prompt $J/\psi$ originating from B meson
decays previously measured by the ALICE Collaboration~\cite{totalbbbara}.

\subsection{Direct photons}\label{sectionvirtualphoton}

Direct photons are defined as photons that do not originate from
hadronic decays. In \pp collisions, they are produced predominantly in hard partonic
interactions and their production rate can be calculated with
perturbative QCD.

The direct-photon cross section can be extracted from the measurement
of real photons detected in the electromagnetic calorimeters or via photon
conversion in the detector material of ALICE~\cite{directphoton2768,directphotonPbPb}. For \pt $<$
5\,GeV/$c$, the extraction of the direct photon signal is difficult
because of a large background from decay photons. An alternative way
to measure direct-photon production is via its internal conversion
into an \ee pair. The advantage of this approach is that the main
background originating from $\pi^{0}$ decays can be suppressed by
selecting \ee pairs with sufficiently large \mee (\mee $>$ $m_{\pi^{0}}$). The drawback is the small internal conversion probability of
$O(10^{-2})$ and the rapidly decreasing cross section as a
function of \mee ($\propto$1/$m_{{\rm ee}}$).

The mass dependence of the virtual-photon production for a given real-photon yield is given by the Kroll-Wada equation~\cite{KrollWada:formula} which can be simplified in
the case of $p_{{\rm T,ee}}^{{\rm 2}}$ $\gg$ $m_{{\rm ee}}^{{\rm 2}}$,
i.e.\,in the limit of quasi-real virtual photons, to:
\begin{equation}
\frac{{\rm d^{2}}N_{{\rm ee}}}{{\rm d}m_{{\rm ee}}{\rm d}p_{{\rm
      T,ee}}}=\frac{2\alpha}{3\pi}\sqrt{1-\frac{4m_{{\rm
        e}}^{2}}{m_{{\rm ee}}^{2}}}
\left(1+\frac{2m_{{\rm e}}^{2}}{m_{{\rm ee}}^{2}}\right)\cdot \frac{1}{m_{{\rm ee}}} \frac{{\rm d}
N_{\gamma}}{{\rm d}p_{{\rm T}}},
\label{KrollWada}
\end{equation}
with $\alpha$, ${\rm d}N_{\gamma}/{\rm d}p_{{\rm T}}$, and $m_{{\rm e}}$ being the fine-structure constant,
the number of real photons at a given \pt\mbox{($=$ $p_{{\rm T,ee}}$)}, and the
electron mass, respectively. To obtain the final expected shape $f_{\rm dir}(m_{{\rm ee}})$ of
the virtual direct photon mass distribution, the decay electrons are smeared by the detector resolution and passed through the acceptance
of the ALICE barrel ($|\eta_{{\rm e}}|$ $<$ 0.8, $p_{{\rm T,e}} >$ 0.2\,GeV/$c$).

\begin{figure}[h]
\centering
\includegraphics[scale=0.4]{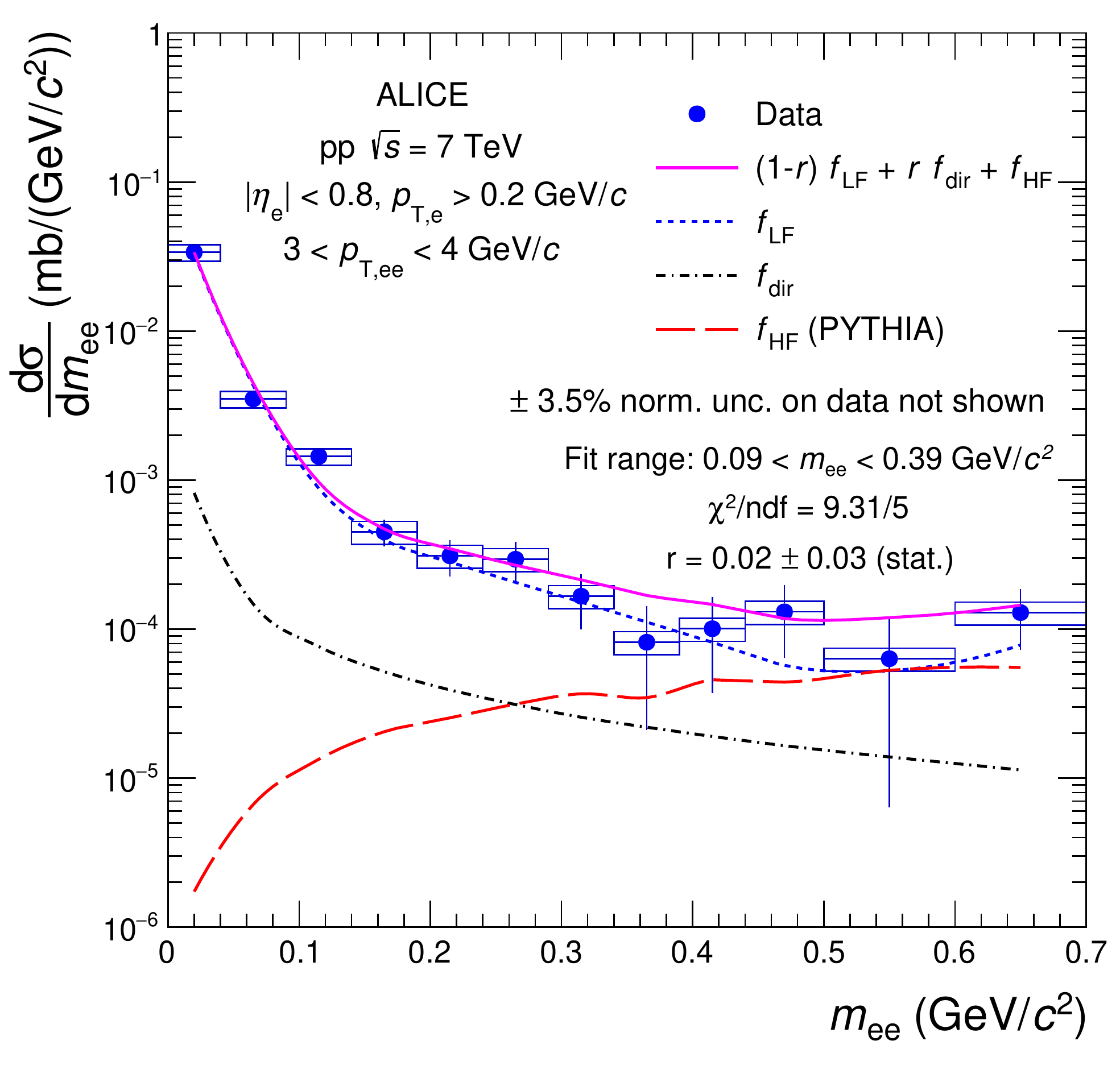}
\caption{Fit of the \ee cross section in \pp collisions at \roots =
  7\,TeV in the ALICE acceptance as a function of \mee in the range 3 $<$ \ptee $<$ 4~\GeVoverc with the three-component function defined by Eq. \eqref{fitfunction}. Statistical and
  systematic uncertainties on the data are shown separately as vertical bars and boxes, respectively.}
\label{fig:virtualphotonaa}
\end{figure}

The measured \mee distributions of e$^{+}$e$^{-}$ pairs are fitted in different \ptee bins with a three-component
function:
\begin{equation}
f(m_{{\rm ee}},r)=(1-r)f_{\rm LF}(m_{{\rm ee}})+ r\,f_{\rm dir}(m_{{\rm
    ee}})+f_{\rm HF}(m_{{\rm ee}}),
\label{fitfunction}
\end{equation}
where $f_{\rm LF}$ is the shape of the mass distribution of the
light-flavour component of the hadronic cocktail and $f_{\rm HF}$ is the \ee cross section of the expected heavy-flavour contribution in
the corresponding \ptee bin. The ratio $r$ of direct to
inclusive photons is the only fit parameter. The fit is limited to
the mass range 0.09 $<$ \mee $<$ 0.39\,\GeVovercs to ensure the condition {\mbox{$p_{{\rm T,ee}}^{\rm 2}$ $\gg$ $m_{{\rm
    ee}}^{\rm 2}$}}. Both $f_{\rm LF}$ and $f_{\rm dir}$ are
normalised such that they separately fit the data for \mee $<$
0.04\,GeV/$c^{2}$, because in this mass region the functional shapes of
$f_{{\rm LF}}$ and $f_{{\rm dir}}$ are essentially identical. In Fig.~\ref{fig:virtualphotonaa}, the measured
 $m_{{\rm ee}}$-differential \ee cross section is shown in the \ptee range \mbox{3
 $<$ \ptee$<$ 4}\,GeV/$c$, together with the fit result. The individual components are plotted
 separately. The systematic uncertainties due to the model dependence of the estimated e$^{+}$e$^{-}$ yield from \ccbar is evaluated by repeating the fit with $f_{\rm HF}$
 computed with the POWHEG generator and found to be below 0.75\% in the full $p_{{\rm T}}$ range.

\begin{figure}[h]
\centering
\includegraphics[scale=0.4]{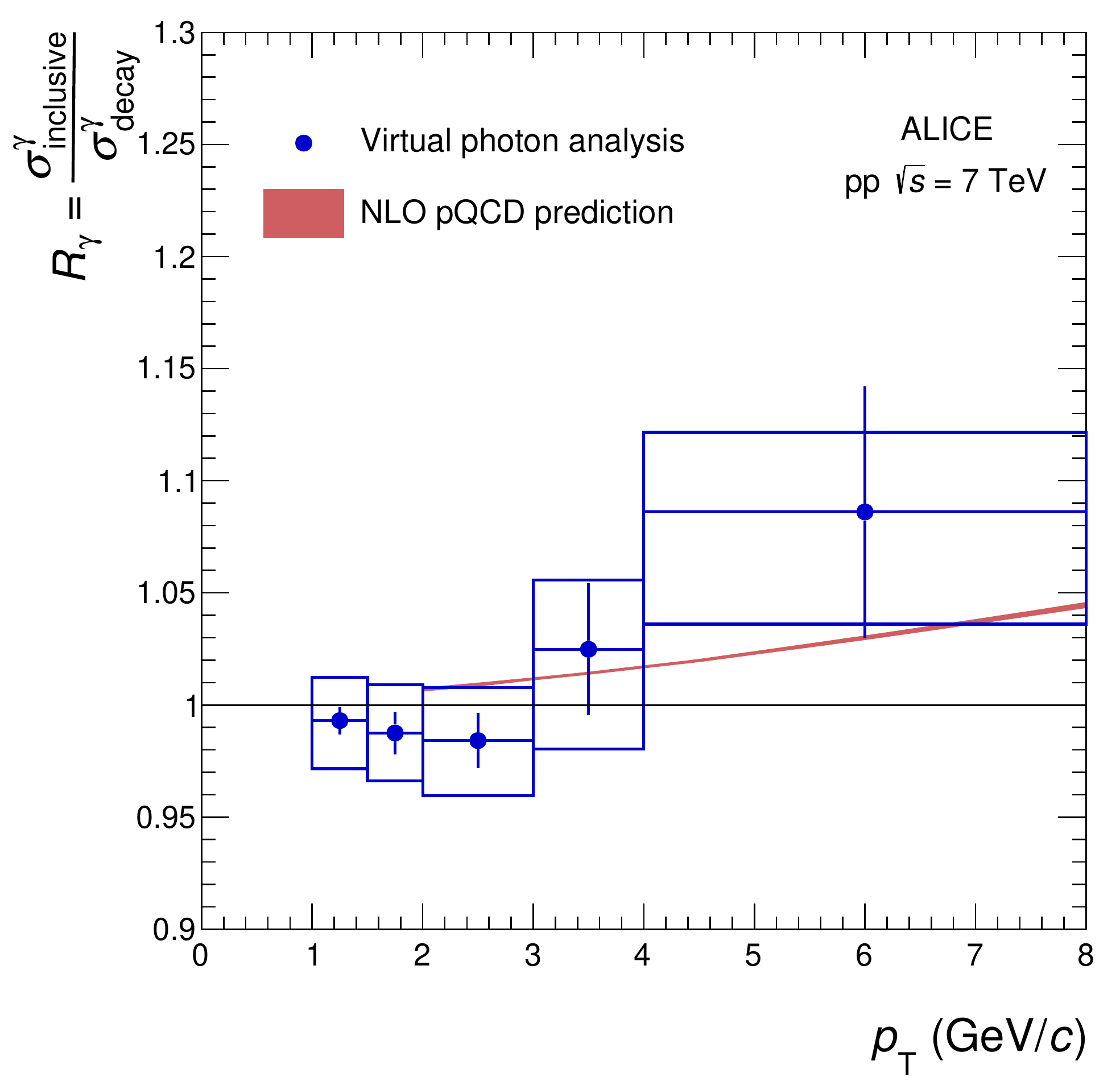}
\caption{Ratio of inclusive to decay photon cross sections extracted
  from the dielectron spectra measured in \pp
  collisions at \roots = 7\,TeV. The results are compared with NLO pQCD
  calculations~\cite{DirectphotonNLOa,Lai:2010vv,Gao:2013xoa,Guzzi:2011sv}. Statistical
  and systematic uncertainties are shown as vertical bars and boxes
  for the data, respectively, and as a band for the NLO pQCD calculations.}
\label{fig:virtualphotona}
\end{figure}

The following sources of systematic uncertainty are considered: (1) the fit
range, (2) the systematic uncertainties of the data, (3) the ones of the hadronic cocktail
components, and (4) the normalisation range. The fit is thus repeated in
different mass intervals between 0.09\,GeV/$c^{2}$ and 0.39\,GeV/$c^{2}$. The
corresponding uncertainty is found to be relevant only in the
$p_{{\rm T}}$ intervals 2--3\,\GeVoverc and 3--4\,GeV/$c$. The uncertainty
arising from the systematic uncertainties of the data is evaluated by shifting all
data points coherently to the upper and lower limits of their systematic
uncertainties and by repeating the fit procedure. The systematic uncertainties from the light-flavour and heavy-flavour cocktail
components are similarly estimated. The contribution of each light-flavour dielectron source
is moved separately to its upper and lower systematic
uncertainties. The largest source of uncertainty originates from
the $\eta$$/$$\pi^{0}$ ratio. In most of the $p_{{\rm T,ee}}$ bins, this is the dominant source of systematic uncertainties. Finally, $f_{\rm LF}$ and $f_{\rm dir}$ are normalised
to the data in the range 0 $<$ \mee $<$ 0.09\,GeV/$c^{2}$ to evaluate
the normalisation uncertainty. All systematic uncertainties are added in quadrature to obtain the total
systematic uncertainty. In Fig.~\ref{fig:virtualphotona}, the
ratio of inclusive to decay photon cross sections,
i.e.\,$R_{\gamma} = \sigma^{\gamma}_{{\rm inclusive}}/\sigma^{\gamma}_{{\rm
    decay}} = 1/(1-r)$, is shown as a function of
$p_{{\rm T}}$.  It is consistent with unity within the statistical and systematic
uncertainties. Perturbative QCD calculations at NLO~\cite{DirectphotonNLOa} performed
with the
CT10 PDFs~\cite{Lai:2010vv,Gao:2013xoa,Guzzi:2011sv}
predict a small ratio of
inclusive to decay photon cross sections over the measured $p_{{\rm
    T}}$ range, compatible with the measurements within uncertainties. The uncertainty band of the calculation is given by the
simultaneous variation of the factorisation, renormalisation, and
fragmentation scales (with $0.5p_{\rm T} < \mu_{\rm F} < 2 p_{\rm T}$
for the factorization scale) used in the calculation. The upper limits at 90\% confidence level (C.L.) on
$R_{\gamma}$, extracted with the Feldman-Cousins
method~\cite{FeldmanCousins}, are summarised in
Table.~\ref{table:upperlimits}. Gaussian distributions are assumed for
statistical and systematic uncertainties, which are treated
independently and summed in quadrature. The results are consistent
with the $R_{\gamma}$ measured in pp collisions at $\sqrt{s} =$
8\,TeV with the real photon analysis performed by the ALICE Collaboration~\cite{directphoton2768}.

\begin{table}[ht!]
\begin{center}
  \centering
  \begin{tabular}{ll}
    \hline \hline
 $p_{{\rm T}}$ interval   & Upper limits at 90\% C.L. \\
               & on $R_{\gamma} = \sigma^{\gamma}_{{\rm
      inclusive}}/\sigma^{\gamma}_{{\rm decay}}$\\\hline
    1 $<$ $p_{{\rm T}}$ $<$ 1.5\,GeV/$c$  & 1.035  \\\hline
    1.5 $<$ $p_{{\rm T}}$ $<$ 2\,GeV/$c$ &  1.027 \\\hline
    2 $<$ $p_{{\rm T}}$ $<$ 3\,GeV/$c$ & 1.030  \\\hline
    3 $<$ $p_{{\rm T}}$ $<$ 4\,GeV/$c$ &  1.096\\\hline
    4 $<$ $p_{{\rm T}}$ $<$ 8\,GeV/$c$ & 1.197 \\
\hline \hline
\end{tabular}
\caption{Upper limits at 90\% C.L. on the ratio of
    inclusive to decay photon cross sections.}
\label{table:upperlimits}
\end{center}
\end{table}
               %%%%%%%%%%% put the body of the article here

\section{Conclusion} \label{section6}

A measurement of \ee pair production at
mid-rapidity ($|\eta_{{\rm e}}|$ $<$ 0.8) in minimum bias \pp collisions at
\mbox{$\sqrt{s}$ = 7\,TeV} with ALICE at the LHC is shown. The results are
presented as a function of the invariant mass \mee \mbox{(0 $<$
\mee $<$ 3.3\,GeV/$c^{2}$)} of the \ee pair, its
transverse momentum \ptee \mbox{(0 $<$ \ptee $<$ 8\,GeV/$c$)}, and the pair transverse impact parameter DCA$_{{\rm ee}}$.

The data are compared with a hadronic cocktail composed of the
expected dielectron cross sections from the known hadronic sources. The contributions from semileptonic
decays of heavy-flavour hadrons are calculated with PYTHIA and POWHEG, and normalised to the measured
total \ccbar and \bbbar cross sections~\cite{totalccbar,totalbbbar}. The shape of the \dcaee
distribution of each source is obtained using a full simulation of the
ALICE detector. The obtained \dcaee templates are normalised to the cocktail
calculations integrated over the same \mee and \ptee range.

Overall good agreement between data and cocktail is observed for
all $m_{{\rm ee}}$, $p_{\rm T,ee}$, and \dcaee intervals considered. In the $\pi^{0}$ mass
region (\mee $<$ 0.14\,GeV/$c^{2}$), the comparison of the measured \dcaee distribution with the
MC templates shows that the detector resolution is well
reproduced in the simulations. In the low-mass region (0.14 $<$ \mee
$<$ 1.1\,GeV/$c^{2}$), prompt and non-prompt contributions can be
separated with the \dcaee observable. In the intermediate-mass region, 1.1 $<$ \mee $<$ 2.7\,GeV/$c^{2}$,
the measured \ee cross section is dominated by correlated \ee pairs
from charm- and beauty-hadron decays. The \ccbar and \bbbar
total cross sections can be extracted from the data  by a
double-differential fit of the measured spectra in ($m_{{\rm ee}}$, $p_{{\rm T,ee}}$) and by fitting the \dcaee
distribution in the IMR. Both fits give consistent results within
statistical and systematic uncertainties. The extracted cross sections
show a large model-dependence between PYTHIA and POWHEG by up to a
factor of two. In the $J/\psi$ mass region \mbox{(2.7 $<$ \mee $<$
3.3\,GeV/$c^{2}$)}, the measured ${\rm DCA}_{{\rm ee}}$- and $p_{{\rm T,ee}}$-differential cross sections are well described by the hadronic cocktail. The \dcaee distribution is
sensitive to the fraction $f_{\rm B}$ of non-prompt $J/\psi$ originating from B-meson
decays. The data are consistent with the previously measured $f_{\rm B}$ by the ALICE Collaboration~\cite{totalbbbara}.

In the quasi-real virtual-photon region,  at low mass (\mee $<$
0.4\,GeV/$c^{2}$) and high \ptee (\ptee $>$ 1\,GeV/$c$), the
contribution of virtual direct photons is extracted from the data by fitting the \mee
distributions in \ptee bins. The extracted ratio of the inclusive-to-decay
photon cross sections is found to be consistent with predictions from pQCD calculations at NLO~\cite{DirectphotonNLOa,Lai:2010vv,Gao:2013xoa,Guzzi:2011sv} within statistical and
systematic uncertainties.

The \ee pair production will be further studied in pp, p$-$Pb
and Pb--Pb collisions with the LHC {\mbox{Run 2}} data, which are currently recorded, as well as with the expected high-statistics data from the LHC {\mbox{Run
3}} starting in 2021~\cite{aliceupgrade,tpcupgrade,tpcupgradebis,itsupgrade}. In particular, the measurement of the
\dcaee spectra in Run 3 will benefit from the new Inner Tracking
System~\cite{itsupgrade} with a smaller material budget and a
resulting higher impact parameter resolution, while the upgrade of the Time Projection
Chamber will provide a significant increase of the statistics~\cite{tpcupgrade,tpcupgradebis}.

%\newpage

%
%
%\input{}               %%%%%%%%%%% put the body of the article here
%
%

%%%%% acknowledgements
\newenvironment{acknowledgement}{\relax}{\relax}
\begin{acknowledgement}
\section*{Acknowledgements}
The ALICE collaboration would like to thank Werner Vogelsang for providing the NLO calculations for direct photon production.
% Version: 2018-04-25

The ALICE Collaboration would like to thank all its engineers and technicians for their invaluable contributions to the construction of the experiment and the CERN accelerator teams for the outstanding performance of the LHC complex.
The ALICE Collaboration gratefully acknowledges the resources and support provided by all Grid centres and the Worldwide LHC Computing Grid (WLCG) collaboration.
The ALICE Collaboration acknowledges the following funding agencies for their support in building and running the ALICE detector:
A. I. Alikhanyan National Science Laboratory (Yerevan Physics Institute) Foundation (ANSL), State Committee of Science and World Federation of Scientists (WFS), Armenia;
Austrian Academy of Sciences and Nationalstiftung f\"{u}r Forschung, Technologie und Entwicklung, Austria;
Ministry of Communications and High Technologies, National Nuclear Research Center, Azerbaijan;
Conselho Nacional de Desenvolvimento Cient\'{\i}fico e Tecnol\'{o}gico (CNPq), Universidade Federal do Rio Grande do Sul (UFRGS), Financiadora de Estudos e Projetos (Finep) and Funda\c{c}\~{a}o de Amparo \`{a} Pesquisa do Estado de S\~{a}o Paulo (FAPESP), Brazil;
Ministry of Science \& Technology of China (MSTC), National Natural Science Foundation of China (NSFC) and Ministry of Education of China (MOEC) , China;
Ministry of Science and Education, Croatia;
Ministry of Education, Youth and Sports of the Czech Republic, Czech Republic;
The Danish Council for Independent Research | Natural Sciences, the Carlsberg Foundation and Danish National Research Foundation (DNRF), Denmark;
Helsinki Institute of Physics (HIP), Finland;
Commissariat \`{a} l'Energie Atomique (CEA) and Institut National de Physique Nucl\'{e}aire et de Physique des Particules (IN2P3) and Centre National de la Recherche Scientifique (CNRS), France;
Bundesministerium f\"{u}r Bildung, Wissenschaft, Forschung und Technologie (BMBF) and GSI Helmholtzzentrum f\"{u}r Schwerionenforschung GmbH, Germany;
General Secretariat for Research and Technology, Ministry of Education, Research and Religions, Greece;
National Research, Development and Innovation Office, Hungary;
Department of Atomic Energy Government of India (DAE), Department of Science and Technology, Government of India (DST), University Grants Commission, Government of India (UGC) and Council of Scientific and Industrial Research (CSIR), India;
Indonesian Institute of Science, Indonesia;
Centro Fermi - Museo Storico della Fisica e Centro Studi e Ricerche Enrico Fermi and Istituto Nazionale di Fisica Nucleare (INFN), Italy;
Institute for Innovative Science and Technology , Nagasaki Institute of Applied Science (IIST), Japan Society for the Promotion of Science (JSPS) KAKENHI and Japanese Ministry of Education, Culture, Sports, Science and Technology (MEXT), Japan;
Consejo Nacional de Ciencia (CONACYT) y Tecnolog\'{i}a, through Fondo de Cooperaci\'{o}n Internacional en Ciencia y Tecnolog\'{i}a (FONCICYT) and Direcci\'{o}n General de Asuntos del Personal Academico (DGAPA), Mexico;
Nederlandse Organisatie voor Wetenschappelijk Onderzoek (NWO), Netherlands;
The Research Council of Norway, Norway;
Commission on Science and Technology for Sustainable Development in the South (COMSATS), Pakistan;
Pontificia Universidad Cat\'{o}lica del Per\'{u}, Peru;
Ministry of Science and Higher Education and National Science Centre, Poland;
Korea Institute of Science and Technology Information and National Research Foundation of Korea (NRF), Republic of Korea;
Ministry of Education and Scientific Research, Institute of Atomic Physics and Romanian National Agency for Science, Technology and Innovation, Romania;
Joint Institute for Nuclear Research (JINR), Ministry of Education and Science of the Russian Federation and National Research Centre Kurchatov Institute, Russia;
Ministry of Education, Science, Research and Sport of the Slovak Republic, Slovakia;
National Research Foundation of South Africa, South Africa;
Centro de Aplicaciones Tecnol\'{o}gicas y Desarrollo Nuclear (CEADEN), Cubaenerg\'{\i}a, Cuba and Centro de Investigaciones Energ\'{e}ticas, Medioambientales y Tecnol\'{o}gicas (CIEMAT), Spain;
Swedish Research Council (VR) and Knut \& Alice Wallenberg Foundation (KAW), Sweden;
European Organization for Nuclear Research, Switzerland;
National Science and Technology Development Agency (NSDTA), Suranaree University of Technology (SUT) and Office of the Higher Education Commission under NRU project of Thailand, Thailand;
Turkish Atomic Energy Agency (TAEK), Turkey;
National Academy of  Sciences of Ukraine, Ukraine;
Science and Technology Facilities Council (STFC), United Kingdom;
National Science Foundation of the United States of America (NSF) and United States Department of Energy, Office of Nuclear Physics (DOE NP), United States of America.
    %%%%%%% done by webmaster team
\end{acknowledgement}

%%%%%%%% Bibliography (In case of using bibtex generate the bbl requested by arXiv)
\bibliographystyle{utphys}   % Remember we use title in the biblio
\bibliography{biblio}
%\input {bibliography.tex}

%%%%%%%%% appendix with author list
\newpage
\appendix
\section{The ALICE Collaboration}
\label{app:collab}
% Collaboration: CERN-LHC-ALICE
% Generation Date is 2018-Apr-25

% How to use:
%%%%%%%%% appendix with author list
%\appendix
%\section{The ALICE Collaboration}
%\label{app:collab}
%\input{Alice_Authorslist_XXXX-Axx-XX.tex}
\begingroup
\small
\begin{flushleft}
S.~Acharya\Irefn{org139}\And 
F.T.-.~Acosta\Irefn{org22}\And 
D.~Adamov\'{a}\Irefn{org94}\And 
J.~Adolfsson\Irefn{org81}\And 
M.M.~Aggarwal\Irefn{org98}\And 
G.~Aglieri Rinella\Irefn{org36}\And 
M.~Agnello\Irefn{org33}\And 
N.~Agrawal\Irefn{org49}\And 
Z.~Ahammed\Irefn{org139}\And 
S.U.~Ahn\Irefn{org77}\And 
S.~Aiola\Irefn{org144}
% ...
\And
\renewcommand\labelenumi{\textsuperscript{\theenumi}~}

\section*{Affiliation notes}
\renewcommand\theenumi{\roman{enumi}}
\begin{Authlist}
\item \Adef{org*}Deceased
\item \Adef{orgI}Dipartimento DET del Politecnico di Torino, Turin, Italy
\item \Adef{orgII}M.V. Lomonosov Moscow State University, D.V. Skobeltsyn Institute of Nuclear, Physics, Moscow, Russia
\item \Adef{orgIII}Department of Applied Physics, Aligarh Muslim University, Aligarh, India
\item \Adef{orgIV}Institute of Theoretical Physics, University of Wroclaw, Poland
\end{Authlist}

\section*{Collaboration Institutes}
\renewcommand\theenumi{\arabic{enumi}~}
\begin{Authlist}
\item \Idef{org1}A.I. Alikhanyan National Science Laboratory (Yerevan Physics Institute) Foundation, Yerevan, Armenia
\item \Idef{org2}Benem\'{e}rita Universidad Aut\'{o}noma de Puebla, Puebla, Mexico
\item \Idef{org3}Bogolyubov Institute for Theoretical Physics, National Academy of Sciences of Ukraine, Kiev, Ukraine
\item \Idef{org4}Bose Institute, Department of Physics  and Centre for Astroparticle Physics and Space Science (CAPSS), Kolkata, India
\item \Idef{org5}Budker Institute for Nuclear Physics, Novosibirsk, Russia
\item \Idef{org6}California Polytechnic State University, San Luis Obispo, California, United States
\item \Idef{org7}Central China Normal University, Wuhan, China
\item \Idef{org8}Centre de Calcul de l'IN2P3, Villeurbanne, Lyon, France
\item \Idef{org9}Centro de Aplicaciones Tecnol\'{o}gicas y Desarrollo Nuclear (CEADEN), Havana, Cuba
\item \Idef{org10}Centro de Investigaci\'{o}n y de Estudios Avanzados (CINVESTAV), Mexico City and M\'{e}rida, Mexico
\item \Idef{org11}Centro Fermi - Museo Storico della Fisica e Centro Studi e Ricerche ``Enrico Fermi', Rome, Italy
\item \Idef{org12}Chicago State University, Chicago, Illinois, United States
\item \Idef{org13}China Institute of Atomic Energy, Beijing, China
\item \Idef{org14}Chonbuk National University, Jeonju, Republic of Korea
\item \Idef{org15}Comenius University Bratislava, Faculty of Mathematics, Physics and Informatics, Bratislava, Slovakia
\item \Idef{org16}COMSATS Institute of Information Technology (CIIT), Islamabad, Pakistan
\item \Idef{org17}Creighton University, Omaha, Nebraska, United States
\item \Idef{org18}Department of Physics, Aligarh Muslim University, Aligarh, India
\item \Idef{org19}Department of Physics, Ohio State University, Columbus, Ohio, United States
\item \Idef{org20}Department of Physics, Pusan National University, Pusan, Republic of Korea
\item \Idef{org21}Department of Physics, Sejong University, Seoul, Republic of Korea
\item \Idef{org22}Department of Physics, University of California, Berkeley, California, United States
\item \Idef{org23}Department of Physics, University of Oslo, Oslo, Norway
\item \Idef{org24}Department of Physics and Technology, University of Bergen, Bergen, Norway
\item \Idef{org25}Dipartimento di Fisica dell'Universit\`{a} 'La Sapienza' and Sezione INFN, Rome, Italy
\item \Idef{org26}Dipartimento di Fisica dell'Universit\`{a} and Sezione INFN, Cagliari, Italy
\item \Idef{org27}Dipartimento di Fisica dell'Universit\`{a} and Sezione INFN, Trieste, Italy
\item \Idef{org28}Dipartimento di Fisica dell'Universit\`{a} and Sezione INFN, Turin, Italy
\item \Idef{org29}Dipartimento di Fisica e Astronomia dell'Universit\`{a} and Sezione INFN, Bologna, Italy
\item \Idef{org30}Dipartimento di Fisica e Astronomia dell'Universit\`{a} and Sezione INFN, Catania, Italy
\item \Idef{org31}Dipartimento di Fisica e Astronomia dell'Universit\`{a} and Sezione INFN, Padova, Italy
\item \Idef{org32}Dipartimento di Fisica `E.R.~Caianiello' dell'Universit\`{a} and Gruppo Collegato INFN, Salerno, Italy
\item \Idef{org33}Dipartimento DISAT del Politecnico and Sezione INFN, Turin, Italy
\item \Idef{org34}Dipartimento di Scienze e Innovazione Tecnologica dell'Universit\`{a} del Piemonte Orientale and INFN Sezione di Torino, Alessandria, Italy
\item \Idef{org35}Dipartimento Interateneo di Fisica `M.~Merlin' and Sezione INFN, Bari, Italy
\item \Idef{org36}European Organization for Nuclear Research (CERN), Geneva, Switzerland
\item \Idef{org37}Faculty of Electrical Engineering, Mechanical Engineering and Naval Architecture, University of Split, Split, Croatia
\item \Idef{org38}Faculty of Engineering and Science, Western Norway University of Applied Sciences, Bergen, Norway
\item \Idef{org39}Faculty of Nuclear Sciences and Physical Engineering, Czech Technical University in Prague, Prague, Czech Republic
\item \Idef{org40}Faculty of Science, P.J.~\v{S}af\'{a}rik University, Ko\v{s}ice, Slovakia
\item \Idef{org41}Frankfurt Institute for Advanced Studies, Johann Wolfgang Goethe-Universit\"{a}t Frankfurt, Frankfurt, Germany
\item \Idef{org42}Gangneung-Wonju National University, Gangneung, Republic of Korea
\item \Idef{org43}Gauhati University, Department of Physics, Guwahati, India
\item \Idef{org44}Helmholtz-Institut f\"{u}r Strahlen- und Kernphysik, Rheinische Friedrich-Wilhelms-Universit\"{a}t Bonn, Bonn, Germany
\item \Idef{org45}Helsinki Institute of Physics (HIP), Helsinki, Finland
\item \Idef{org46}Hiroshima University, Hiroshima, Japan
\item \Idef{org47}Hochschule Worms, Zentrum  f\"{u}r Technologietransfer und Telekommunikation (ZTT), Worms, Germany
\item \Idef{org48}Horia Hulubei National Institute of Physics and Nuclear Engineering, Bucharest, Romania
\item \Idef{org49}Indian Institute of Technology Bombay (IIT), Mumbai, India
\item \Idef{org50}Indian Institute of Technology Indore, Indore, India
\item \Idef{org51}Indonesian Institute of Sciences, Jakarta, Indonesia
\item \Idef{org52}INFN, Laboratori Nazionali di Frascati, Frascati, Italy
\item \Idef{org53}INFN, Sezione di Bari, Bari, Italy
\item \Idef{org54}INFN, Sezione di Bologna, Bologna, Italy
\item \Idef{org55}INFN, Sezione di Cagliari, Cagliari, Italy
\item \Idef{org56}INFN, Sezione di Catania, Catania, Italy
\item \Idef{org57}INFN, Sezione di Padova, Padova, Italy
\item \Idef{org58}INFN, Sezione di Roma, Rome, Italy
\item \Idef{org59}INFN, Sezione di Torino, Turin, Italy
\item \Idef{org60}INFN, Sezione di Trieste, Trieste, Italy
\item \Idef{org61}Inha University, Incheon, Republic of Korea
\item \Idef{org62}Institut de Physique Nucl\'{e}aire d'Orsay (IPNO), Institut National de Physique Nucl\'{e}aire et de Physique des Particules (IN2P3/CNRS), Universit\'{e} de Paris-Sud, Universit\'{e} Paris-Saclay, Orsay, France
\item \Idef{org63}Institute for Nuclear Research, Academy of Sciences, Moscow, Russia
\item \Idef{org64}Institute for Subatomic Physics, Utrecht University/Nikhef, Utrecht, Netherlands
\item \Idef{org65}Institute for Theoretical and Experimental Physics, Moscow, Russia
\item \Idef{org66}Institute of Experimental Physics, Slovak Academy of Sciences, Ko\v{s}ice, Slovakia
\item \Idef{org67}Institute of Physics, Bhubaneswar, India
\item \Idef{org68}Institute of Physics of the Czech Academy of Sciences, Prague, Czech Republic
\item \Idef{org69}Institute of Space Science (ISS), Bucharest, Romania
\item \Idef{org70}Institut f\"{u}r Kernphysik, Johann Wolfgang Goethe-Universit\"{a}t Frankfurt, Frankfurt, Germany
\item \Idef{org71}Instituto de Ciencias Nucleares, Universidad Nacional Aut\'{o}noma de M\'{e}xico, Mexico City, Mexico
\item \Idef{org72}Instituto de F\'{i}sica, Universidade Federal do Rio Grande do Sul (UFRGS), Porto Alegre, Brazil
\item \Idef{org73}Instituto de F\'{\i}sica, Universidad Nacional Aut\'{o}noma de M\'{e}xico, Mexico City, Mexico
\item \Idef{org74}iThemba LABS, National Research Foundation, Somerset West, South Africa
\item \Idef{org75}Johann-Wolfgang-Goethe Universit\"{a}t Frankfurt Institut f\"{u}r Informatik, Fachbereich Informatik und Mathematik, Frankfurt, Germany
\item \Idef{org76}Joint Institute for Nuclear Research (JINR), Dubna, Russia
\item \Idef{org77}Korea Institute of Science and Technology Information, Daejeon, Republic of Korea
\item \Idef{org78}KTO Karatay University, Konya, Turkey
\item \Idef{org79}Laboratoire de Physique Subatomique et de Cosmologie, Universit\'{e} Grenoble-Alpes, CNRS-IN2P3, Grenoble, France
\item \Idef{org80}Lawrence Berkeley National Laboratory, Berkeley, California, United States
\item \Idef{org81}Lund University Department of Physics, Division of Particle Physics, Lund, Sweden
\item \Idef{org82}Nagasaki Institute of Applied Science, Nagasaki, Japan
\item \Idef{org83}Nara Women{'}s University (NWU), Nara, Japan
\item \Idef{org84}National and Kapodistrian University of Athens, School of Science, Department of Physics , Athens, Greece
\item \Idef{org85}National Centre for Nuclear Research, Warsaw, Poland
\item \Idef{org86}National Institute of Science Education and Research, HBNI, Jatni, India
\item \Idef{org87}National Nuclear Research Center, Baku, Azerbaijan
\item \Idef{org88}National Research Centre Kurchatov Institute, Moscow, Russia
\item \Idef{org89}Niels Bohr Institute, University of Copenhagen, Copenhagen, Denmark
\item \Idef{org90}Nikhef, National institute for subatomic physics, Amsterdam, Netherlands
\item \Idef{org91}NRC Kurchatov Institute IHEP, Protvino, Russia
\item \Idef{org92}NRNU Moscow Engineering Physics Institute, Moscow, Russia
\item \Idef{org93}Nuclear Physics Group, STFC Daresbury Laboratory, Daresbury, United Kingdom
\item \Idef{org94}Nuclear Physics Institute of the Czech Academy of Sciences, \v{R}e\v{z} u Prahy, Czech Republic
\item \Idef{org95}Oak Ridge National Laboratory, Oak Ridge, Tennessee, United States
\item \Idef{org96}Petersburg Nuclear Physics Institute, Gatchina, Russia
\item \Idef{org97}Physics department, Faculty of science, University of Zagreb, Zagreb, Croatia
\item \Idef{org98}Physics Department, Panjab University, Chandigarh, India
\item \Idef{org99}Physics Department, University of Jammu, Jammu, India
\item \Idef{org100}Physics Department, University of Rajasthan, Jaipur, India
\item \Idef{org101}Physikalisches Institut, Eberhard-Karls-Universit\"{a}t T\"{u}bingen, T\"{u}bingen, Germany
\item \Idef{org102}Physikalisches Institut, Ruprecht-Karls-Universit\"{a}t Heidelberg, Heidelberg, Germany
\item \Idef{org103}Physik Department, Technische Universit\"{a}t M\"{u}nchen, Munich, Germany
\item \Idef{org104}Research Division and ExtreMe Matter Institute EMMI, GSI Helmholtzzentrum f\"ur Schwerionenforschung GmbH, Darmstadt, Germany
\item \Idef{org105}Rudjer Bo\v{s}kovi\'{c} Institute, Zagreb, Croatia
\item \Idef{org106}Russian Federal Nuclear Center (VNIIEF), Sarov, Russia
\item \Idef{org107}Saha Institute of Nuclear Physics, Kolkata, India
\item \Idef{org108}School of Physics and Astronomy, University of Birmingham, Birmingham, United Kingdom
\item \Idef{org109}Secci\'{o}n F\'{\i}sica, Departamento de Ciencias, Pontificia Universidad Cat\'{o}lica del Per\'{u}, Lima, Peru
\item \Idef{org110}Shanghai Institute of Applied Physics, Shanghai, China
\item \Idef{org111}Stefan Meyer Institut f\"{u}r Subatomare Physik (SMI), Vienna, Austria
\item \Idef{org112}SUBATECH, IMT Atlantique, Universit\'{e} de Nantes, CNRS-IN2P3, Nantes, France
\item \Idef{org113}Suranaree University of Technology, Nakhon Ratchasima, Thailand
\item \Idef{org114}Technical University of Ko\v{s}ice, Ko\v{s}ice, Slovakia
\item \Idef{org115}Technische Universit\"{a}t M\"{u}nchen, Excellence Cluster 'Universe', Munich, Germany
\item \Idef{org116}The Henryk Niewodniczanski Institute of Nuclear Physics, Polish Academy of Sciences, Cracow, Poland
\item \Idef{org117}The University of Texas at Austin, Austin, Texas, United States
\item \Idef{org118}Universidad Aut\'{o}noma de Sinaloa, Culiac\'{a}n, Mexico
\item \Idef{org119}Universidade de S\~{a}o Paulo (USP), S\~{a}o Paulo, Brazil
\item \Idef{org120}Universidade Estadual de Campinas (UNICAMP), Campinas, Brazil
\item \Idef{org121}Universidade Federal do ABC, Santo Andre, Brazil
\item \Idef{org122}University College of Southeast Norway, Tonsberg, Norway
\item \Idef{org123}University of Cape Town, Cape Town, South Africa
\item \Idef{org124}University of Houston, Houston, Texas, United States
\item \Idef{org125}University of Jyv\"{a}skyl\"{a}, Jyv\"{a}skyl\"{a}, Finland
\item \Idef{org126}University of Liverpool, Department of Physics Oliver Lodge Laboratory , Liverpool, United Kingdom
\item \Idef{org127}University of Tennessee, Knoxville, Tennessee, United States
\item \Idef{org128}University of the Witwatersrand, Johannesburg, South Africa
\item \Idef{org129}University of Tokyo, Tokyo, Japan
\item \Idef{org130}University of Tsukuba, Tsukuba, Japan
\item \Idef{org131}Universit\'{e} Clermont Auvergne, CNRS/IN2P3, LPC, Clermont-Ferrand, France
\item \Idef{org132}Universit\'{e} de Lyon, Universit\'{e} Lyon 1, CNRS/IN2P3, IPN-Lyon, Villeurbanne, Lyon, France
\item \Idef{org133}Universit\'{e} de Strasbourg, CNRS, IPHC UMR 7178, F-67000 Strasbourg, France, Strasbourg, France
\item \Idef{org134} Universit\'{e} Paris-Saclay Centre d¿\'Etudes de Saclay (CEA), IRFU, Department de Physique Nucl\'{e}aire (DPhN), Saclay, France
\item \Idef{org135}Universit\`{a} degli Studi di Foggia, Foggia, Italy
\item \Idef{org136}Universit\`{a} degli Studi di Pavia, Pavia, Italy
\item \Idef{org137}Universit\`{a} di Brescia, Brescia, Italy
\item \Idef{org138}V.~Fock Institute for Physics, St. Petersburg State University, St. Petersburg, Russia
\item \Idef{org139}Variable Energy Cyclotron Centre, Kolkata, India
\item \Idef{org140}Warsaw University of Technology, Warsaw, Poland
\item \Idef{org141}Wayne State University, Detroit, Michigan, United States
\item \Idef{org142}Westf\"{a}lische Wilhelms-Universit\"{a}t M\"{u}nster, Institut f\"{u}r Kernphysik, M\"{u}nster, Germany
\item \Idef{org143}Wigner Research Centre for Physics, Hungarian Academy of Sciences, Budapest, Hungary
\item \Idef{org144}Yale University, New Haven, Connecticut, United States
\item \Idef{org145}Yonsei University, Seoul, Republic of Korea
\end{Authlist}
\endgroup
  %%%%%%% done by webmaster team
\end{document}